\documentclass[aps,prb,reprint,amsfonts,amssymb,amsmath,superscriptaddress,longbibliography,floatfix]{revtex4-2}

\pdfoutput=1

\usepackage{amssymb,amsmath,bm}
\usepackage{braket}
\usepackage{fullpage}
\usepackage{graphicx}
\usepackage{bibunits}
\usepackage{amsmath}
\usepackage{wrapfig}
\usepackage{mathtools}
\usepackage{floatflt}
\usepackage{color,soul}
\usepackage{xspace}
\usepackage{dsfont}
\usepackage[official,right]{eurosym}
\usepackage[top=0.8in, bottom=0.8in, left=0.75in, right=0.75in]{geometry}
\usepackage[utf8]{inputenc}
\usepackage{natbib}
\usepackage[colorlinks=true,urlcolor=blue,citecolor=blue,linkcolor=blue,bookmarks=false]{hyperref}
\usepackage[flushleft]{threeparttable}
\usepackage{array}
\usepackage{booktabs,makecell,tabularx}
\usepackage{tabularray}
\newcolumntype{Y}{>{\centering\arraybackslash}X}
\usepackage{csquotes}
\MakeOuterQuote{"}
\usepackage{cleveref}
\usepackage[T1]{fontenc}

\newcommand{\dmnrc}{(CD$_3$ND$_3$)$_2$NaRuCl$_6$\xspace}
\newsavebox{\mstrut}
\newcommand{\dbraket}[1]{%
	\sbox{\mstrut}{\(#1\)}%
	\mathinner{\left\langle\kern-0.4\ht\mstrut\left\langle{#1}\right\rangle\kern-0.4\ht\mstrut\right\rangle}%
}

\makeatletter
\newcommand*{\hyperlinkcite}[1]{\hyper@link{cite}{cite.#1}}
\makeatother

\newcommand{\be}{\begin{equation}}
	\newcommand{\ee}{\end{equation} }
\newcommand{\bea}{\begin{eqnarray} }
	\newcommand{\eea}{\end{eqnarray} }

\begin{document}
	
	\title{$\mathbb{Z}_2$ Vortex Crystal Candidate in the Triangular $S=1/2$ Quantum Antiferromagnet}
	
	\author{Jakob~Nagl}
	\email{jnagl@ethz.ch}
	\affiliation{Laboratory for Solid State Physics, ETH Z{\"u}rich, 8093 Z{\"u}rich, Switzerland}
	
	\author{Kirill~Yu.~Povarov}
	\affiliation{Dresden High Magnetic Field Laboratory (HLD-EMFL) and W\"urzburg-Dresden Cluster of Excellence ct.qmat, Helmholtz-Zentrum Dresden-Rossendorf (HZDR), 01328 Dresden, Germany}
	
	\author{Benjamin~Duncan}
	\affiliation{Laboratory for Solid State Physics, ETH Z{\"u}rich, 8093 Z{\"u}rich, Switzerland}
	\author{Catharina~N{\"a}ppi}
	\affiliation{Laboratory for Solid State Physics, ETH Z{\"u}rich, 8093 Z{\"u}rich, Switzerland}
	
	\author{Dmitry~Khalyavin}
	\affiliation{ISIS Facility, Rutherford Appleton Laboratory, Chilton, Didcot, Oxon OX11 0QX, United Kingdom}
	\author{Pascal~Manuel}
	\affiliation{ISIS Facility, Rutherford Appleton Laboratory, Chilton, Didcot, Oxon OX11 0QX, United Kingdom}
	\author{Fabio~Orlandi}
	\affiliation{ISIS Facility, Rutherford Appleton Laboratory, Chilton, Didcot, Oxon OX11 0QX, United Kingdom}
	
	\author{Jeremy~Sourd}
	\affiliation{Dresden High Magnetic Field Laboratory (HLD-EMFL) and W\"urzburg-Dresden Cluster of Excellence ct.qmat, Helmholtz-Zentrum Dresden-Rossendorf (HZDR), 01328 Dresden, Germany}
	\author{Beat~Valentin~Schwarze}
	\affiliation{Dresden High Magnetic Field Laboratory (HLD-EMFL) and W\"urzburg-Dresden Cluster of Excellence ct.qmat, Helmholtz-Zentrum Dresden-Rossendorf (HZDR), 01328 Dresden, Germany}
	\author{Freya~Husstedt}
	\affiliation{Dresden High Magnetic Field Laboratory (HLD-EMFL) and W\"urzburg-Dresden Cluster of Excellence ct.qmat, Helmholtz-Zentrum Dresden-Rossendorf (HZDR), 01328 Dresden, Germany}
	\affiliation{Institut f\"ur Festk\"orper- und Materialphysik, Technische Universit\"at Dresden, 01062 Dresden, Germany}
	\author{Sergei~A.~Zvyagin}
	\affiliation{Dresden High Magnetic Field Laboratory (HLD-EMFL) and W\"urzburg-Dresden Cluster of Excellence ct.qmat, Helmholtz-Zentrum Dresden-Rossendorf (HZDR), 01328 Dresden, Germany}
	
	\author{Oksana~Zaharko}
	\affiliation{PSI Center for Neutron and Muon Sciences, Forschungsstrasse 111, 5232 Villigen, PSI, Switzerland}
	
	\author{Paul~Steffens}
	\affiliation{Institut Laue-Langevin, 71 avenue des Martyrs, CS 20156, 38042 Grenoble Cedex 9, France}
	\author{Arno~Hiess}
	\affiliation{Institut Laue-Langevin, 71 avenue des Martyrs, CS 20156, 38042 Grenoble Cedex 9, France}
	\affiliation{European Spallation Source ERIC, P.O. Box 176, 22100 Lund, Sweden}
	
	\author{David~R.~Allan}
	\affiliation{Diamond Light Source, Harwell Science and Innovation Campus, Didcot, Oxfordshire OX11 0DE, UK}
	\author{Sarah~A.~Barnett}
	\affiliation{Diamond Light Source, Harwell Science and Innovation Campus, Didcot, Oxfordshire OX11 0DE, UK}
	
	\author{Zewu~Yan}
	\affiliation{Laboratory for Solid State Physics, ETH Z{\"u}rich, 8093 Z{\"u}rich, Switzerland}
	
	\author{Severian~Gvasaliya}
	\affiliation{Laboratory for Solid State Physics, ETH Z{\"u}rich, 8093 Z{\"u}rich, Switzerland}
	
	\author{Andrey~Zheludev}
	\email{zhelud@ethz.ch}
	\homepage{http://www.neutron.ethz.ch/}
	\affiliation{Laboratory for Solid State Physics, ETH Z{\"u}rich, 8093 Z{\"u}rich, Switzerland}
	
	\date{\today}

	\begin{abstract}
		The prospect of merging the paradigms of geometric frustration on a triangular lattice and bond anisotropies in the strong spin-orbit coupling limit holds tremendous promise in the search for exotic quantum materials. 
		Here we identify a new candidate system to realize such physics, the organic quantum antiferromagnet \dmnrc.
		We report a combination of thermodynamic, magneto-elastic and neutron scattering experiments on single-crystals to determine the phase diagram in axial magnetic fields $\mathbf{H \parallel c}$ and propose a minimal model Hamiltonian.
		\dmnrc displays an ideal triangular arrangement of Ru$^{3+}$ ions adopting the spin-orbital entangled $j_{\rm eff} = 1/2$ state.
		It hosts residual magnetic order below $T_{\rm N} = 0.23$ K and a highly unusual $H-T$ phase diagram including three different incommensurate states.
		Spin-waves in the high-field polarized regime are described by a Heisenberg triangular lattice Hamiltonian with a potential sub-leading bond dependent anisotropy term $J_{\pm\pm}$.
		We argue that the multi-$\mathbf{q}$ ground state in zero magnetic field is a prime candidate for hosting the $\mathbb{Z}_2$ vortex crystal proposed on the triangular Heisenberg-Kitaev model.
		\dmnrc is the first member in an extended family of quantum triangular lattice magnets, providing a new playground to study the interplay of geometric frustration and spin-orbit effects.
	\end{abstract}
	
	\maketitle
	
	\section{Introduction}
	
	Quantum antiferromagnets on a $S=1/2$ frustrated triangular lattice pose one of the oldest and most influential paradigms in quantum magnetism \cite{wannierAntiferromagnetismTriangularIsing1950, balentsSpinLiquidsFrustrated2010, lacroixIntroductionFrustratedMagnetism2011}. 
	Even the basic Heisenberg model continues to draw considerable interest, as good material prototypes have only recently become available, finally allowing for concrete comparison to theoretical predictions of exchange renormalization and magnon decay \cite{chernyshevSpinWavesTriangular2009, itoStructureMagneticExcitations2017, xieCompleteFieldinducedSpectral2023a}. 
	Away from the isotropic limit, the possibility of a quantum spin liquid ground state remains an intensely debated issue \cite{liSpinLiquidsGeometrically2020, maksimovAnisotropicExchangeMagnetsTriangular2019}, with numerous theoretical predictions and only a handful of promising material candidates. 
	Recent experimental work has focused mostly on Co$^{2+}$ \cite{itoStructureMagneticExcitations2017, zhuWannierStatesSpin2025, woodlandContinuumExcitationsSharp2025} and Yb$^{3+}$ \cite{xieCompleteFieldinducedSpectral2023a, scheieNonlinearMagnonsExchange2024, zhuDisorderInducedMimicrySpin2017} systems, with the heavy $4d$ and $5d$ transition metals going largely unexplored despite their prominence in the context of bond dependent interactions in Kitaev-honeycomb systems \cite{jackeliMottInsulatorsStrong2009, trebstKitaevMaterials2022}. Thus, combining such bond anisotropies with a frustrated lattice topology poses a promising route to realizing new emergent quantum phases.
	
	One such class of exotic ground states - often stabilized by anisotropy and exchange competition - are the topological spin textures \cite{zhouTopologicalSpinTextures2025a}.
	The triangular Heisenberg antiferromagnet has long been known to host $\mathbb{Z}_2$ vortices as topological excitations \cite{kawamuraPhaseTransitionTwoDimensional1984}, particle-like "twists" in the familiar $120^\circ$ three-sublattice order.
	Recently it was realized that they become thermodynamically stable when Kitaev interactions are invoked \cite{rousochatzakisKitaevAnisotropyInduces2016}.
	Indeed, even a tiny Kitaev term $|K|/J \ll 1$ is enough 
	to condense them into the ground state, forming an incommensurate (IC) triangular superlattice of $\mathbb{Z}_2$ vortices.
	This is analogous to the formation of skyrmions in chiral helimagnets \cite{skyrmeUnifiedFieldTheory1962, bogdanovThermodynamicallyStableVortices1989, muhlbauerSkyrmionLatticeChiral2009a, sekiObservationSkyrmionsMultiferroic2012} or Abrikosov vortices in type-II superconductors \cite{abrikosovMagneticPropertiesSuperconductors1957, essmannDirectObservationIndividual1967, hessScanningTunnelingMicroscopeObservationAbrikosov1989}.
	These predictions have garnered intense theoretical interest \cite{beckerSpinorbitPhysicsJ12015, catuneanuMagneticOrdersProximal2015, wangThreeDimensionalCrystallizationVortex2015, yaoStabilizationModulationTopological2018, kishimotoGroundStatePhase2018, liCollectiveSpinDynamics2019, osorioSkyrmions$mathbbZ_2$Vortices2019, seabrook$mathbbZ_2$VorticesGround2020, shinjoDensityMatrixRenormalizationGroup2016}, in part because this phase	appears to be quite robust in the Heisenberg-Kitaev model and may not require excessive fine-tuning of interactions as in the elusive spin-liquids \cite{maksimovAnisotropicExchangeMagnetsTriangular2019}.
	An experimental implementation of the $\mathbb{Z}_2$ vortex crystal currently remains open: Only a scant few $4d/5d$ compounds have been proposed to realize the parent Hamiltonian \cite{beckerSpinorbitPhysicsJ12015, catuneanuMagneticOrdersProximal2015, bhattacharyyaNaRuO2KitaevHeisenbergExchange2023} and none are available in single crystalline form, keeping studies in their infancy.

	\begin{figure}[tbp]
		\includegraphics[scale=1]{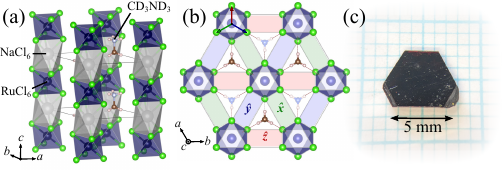}
		\caption{Crystal structure of \dmnrc. (a) Schematic view of the chemical structure, with AA-stacked triangular planes of Ru$^{3+}$ ions. (b) Top-down view of the triangular lattice, emphasizing the three different bond types with parallel-edge RuCl$_6$ octahedra. (c) A typical single-crystal sample of \dmnrc.}
		\label{fig:Crystal_structure}
	\end{figure}
	
	We address this problem by identifying an entirely new candidate material to host such physics, namely the recently discovered organic quantum antiferromagnetic insulator \dmnrc \cite{vishnoiStructuralDiversityMagnetic2020}, crystallizing in a trigonal $P\bar{3}m1$ structure (Fig.~\ref{fig:Crystal_structure}a).
	Its magnetism originates from the low-spin $4d^5$ Ru$^{3+}$ ions with nearly perfect octahedral coordination, leaving an unquenched $l = 1$ orbital moment that should engender spin-orbital entangled $j_{\rm eff} = 1/2$ degrees of freedom at low temperatures.
	The material realizes an AA-stacked triangular lattice of RuCl$_6$ octahedra in the crystallographic $ab$ plane (Fig.~\ref{fig:Crystal_structure}b), with comparable $in$-plane and $inter$-plane distances $a = 7.14$~\AA~and $c = 6.74$~\AA.
	\nocite{supplement}
	As described in the Supplemental Material (\hyperlinkcite{supplement}{S1-4}, and \hyperlinkcite{supplement}{T1-3}), it undergoes a subtle structural transition to $P\bar{3}$ at $T^\star \simeq 118$~K.
	However, the ideal triangular coordination of Ru is preserved and Dzyaloshinskii-Moriya terms remain forbidden by symmetry down to the lowest temperatures.
	The bonds connecting RuCl$_6$ octahedra are not of edge-sharing type as in the famous Kitaev-type honeycomb materials \cite{jackeliMottInsulatorsStrong2009}. 
	Instead they display a parallel-edge geometry, nevertheless with two interfering superexchange pathways and three orthogonal bond types (Fig.~\ref{fig:Crystal_structure}b). A significant Kitaev-term has been proposed for such configurations by several authors \cite{trebstKitaevMaterials2022, beckerSpinorbitPhysicsJ12015, catuneanuMagneticOrdersProximal2015}, but these claims have yet to be verified experimentally due to a lack of material candidates available in single crystalline form.

	In this work, we present a thorough investigation of the magnetic ground states and excitations on single crystals of \dmnrc through a combination of neutron scattering and bulk thermodynamic probes.
	We show that the pseudospin wavefunctions are indeed dominated by the ideal $j_{\rm eff} = 1/2$ state.
	Below $T_{\rm N1} \approx 0.23$~K, \dmnrc exhibits a two-step transition into a long-range ordered ground state.
	The latter turns out to be a complex multi-$\mathbf{q}$ order with sets of three IC Bragg peaks around the K and H-points, making it a strong contender for the $\mathbb{Z}_2$ vortex crystal phase - with the additional complication of weak AFM inter-plane interactions.
	In an axial magnetic field $\mathbf{H \parallel c}$, we observe a cascade of \textit{seven} field-induced ordered phases with several commensurate-incommensurate transitions.
	While the commensurate states in the center of the phase diagram may be understood through order-by-disorder selection on a semi-classical Heisenberg model, the three IC phases towards low/high fields point to the relevance of anisotropies.
	The spin-wave dispersion in the field-polarized regime is captured by a quasi-2D nearest neighbor Heisenberg triangular lattice model with a potential sub-leading bond dependent term $J_{\pm\pm}$.
	\dmnrc is the first member in an entirely new family of quantum triangular lattice magnets based on heavy $4d/5d$ transition metals, providing fertile grounds to explore the incommensurate/topological phases emerging from the interplay between geometric frustration and spin-orbit coupling.

	\section{Results}
	
	\subsection{Basic Properties \& Single-Ion Physics}
	
\begin{figure}[tbp]
		\includegraphics[scale=1]{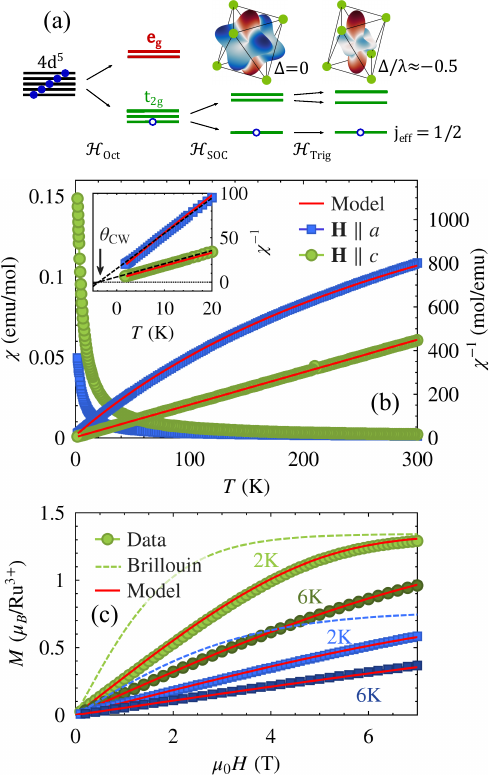}
		\caption{Single-ion physics and mean-field correlations in \dmnrc.
		(a) Schematic of the energy spectrum for a single Ru$^{3+}$ ion. 
		The free-ion multiplet is split by the cubic crystal electric field, the spin-orbit coupling $\lambda$ and a trigonal distortion $\Delta$.
		An inset visualizes the spatial shape of the pseudospin wavefunctions, both for the ideal cubic case ($\Delta = 0$) and for the parameters $\Delta/\lambda \approx -0.5$ relevant to \dmnrc.
		Red/blue color indicates spin up/down of the hole, mixed together in the spin-orbital entangled $j_{\rm eff} = 1/2$ ground state.
		(b) Magnetic susceptibility for a small probing field $\mu_0 H = 0.1$~T along the principal crystallographic axes. 
		The inset shows a low-temperature Curie-Weiss fit.
		(c) Magnetization curves at various temperatures (markers), along with Brillouin function expected in absence of two-ion correlations (dashed lines).
		Red lines in (b,c) represent the calculated magnetic response based on the single-ion + mean-field model discussed in the text.}
		\label{fig:Single_Ion}
	\end{figure}

	In Fig.~\ref{fig:Single_Ion}(b) we present the uniform magnetic susceptibility of \dmnrc along the principal crystallographic axes.	
	For $\mathbf{H \parallel c}$ we observe a linear Curie-Weiss behavior up to room temperature in the inverse susceptibility, whereas for in-plane fields, $\chi_{ab}^{-1}(T)$ exhibits a strong non-linear bending, indicative of crystal-field effects.
	The data exhibit a pronounced easy-axis anisotropy, reaching $\chi_c / \chi_{ab} \sim 2.3$ in the low-temperature limit.
	This is likewise reflected in the effective moments $\mu_{\rm eff} = g \sqrt{S(S+1)} \mu_{\rm B}$, pointing to anisotropic $S_{\rm eff} = 1/2$ degrees of freedom with $g_c / g_{ab} \sim 1.6$, typical for systems with an unquenched orbital contribution.
	A simple Curie-Weiss analysis in the range $10-40$~K results in an average Weiss temperature of $\theta_{\rm CW} \approx -3.8$~K (see Table~\ref{table:Curie} for details), indicative of moderate antiferromagnetic interactions.
	Note the limited fitting range (chosen empirically to remain below the temperature at which curvature in $\chi_{ab}^{-1}$ becomes apparent), which permits only a rough estimate of $\theta_{\rm CW}$ and highlights the need for a comprehensive treatment of the single-ion physics (see below).
	Magnetometry data on the same sample [see Fig.~\ref{fig:Single_Ion}(c)] begin to show significant deviations from a non-interacting Brillouin function below $\lesssim 10$~K, consistent with an onset of short-range spin correlations.
	
	\begin{table}[b]
		\caption{Fit results for a Curie--Weiss analysis in the temperature range 10 K $\leq T \leq$ 40 K.}
		\centering
		\begin{tblr}{
				width = 0.45\textwidth,
				colspec = {X[c] X[c] X[c] X[c]},
				rowsep = 0.6ex,
				hline{1,Z} = {1}{-}{solid}, 
				hline{2},
				hline{1,Z} = {2}{-}{solid}
			}
			& $\theta_{\rm CW}$ (K) & $\mu_{\rm eff}$ ($\mu_{\rm B}$) & $g$ \\
			$\mathbf{H}\parallel\mathit{a}$ & -3.82 & 1.42 & 1.64 \\
			$\mathbf{H}\parallel\mathit{c}$ & -3.93 & 2.32 & 2.68 \\
		\end{tblr}
		\label{table:Curie}
	\end{table}
	
	\begin{figure}[tbp]
		\includegraphics[scale=1]{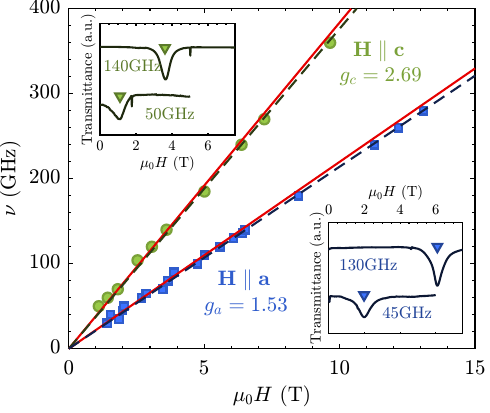}
		\caption{Frequency-field diagram of ESR excitations measured at $T = 1.5$~K for both $\mathbf{H \parallel c}$ (green) and $\mathbf{H \parallel a}$ (blue) field orientations. 
		Dashed lines represent linear fits to the function $h\nu = g\mu_{\rm B} \mu_0 H$, while solid red lines show calculations based on the single-ion + mean-field model discussed in the text.
		ESR spectra at selected frequencies are presented as insets.
		Characteristic dips in transmittance correspond to the Larmor resonant field at given frequency, while sharp features are caused by the $g = 2.00$ DPPH marker.}
		\label{fig:ESR}
	\end{figure}
		
	More precise $g$-factor values are obtained from electron spin resonance (ESR) measurements. 
	For both longitudinal and transverse field geometries, we observe a single  resonance line at $T=1.5$~K, as illustrated in Fig.~\ref{fig:ESR} - a rare example of resonance in a strongly correlated system of Ru$^{3+}$ magnetic ions~\cite{ponomaryov_PRL_NatureMagExc2013}.
	The frequency-field dependence is linear, suggesting simple Larmor precession at these temperatures.
	Fits to $h \nu = g\mu_{\rm B} \mu_0 H$ yield $g_{ab} = 1.53(1)$ and $g_c = 2.69(1)$, nicely matching our estimates from susceptibility.
	We detect no apparent temperature dependence of the $g$-factors, as confirmed by ESR data up to $T \sim 90$~K (see Supplemental Material \hyperlinkcite{supplement}{S6}).
	Likewise, the spin gap in the field-polarized regime - extracted from heat capacity and inelastic neutron scattering - agrees well with $g_c$ down to dilution temperatures (see below).
		
	For Ru$^{3+}$ transition metal ions with $4d^5$ electronic configuration, the octahedral crystal field results in a single hole residing in the $t_{2g}$ manifold, forming a spin $S = 1/2$ and an effective orbital moment $l_{\rm eff} = 1$.
	These states are further split by the strong spin-orbit coupling $\lambda \sim 0.15$~eV \cite{geschwindSpinResonanceTransition1962} and by a potential trigonal distortion $\Delta$. 
	This is borne out in the following local single-ion Hamiltonian
	\begin{equation}
		\mathcal{H} = \lambda \hat{\boldsymbol{\ell}} \cdot \mathbf{\hat{S}} + \Delta \hat{\ell}^2_z - \mu_{\rm B} \mathbf{H} \cdot \hat{\mathbf{M}}
	\end{equation}
	where $\mathbf{\hat{M}} = 2\mathbf{\hat{S}} - k \boldsymbol{\hat{\ell}}$ is the on-site magnetization of the hole and $k \lesssim 1$ accounts for a small reduction in orbital moment due to anion hybridization \cite{abragamElectronParamagneticResonance}. 
	In the cubic limit $|\Delta / \lambda| \ll 1$, the ground state is a spin-orbital entangled $j_{\rm eff} = 1/2$ Kramers doublet with equal weights on all three $t_{2g}$ orbitals \cite{liuExchangeInteractionsD52022}.
	
	The Hamiltonian parameters can be extracted from a global fit to the susceptibility, magnetization and ESR data.
	In order to account for incipient two-ion correlations, we introduce a Weiss molecular field $\mathbf{H} \rightarrow \mathbf{H}_{\rm ext} + \mathbf{H}_{\rm MF}$ and solve the resulting eigenvalue problem self-consistently.
	We obtain $\lambda = 153.3$~meV, $\Delta = -77.9$~meV and $J_{\rm MF} = 1.71$~K (for details we refer to Supplemental Material \hyperlinkcite{supplement}{S5} and \hyperlinkcite{supplement}{T4}), where the latter is a mean-field exchange parameter setting the strength of the molecular field.
	Our fit results are shown in Fig.~\ref{fig:Single_Ion} as red lines, confirming excellent agreement among different probes.
	With $\Delta / \lambda \simeq -0.51$, we are still within the spin-orbital entangled regime.
	The ground state wavefunctions remain dominated by the ideal $j_{\rm eff} = 1/2$ state, though the magnetization density acquires a notable easy-axis character through the trigonal distortion, with slightly more weight on the $\ell_z = \pm 1$ states (see Fig.~\ref{fig:Single_Ion}(a)).
	We note such a distortion corresponds to a slight elongation of octahedra along the $c$-axis within a point-charge picture, in agreement with the refined chemical structure \cite{supplement}.
	The data are well described by a mean-field exchange constant $J_{\rm MF}$ of Heisenberg type - which would correspond to $\theta_{\rm CW} \approx -2.6$~K - with no signs of XXZ anisotropy within estimated standard deviations.
	Given the above trigonal distortion, we expect the dominant superexchange scale to be of Heisenberg type \cite{winterModelsMaterialsGeneralized2017}, with potential bond anisotropies only taking a sub-leading role.
	But as we discuss below, even a small bond-dependent term can induce qualitatively new physics on the triangular lattice \cite{beckerSpinorbitPhysicsJ12015, rousochatzakisKitaevAnisotropyInduces2016}.

	\subsection{Calorimetry}
	
\begin{figure}[tbp]
		\includegraphics[scale=1]{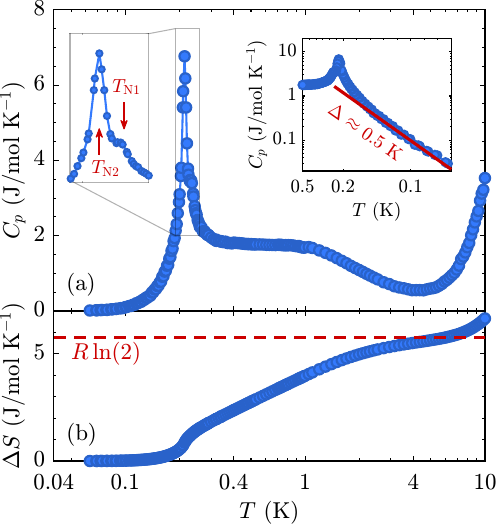}
		\caption{Temperature dependence of specific heat in \dmnrc. 
			(a) Measured heat capacity $C_p(T)$ in zero magnetic field, exhibiting a two-step transition to magnetic long-range order. 
			Left inset: Zoom-in on the critical region, showing two separate anomalies at $T_{\rm N1}$ and $T_{\rm N2}$ K. 
			Right inset: Arrhenius plot depicting the activated behavior in the ordered state.
			(b) Total entropy change $\Delta S (T)$ obtained through numerical integration of the specific heat.
			Note the plateau at $R\ln(2)$, reflecting the separation of energy scales between magnetic and lattice degrees of freedom.
		}
		\label{fig:HC_zf}
	\end{figure}

	In Fig.~\ref{fig:HC_zf}(a) we show the temperature dependence of the specific heat capacity in \dmnrc (note the logarithmic scale). 
	We focus on the low-temperature regime $T \lesssim 4$~K, where the lattice is frozen and only magnetic degrees of freedom remain.
	There is a broad upturn in $C_p(T)$ around $\theta_{\rm CW}$, which we associate with the onset of short-range magnetic correlations.
	It develops into a plateau between 1~K and 0.2~K, followed by a sharp anomaly indicative of a phase transition to magnetic long-range order.
	The latter takes place in two steps: Zooming in close to the critical region reveals both a shoulder feature at $T_{\rm N1} = 0.23$~K and a sharp peak at $T_{\rm N2} = 0.215$~K.
	In the ordered phase, the heat capacity fits either to a (gapless) power-law $C_p \propto T^x$ with $x \approx 4.5$, or to an activated behavior $C_p \propto e^{-\Delta/T}$ with  $\Delta \approx 0.5$~K. 
	While such a strong power-law exponent seems rather unphysical, a small excitation gap of order $\sim 0.5$~K is consistent with the magnetic energy scale.
	This points to a gapped ground state, perhaps hinting at some anisotropy terms that break the $\mathcal{O}(3)$ symmetry in spin-space.
	The total change in entropy $\Delta S = \int C_p/T$ [see Fig.~\ref{fig:HC_zf}(b)] is extracted through numerical integration of the heat capacity data.
	It reaches a plateau around 4~K, nicely extrapolating to $R \ln(2)$ as expected for $j_{\rm eff} = 1/2$ degrees of freedom.

\begin{figure}[tbp]
		\includegraphics[scale=1]{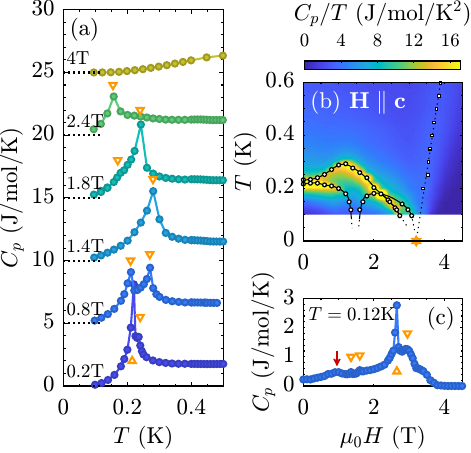}
		\caption{Specific heat of \dmnrc in axial magnetic fields $\mathbf{H \parallel c}$.
			(a) Exemplary $C_p(T)$ scans for various fixed fields $H$.
			(b) Preliminary phase diagram. Color shows $C_p/T$ data, while circles identify sharp anomalies in heat capacity and squares denote the Schottky maximum above $H_{\rm sat}$.
			(c) Exemplary $C_p(H)$ scan at $T = 0.12$~K.
			Orange triangles highlight the peak positions in the data.
			Red arrow marks a broad maximum in $C_p$ discussed in the text.}
		\label{fig:HC_field}
	\end{figure}

	In an axial magnetic field $\mathbf{H \parallel c}$, the two transitions at $T_{\rm N1}$ and $T_{\rm N2}$ begin to separate [see temperature scans in Fig.~\ref{fig:HC_field}(a)].
	Magnetic order is enhanced by small applied fields, creating a reentrant bulge in the phase diagram with a maximal ordering temperature $T_{\rm N1}^{\rm max} \approx 0.3$~K.
	Meanwhile, $T_{\rm N2}$ moves towards lower temperatures and becomes suppressed, disappearing completely around 1.3~T.
	A small anomaly re-emerges above 1.6~T and seems to meet the main transition line at a bicritical point near $\mu_0 H_{\rm bc} \approx 2.2$~T.
	At higher fields, we observe an additional pre-saturation phase, evident in field scans such as Fig.~\ref{fig:HC_field}(c) as two consecutive anomalies around 2.65 and 2.95~T.
	Finally, above 3.2~T we enter the pseudospin polarized regime where specific heat becomes gapped.
	We summarize this in the preliminary phase diagram shown in Fig.~\ref{fig:HC_field}(b), reminiscent of the 2D Heisenberg TLAF \cite{chubukovQuantumTheoryAntiferromagnet1991a, seabraPhaseDiagramClassical2011, yamamotoQuantumPhaseDiagram2014} except for the extra pre-saturation phase.
	The spin gap increases linearly with field [white squares, denoting the Schottky maximum positions in $C_p(T,H)$].
	A fit to $T_{\rm max} \propto (H - H_{\rm sat})^{z\nu}$ yields a saturation field $H_{\rm sat} = 3.20(2)$~T and crossover exponent $z\nu = 1.15(5)$ for the proximate quantum critical point.
	
	\subsection{Magnetoelastic Effects}
	
	\begin{figure}[tbp]
		\includegraphics[scale=1]{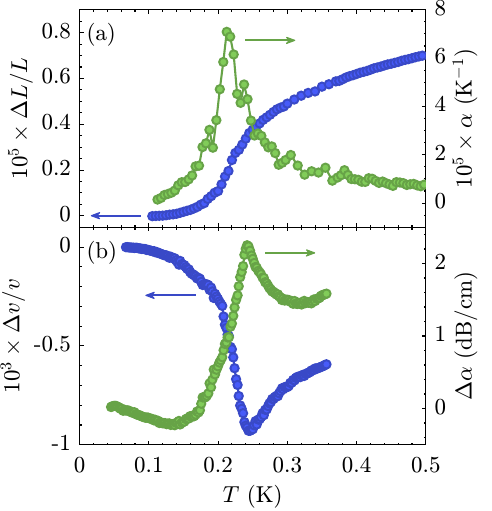}
		\caption{Magnetoelastic effects in \dmnrc. (a) Temperature dependent sample dilation $\Delta l/l$ along the $\mathbf{a^\star}$-axis, along with its numerical derivative, the thermal expansion coefficient $\alpha = \frac{1}{l} \frac{\partial \Delta l}{\partial T}$. (b) Relative changes in sound velocity and attenuation for the longitudinal $c_{11}$-mode.}
		\label{fig:ThermExp}
	\end{figure}
	
	\begin{figure*}[tbp]
		\includegraphics[scale=1]{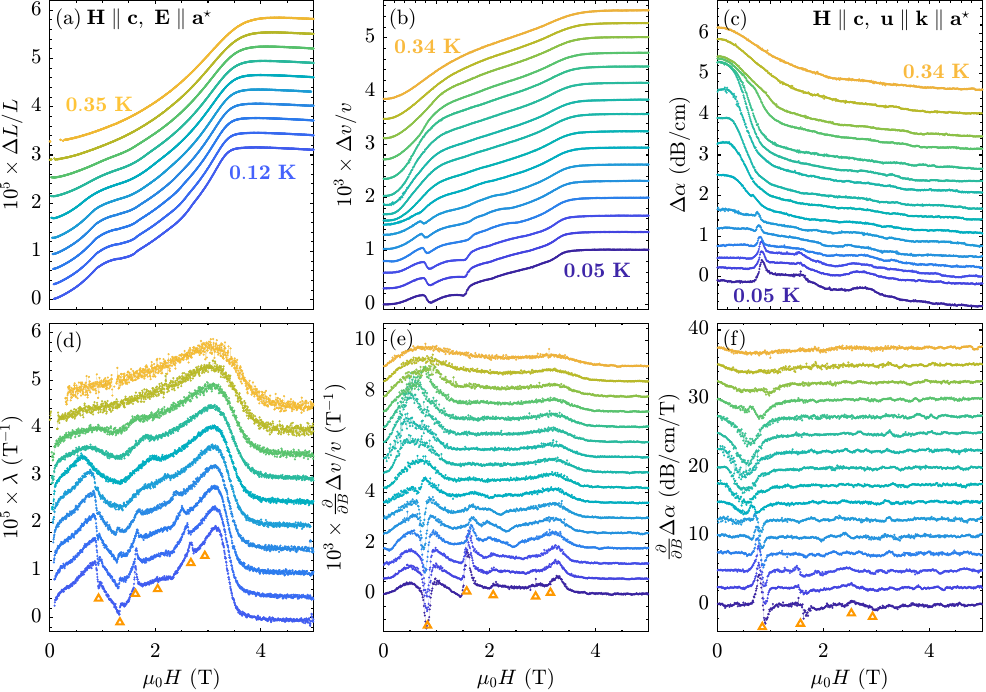}
		\caption{Field-induced magnetoelastic changes (top panels) and their numerical derivatives (bottom) for $\mathbf{H \parallel c}$ configuration. We show magnetostrictive changes in sample dilation (a), sound velocity (b) and attenuation (c) in the triangular plane. Various transition fields and measurement temperatures are indicated in the figure. Curves at different temperatures are offset for visibility.}
		\label{fig:SoundStrict_field}
	\end{figure*}

	Given the spin-orbital entangled ground state doublet, the onset of magnetic order may have significant effects on the lattice (and vice versa).
	To probe such magneto-elastic coupling, we employ a combination of dilatometry and ultrasound measurements.
	In Fig.~\ref{fig:ThermExp}(a) we show the relative change in sample length $\Delta l/l$ along $\mathbf{a^\star}$ against temperature.
	The lattice contracts significantly below $T_{\rm N}$, which may be understood through an exchange-striction mechanism \cite{supplement, linesElasticPropertiesMagnetic1979}:
	Exchange interactions depend on inter-atomic distances, so each bond acquires an extra stiffness contribution $\propto \dbraket{\mathbf{\hat{S}}_i \cdot \mathbf{\hat{S}}_j}$ in the ordered state and the crystal shrinks to minimize the free energy.
	Particularly interesting is the numerical derivative of this length change, i.e. the thermal expansion coefficient $\alpha(T) = \frac{1}{l} \frac{\partial \Delta l}{\partial T}$.
	Through Ehrenfest relations, the discontinuity in thermal expansion and heat capacity at $T_{\rm N}$ can be combined to estimate the initial pressure dependence of the transition temperature $\partial T_{\rm N} / \partial p = V_m T_{\rm N} \frac{\Delta \alpha^{\rm crit}}{\Delta C_p^{\rm crit}}$.
	Here $\Delta \alpha^{\rm crit}$ and $\Delta C_p^{\rm crit}$ represent the height of the critical anomaly in thermal expansion and heat capacity respectively, while $V_m$ is the molar volume.
	We obtain a {\it huge} uniaxial pressure dependence $\partial \ln(T_{\rm N2}) / \partial p_{100} \approx 250\%$/GPa, indicating that the magnetic structure is highly sensitive to in-plane strains.
	This is on par with other readily pressure-tunable spin systems like TlCuCl$_3$ \cite{johannsenMagnetoelasticCoupling2005} or $\alpha$-RuCl$_3$ \cite{kocsisMagnetoelasticCoupling2022}.
	Analogous thermal expansion effects along the $\mathbf{c}$-axis are an order of magnitude weaker (see Supplemental Material \hyperlinkcite{supplement}{S7}), consistent with a quasi-2D hierarchy of superexchange interactions.
	
	Complementary information can be gleaned from ultrasound, where we measure relative changes in sound velocity $\Delta v/v$ and attenuation $\Delta \alpha$ through the transition [see Fig.~\ref{fig:ThermExp}(b)].
	We focus on the longitudinal $c_{11}$-mode, propagating in the triangular plane with a sound velocity $v_{11} \simeq 2.1$ km/s.
	Both probes exhibit strong anomalies at $T_{\rm N}$: 
	Whereas the phonon frequencies soften near the onset of magnetic correlations (similar to Kohn anomaly), the attenuation - like all transport coefficients \cite{hohenbergTheoryDynamicCritical1977} - diverges at the transition, as sound waves become damped by critical fluctuations.
	Assuming an exchange-striction framework, we can estimate the magnetoelastic coupling constant $g_\Gamma \equiv \frac{\partial J}{\partial \varepsilon_\Gamma}$ for a particular strain component $\varepsilon_\Gamma$.
	For the A$_1$ "breathing" distortion of the triangular plane that we probe experimentally, we obtain $g_{\rm A1} \sim 40$~K (see Supplemental Material \hyperlinkcite{supplement}{T5}).
	This is an order of magnitude larger than in similar triangular lattice systems like Ba$_3$CoSb$_2$O$_9$ \cite{liMagnetoelasticCouplingMagnetization2019a} or Cs$_2$CuCl$_4$ \cite{congMagnetoacousticStudyQuantum2016, thallapakaMagnetoStructuralPropertiesLayered2019a}, signaling a strong coupling between pseudospin and lattice degrees of freedom.

	This is borne out in additional field-induced measurements summarized in Fig.~\ref{fig:SoundStrict_field}, revealing a much richer phase evolution than anticipated from heat capacity.
	The most pronounced anomalies appear in the region between 0.9~T and 1.6~T, where the sample length exhibits a plateau-like feature, while the sound velocity and attenuation show clear dips and peaks, respectively.
	The first of these features neatly matches a broad maximum observed in $C_p(H)$ around 0.9~T [red arrow in Fig.~\ref{fig:HC_field}(c)].
	Further anomalies become visible especially in the magnetostriction coefficient $\lambda(H) = \frac{1}{\mu_0 l} \frac{\partial \Delta l}{\partial H}$ [Fig.~\ref{fig:SoundStrict_field}(d)] or the analogous numerical derivatives of $\Delta v/v$ and $\Delta \alpha$, revealing a cascade of field-induced phase transitions.
	In the low-temperature limit, there are six phases clearly identified by several probes, with transition fields near 0.9, 1.35, 1.6, 2.1, 2.6 and 3.2~T.
	With increasing temperature, all features gradually become broadened and disappear, leaving behind only paramagnetic behavior above $T_{\rm N1}^{\rm max} \approx 0.3$~K.
	
	By combining the anomalies from all probes described above, we obtain the $H-T$ phase diagram depicted in Fig.~\ref{fig:PD_mag}(a).
	Note the presence of another slim presaturation phase between 2.6 and 2.9~T, stable only below $T \sim 80$~mK.
	The latter is clearly identified only by neutron diffraction (see below), though our ultrasound data do show some subtle features in this regime.

	\subsection{Magnetization Plateau}
	
	\begin{figure}[tbp]
		\includegraphics[scale=1]{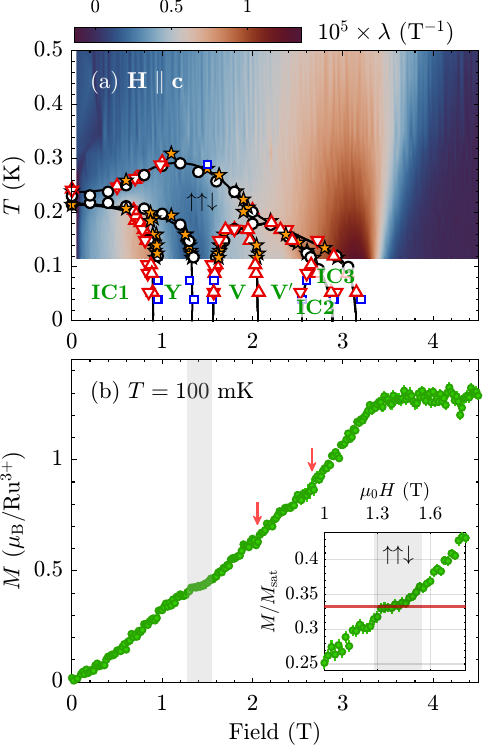}
		\caption{(a) Magnetic phase diagram of \dmnrc for the $\mathbf{H \parallel c}$ field direction. Color indicates the magnetostriction coefficient $\lambda(H)$. White circles represent lambda anomalies from heat capacity. Orange stars are transitions extracted from magnetostriction. Red triangles (up/down facing) indicate anomalies in sound velocity and attenuation. Blue squares are transitions seen in neutron diffraction.
			(b) Magnetization curve along $\mathbf{H \parallel c}$ at $T = 100$ mK, measured with the Faraday balance technique. The inset shows a zoom-in on the "$\uparrow\uparrow\downarrow$"-phase, displaying a plateau at one third of the saturation magnetization. Additional kinks matching the phase boundaries in (a) are marked by red arrows.}
		\label{fig:PD_mag}
	\end{figure}
	
	While the sequence of phases in Fig.~\ref{fig:PD_mag}(a) looks relatively complicated, there is an obvious starting point for understanding the magnetic behavior: The third phase, labeled "$\uparrow \uparrow \downarrow$".
	Its reentrant shape in the center of the $H - T$ phase diagram and the second order transition line directly into the paramagnetic state are both typical features of the $m = 1/3$ magnetization plateau expected to emerge in an ideal Heisenberg triangular lattice system \cite{chubukovQuantumTheoryAntiferromagnet1991a, seabraPhaseDiagramClassical2011, yamamotoQuantumPhaseDiagram2014}.
	Being a collinear state, it is most strongly preferred by order-by-disorder and is therefore relatively stable with respect to most Hamiltonian perturbations.
	To confirm this, we carried out magnetization measurements at $T = 100$ mK using a home-made Faraday balance setup \cite{blosserMiniatureCapacitiveFaraday2020}, shown in Fig.~\ref{fig:PD_mag}(b).
	Indeed, between 1.3~T and 1.55~T we observe a weak but distinctive kink in the magnetization curve, which is otherwise mostly convex up to saturation.
	In this regime, the magnetization reaches exactly one third of its saturation value $M_{\rm sat} = 1.29(1)~\mu_{\rm B}/$Ru$^{3+}$, corroborating our identification as a $m = 1/3$ magnetization plateau.
	Interestingly, the magnetostriction $\Delta l (H)/l$ [see Fig.~\ref{fig:SoundStrict_field}(a)] seems to mimic the magnetization curve in the entire field range.
	This gives credence to the exchange-striction picture discussed above, in which the sample dilation would simply mimic the magnetic bond energies along the probing direction, i.e. $\Delta l/l \sim \frac{g_{\rm A1}}{k} \sum_{\langle ij \rangle \parallel \mathbf{a^\star}} \dbraket{\mathbf{\hat{S}}_i \cdot \mathbf{\hat{S}}_j}$ where $k$ is a stiffness constant.
	Note also the presence of weaker kink features around 2.05~T and 2.6~T, matching the critical fields separating the V/V' and V'/IC3 phases, respectively.
	Finally, we remark on the cusp in magnetization at $H_{\rm sat}$:
	The latter appears to be rounded significantly more than the $2 k_{\rm B}T \sim 0.1$~T expected from thermal broadening.
	Such behavior is typical of broken $SU(2)$ symmetry in spin-orbit coupled systems, perhaps hinting at the presence of some non-trivial anisotropy terms \cite{janssen_JPhysCondMat_HKMag2019}.

	\subsection{Neutron Diffraction}
	\label{sec:diff}
	
	\begin{figure*}[tbp]
		\includegraphics[scale=1]{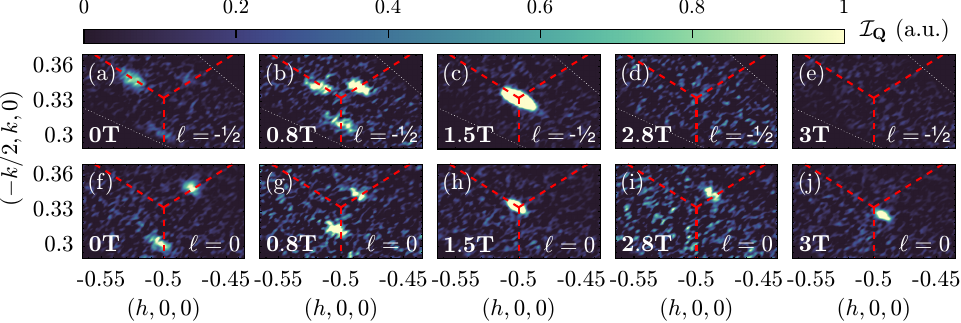}
		\caption{Magnetic neutron diffraction in \dmnrc for longitudinal fields $\mathbf{H \parallel c}$.
		Elastic scattering intensity at $T \lesssim 50$~mK near the H-point $\mathbf{Q} \approx (-2/3,1/3,-1/2)$ (a-e) and K-point $\mathbf{Q} \approx (-2/3,1/3,0)$ (f-j) for various magnetic field strengths.
		A paramagnetic background taken at $T = 0.5$~K has been subtracted point-by-point from all panels.
		Red dashed lines indicate the Brillouin zone boundaries and white dotted lines in (a-e) denote the detector edges.}
		\label{fig:Diffraction}
	\end{figure*}
	
	\begin{figure}[tbp]
		\includegraphics[scale=1]{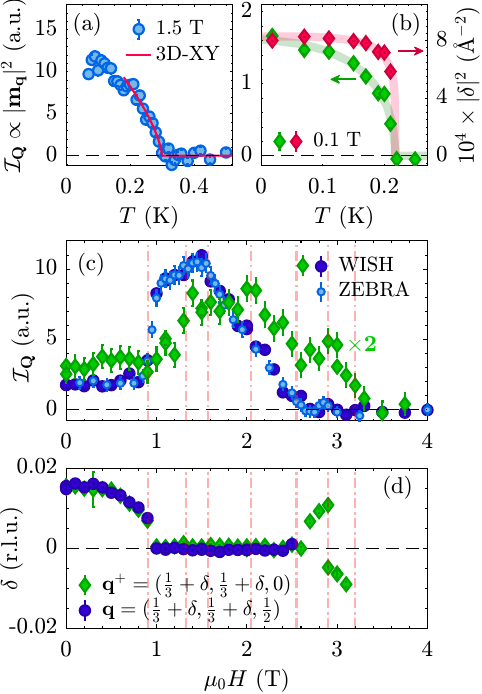}
		\caption{Temperature and field evolution of magnetic order in \dmnrc for longitudinal fields $\mathbf{H \parallel c}$.
			(a) Temperature dependence of the integrated intensity from the $\mathbf{Q} = (-2/3,1/3,1/2)$ reflection, taken in the plateau phase at $\mu_0 H = 1.5$~T.
			(b) Evolution of the magnetic Bragg intensity and incommensurability of the $\mathbf{q}^+ = (1/3+\delta,1/3+\delta,0)$ harmonics against temperature in the IC1 phase at $\mu_0 H = 0.1$~T.
			Note the difference in curvature, showing that $|\mathbf{m}_\mathbf{q}| \not\propto |\delta|$.
			(c,d) Field dependence of the magnetic Bragg intensity and incommensurability $\delta (H)$ for both types of ordering vectors $\mathbf{q}$ and $\mathbf{q}^+$.
			Vertical lines denote the transition fields from Fig.~\ref{fig:PD_mag}(a).}
		\label{fig:DiffInt}
	\end{figure}
	
	We carried out single-crystal neutron diffraction experiments to investigate the magnetic order realized across the entire phase diagram in $\mathbf{H \parallel c}$ orientation, the results of which are summarized in Fig.~\ref{fig:Diffraction} and Fig.~\ref{fig:DiffInt}.	
	As above, we start from the simplest phase, the collinear magnetization plateau.
	At $\mu_0 H = 1.5$~T we observe sharp magnetic Bragg peaks with a power-law onset of the order parameter below $T_{\rm N} \sim 0.29$~K [see Fig.\ref{fig:DiffInt}(a)].
	This is broadly consistent with the $2\beta \approx 0.7$ exponent expected for 3D XY universality - same class as the three-state Potts model in three dimensions, governing collinear $\uparrow \uparrow \downarrow$ order with a spontaneous choice of the "$\downarrow$"-sublattice \cite{pelissettoCriticalPhenomena2002}.
	The propagation vector is $\mathbf{q} = (1/3,1/3,1/2)$, pointing to an antiferromagnetic nearest neighbor inter-plane coupling $J_c > 0$ [Fig.\ref{fig:Diffraction}(c)].
	Additional reflections are visible in the $l = 0$ plane, corresponding to the second harmonic $2\mathbf{q} \equiv (1/3,1/3,0)$ [Fig.\ref{fig:Diffraction}(h)].
	Due to the non-bipartite three-sublattice order in each triangular layer, the inter-plane coupling is frustrated by the field.
	The Zeeman term prefers to cant {\it all} layers uniformly towards $\mathbf{H \parallel c}$, leaving one in every three $J_c$-bonds unsatisfied.
	Our magnetization plateau comprises alternating layers of "$\uparrow \uparrow \downarrow$" and "$\uparrow \downarrow \uparrow$" order [see Fig.~\ref{fig:MagStr} and discussion in Sec.~\ref{sec:disc}].
	This results in a mixed FM-AFM inter-layer stacking involving three Fourier components $\mathbf{k} = \mathbf{q}$, $\mathbf{k} = 2\mathbf{q}$ and $\mathbf{k} = 0$ (i.e. the uniform magnetization), all of which are necessary to construct a semi-classical equal moment structure.

	Towards lower fields, there is a commensurate-incommensurate lock-in transition at $\mu_0 H_c \approx 0.95$~T.
	Below this point, the magnetic Bragg peaks split into sets of three incommensurate (IC) reflections, equally spaced around the H and K-points [Fig.\ref{fig:Diffraction}(a,b,f,g)].
	They move along the edges of the Brillouin zone and may be indexed by a propagation vector $\mathbf{q} = (q_{\rm IC},q_{\rm IC},1/2)$, where $q_{\rm IC} \equiv 1/3 + \delta \simeq 0.35$~r.l.u. and $\delta$ is a field and temperature dependent IC shift.
	The magnetic texture spans 20 chemical cells in the triangular plane, corresponding to $a_{\rm m} \sim 143$~\AA.
	Symmetry imposes three degenerate arms of the $k$-vector star
	\begin{equation}
		\begin{Bmatrix*}[l]
			\mathbf{q}_1 = (q_{\rm IC}, q_{\rm IC}, 1/2) \\
			\mathbf{q}_2 = (q_{\rm IC}, -2q_{\rm IC}, 1/2) \\
			\mathbf{q}_3 = (-2q_{\rm IC}, q_{\rm IC}, 1/2),
		\end{Bmatrix*}
	\end{equation}
	resulting in triplets of incommensurate peaks that could indicate either a multi-$\mathbf{q}$ structure or a trivial population of three-fold degenerate domains.
	
	As seen in Fig~\ref{fig:Diffraction}(f,g), we also observe a secondary modulation, associated with a propagation vector $\mathbf{q}^+ = (q_{\rm IC},q_{\rm IC},0)$.
	Both ordering vectors adopt the exact same IC shift $\delta(H,T)$ [Fig.~\ref{fig:DiffInt}(d)], clearly indicating that these are Fourier components present in the same magnetic phase.
	We note that for finite $\delta$, the $l = 0$ reflections are distinct from the second harmonics seen in the commensurate phases, i.e. $\mathbf{q}^+_i \not\equiv 2\mathbf{q}_i$ where $i \in \{1,2,3\}$ enumerates the arms of the star.
	Instead, these peaks may be indexed as coupled harmonics of the form $\mathbf{q}^+_i \equiv \mathbf{q}_j + \mathbf{q}_k$.
	They represent direct interference terms between {\it different} arms of the propagation vector star, which cannot arise from separate magnetic domains. 
	Therefore, the ground state must be a multi-$\mathbf{q}$ structure.
	Fig.~\ref{fig:DiffInt}(b) depicts the temperature dependence of the order parameter and the incommensurability in the IC1 phase.
	Note that these quantities are not proportional to one another.
	We detect no changes in scattering intensity upon employing a field cooling or zero-field cooling protocol.
	
	In the high-field limit, we observe a similar C-IC transition.
	The commensurate $l = 1/2$ peaks seen in the $m = 1/3$ plateau gradually lose intensity and disappear completely above 2.6~T.
	This leaves only the $l = 0$ reflections which move away from the K-point, resulting in a long-period incommensurate structure with simple ferromagnetic inter-plane stacking [Fig.\ref{fig:Diffraction}(d,e,i,j)].
	We clearly observe {\it two} distinct IC presaturation phases:
	First the peaks shift towards the M points, then around 2.9~T the trend reverses and the ordering vector moves towards $\Gamma$.
	In these high-field phases, only one satellite remains around each commensurate position, indicating a simple single-$\mathbf{q}$ magnetic order.
	The complete field dependence of the integrated intensity and incommensurability for both ordering vectors $\mathbf{q}$ and $\mathbf{q}^+$ is shown in Fig.~\ref{fig:DiffInt}(c,d), in good agreement with the phase boundaries extracted from thermodynamics.

	\subsection{Inelastic Neutron Scattering}
	
\begin{figure*}[tbp]
		\includegraphics[scale=1]{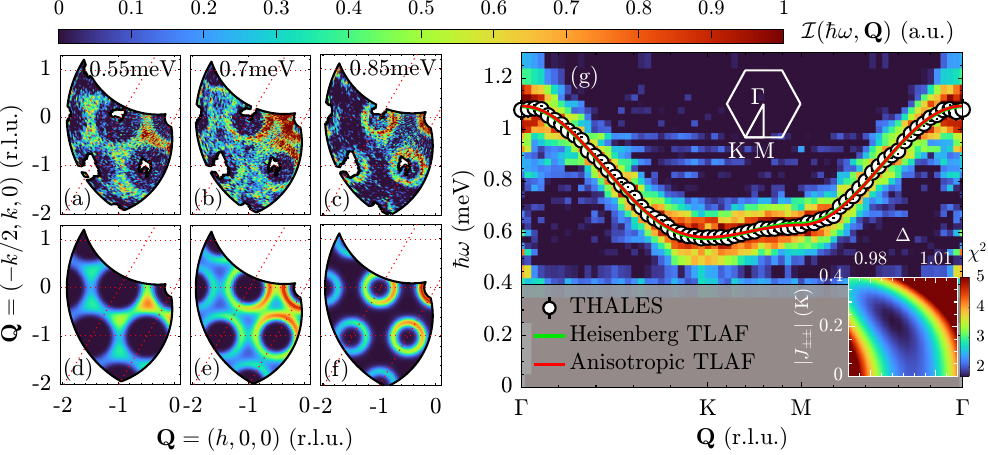}
		\caption{Inelastic spectra of \dmnrc in the fully polarized state at $\mu_0 H = 7$~T. 
		(a-c) False color plots of INS intensity for 2D slices in momentum space taken at various constant energies. The white regions mask areas affected by Currat-Axe spurious scattering.
		(d-f) Simulated LSWT intensity in the same range using Eq.~\ref{Hamilt}, convoluted with experimental resolution and form factor.
		(g) Dispersion relation along high symmetry directions. Markers show extracted peak positions, green/red lines are fits to a 2D Heisenberg/anisotropic model discussed in the text.
		The white hexagon represents a sketch of the Brillouin zone, where $\Gamma$, K and M label the high symmetry points in reciprocal space.
		Inset shows the $\chi^2$ loss function for anisotropic exchange parameters in the fit to Eq.~\ref{Hamilt}.}
		\label{fig:INS}
	\end{figure*}

	In Fig.~\ref{fig:INS} we present the magnetic excitation spectrum of \dmnrc in the triangular $(hk0)$ plane, obtained from single-crystal neutron spectroscopy in the $\mathbf{H \parallel c}$ field-polarized regime at $\mu_0 H = 7$~T.
	As expected, there is one resolution-limited spin-wave branch with a bandwidth $9JS \approx 0.50$~meV.
	The spin gap $\Delta = 0.58(1)$~meV to the lowest excitation at the K-point is in excellent agreement with the value $\Delta = g_c \mu_{\rm B} \mu_0 (H - H_{\rm sat}) = 0.591(8)$~meV expected from ESR and thermodynamics.

	Spin-wave theory is exact for the fully polarized ferromagnet \cite{coldeaDirectMeasurementSpin2002}, so a fit to the extracted dispersion relation $\hbar\omega_\mathbf{Q}$ may be used to determine the leading terms in the low-energy effective spin Hamiltonian.
	Since the $g$-factor and external field are known to high precision, we keep the Zeeman term fixed throughout this procedure.
	The data are well described by a simple nearest neighbor Heisenberg triangular lattice model with $J = 1.32(1)$~K (Fig.~\ref{fig:INS}(g), green line).
	Potential further neighbor interactions $J_2$ or $J_3$ do not improve the fit
	and can be excluded at the $\sim 2\%$ level.
	We cannot probe the inter-plane couplings directly due to the scattering geometry $(hk0)$ in this experiment.
	However, the classical saturation field $\mu_0 H_{\rm sat} = 9JS/g_c \mu_{\rm B} = 3.29(4)$~T nicely matches the $\mu_0 H_{\rm sat} = 3.20(5)$~T observed experimentally.
	This places strong constraints on the leading AFM inter-plane coupling $J_c$.
	We estimate $J_c \lesssim 0.1J$, clearly pointing to a quasi-2D system.
	
	Given the spin-orbital entangled wavefunctions, we also attempt to fit a more general anisotropic Hamiltonian
	\begin{gather}
		\mathcal{\hat{H}} =
		\sum_{\langle ij \rangle} J(S^x_i S^x_j + S^y_i S^y_j + \Delta S^z_i 	S^z_j) \label{Hamilt} \\
		+ 2J_{\pm\pm} [(S^x_i S^x_j - S^y_i S^y_j) c_\alpha - (S^x_i S^y_j + 	S^y_i S^x_j) s_\alpha] \nonumber\\
		+ J_{z\pm} [(S^y_i S^z_j + S^z_i S^y_j) c_\alpha - (S^x_i S^z_j + S^z_i S^x_j) s_\alpha].\nonumber
	\end{gather}
	where $c(s)_\alpha = \cos(\sin) \varphi_\alpha$.
	Here the phases $\varphi_\alpha \in (0,2\pi/3,4\pi/3)$ associated with the  translation vectors $\mathbf{a, b, a+b}$ give rise to a bond dependence on the pseudo-dipolar couplings $J_{\pm\pm}$ and $J_{z\pm}$.
	In practice, we are not sensitive to $J_{z\pm}$, as it only enters the dispersion relation for a finite in-plane magnetic moment.
	This model results in a marginally better fit ($\chi^2 \approx 1.7$ vs. $\chi^2 \approx 2$), with $J = 1.31(2)$~K, $\Delta = 1.00(2)$ and $|J_{\pm\pm}| = 0.31^{+0.12}_{-0.23}$~K.
	The system is clearly Heisenberg-like, with no signs of XXZ anisotropy.
	Our data do favor a strong $J_{\pm\pm}$ term but with significant uncertainties, leaving neither one of the bond dependent terms well constrained.
	We leave this issue to future investigations in a transverse-field geometry \cite{woodlandContinuumExcitationsSharp2025}.
	A direct comparison between experimental data and simulated spin-wave intensities is shown in Fig.~\ref{fig:INS}(a-f) through representative constant-energy cuts.
	Given the already excellent fit and limited sensitivity to these anisotropic terms, extending this model (e.g. with further neighbor interactions and their anisotropies) would risk overparameterization.

	\section{Discussion}
	\label{sec:disc}
	
	The novel quantum material \dmnrc exhibits a perfect triangular lattice of Ru$^{3+}$ ions realizing the spin-orbital entangled $j_{\rm eff} = 1/2$ state.
	Perhaps its most striking property is the highly complex phase diagram in $\mathbf{H \parallel c}$ composed of seven distinct spin states, three of which are incommensurate with the lattice.
	The spin Hamiltonian is described at leading order by a purely 2D isotropic Heisenberg model with nearest neighbor coupling $J \approx 1.3$~K.
	However, in the following we argue for three important perturbations to this picture, which drive the physics, resulting in the rich sequence of field-induced transitions:
	inter-plane coupling, exchange anisotropy and magnetoelastic effects.
	
	\begin{figure}[tbp]
		\includegraphics[scale=1]{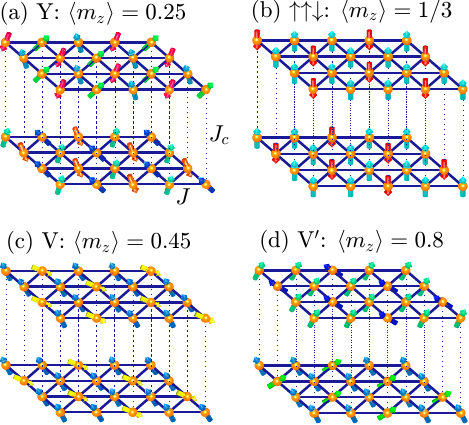}
		\caption{Sketch of the proposed magnetic structures in the four commensurate phases (a) Y, (b) $\uparrow\uparrow\downarrow$, (c) V and (d) V$^\prime$.}
		\label{fig:MagStr}
	\end{figure}
	
	At least the four commensurate states in the center of the phase diagram may be understood semi-classically by considering only the first of these terms, the inter-plane exchange.
	\dmnrc realizes a direct AA-type stacking of triangular planes with an AFM nearest neighbor coupling $J_c \lesssim 0.1 J$.
	Its competition with the magnetic field $H$ results in a mixed FM-AFM inter-layer pattern which contains both $\mathbf{q} = (1/3,1/3,1/2)$ and $2\mathbf{q}$ Fourier components.
	In Fig.~\ref{fig:MagStr} we sketch the proposed spin configurations in these phases, obtained through classical energy minimization of all possible six-sublattice groundstates with real-space perturbation corrections \cite{zhitomirskyRealspacePerturbationTheory2015} to model the effects of quantum order-by-disorder (for details we refer the reader to Supplemental Material \hyperlinkcite{supplement}{S10}).
	The simplest is the $m = 1/3$ magnetization plateau, a collinear state forming alternating layers of $\uparrow \uparrow \downarrow$ and $\uparrow \downarrow \uparrow$ moments.
	This is followed by the well-known "Y" and "V" phases at slightly lower/higher fields, where the finite $J_c$ results in an AFM canting of the in-plane spin components.
	Finally, a fourth phase is generally expected to arise in AA-stacked triangular lattice magnets with $J_c > 0$, often labeled "V$^\prime$" \cite{yamamotoMicroscopicModelCalculations2015, liEffectsInterlayerBiquadratic2020}.
	It is a simple consequence of the $J_c$ vs. $H$ competition, which favors a different inter-plane stacking close to the plateau and near saturation: In the former case the choice of minority spin is staggered between layers, whereas in the latter it is not.
	A first-order transition between these states is predicted at $H_{\rm c} \approx 0.7 H_{\rm sat}$ in the Heisenberg limit \cite{yamamotoMicroscopicModelCalculations2015}, only weakly depending on $J_c$.
	This is in good agreement with the $H_{\rm c} \approx 2.1$~T (corresponding to $\sim 0.65 H_{\rm sat}$) seen in experiment.
	An analogous transition has been identified in Ba$_3$CoSb$_2$O$_9$ \cite{okadaFieldorientationDependenceQuantum2022, koutroulakisQuantumPhaseDiagram2015}, the closest proxy to an ideal Heisenberg triangular lattice antiferromagnet known to date.

	Let us turn to the IC behavior observed towards the edges of the phase diagram:
	The propagation vector $\mathbf{q} = (q_{\rm IC}, q_{\rm IC}, 1/2)$ cannot be explained by a Heisenberg model on a perfect triangular lattice and necessarily imposes some in-plane anisotropy term, either in spin-space or in real space.
	This anisotropy energy will tend to stabilize some complex non-coplanar order at all fields, but it has to compete with quantum order-by-disorder selection.
	The latter prefers more collinear/coplanar states due to their softer excitation spectrum, i.e. the commensurate Y, $\uparrow \uparrow \downarrow$ and V phases.
	Order-by-disorder corrections are strongest towards the center of the phase diagram (particularly around $\sim H_{\rm sat}/3$), but completely disappear for the fully polarized ferromagnet as $H \rightarrow H_{\rm sat}$.
	This provides a natural mechanism for IC order to develop, both in zero-field and just below saturation.
	
	The simplest way to explain $q_{\rm IC}$ would be a structural transition, giving rise to a "static" $J - J'$ isosceles distortion of the triangular plane that discriminates between chain-like $J$ and zig-zag $J'$ interactions.
	This model has been discussed thoroughly in the literature \cite{chenGroundStatesSpin12013, grisetDeformedTriangularLattice2011, tayVariationalStudiesTriangular2010}, mostly in the context of the distorted triangular lattice magnets Cs$_2$CuCl$_4$ \cite{coldeaExperimentalRealization2D2001, tokiwaMagneticPhaseTransitions2006}, Cs$_2$CuBr$_4$ \cite{onoMagnetizationPlateaux12004, fortuneCascadeMagneticFieldInducedQuantum2009} or Ca$_3$ReO$_5$Cl$_2$ \cite{hirai_ACS_AnisotropicTriangular2020, nawa_PRR_BoundSpinon2020}.
	But a "clean" experimental realization in the limit $(J'-J) \ll J$ has not been reported to date.
	The classical energy is minimized by a propagation vector $q_{\rm IC} = \arccos(-J'/2J)/2\pi$, resulting in an IC spiral groundstate with three degenerate structural domains.
	After accounting for quantum renormalization, the detected $q_{\rm IC} \approx 0.35$~r.l.u. would correspond to an exchange ratio of $J'/J \approx 1.06$ \cite{weihongPhaseDiagramClass1999, thesbergExactDiagonalizationStudy2014, hasikIncommensurateOrderTranslationally2024}.
	The expected $H-T$ phase diagram \cite{chenGroundStatesSpin12013, grisetDeformedTriangularLattice2011, tayVariationalStudiesTriangular2010} looks strikingly similar to experiment, with IC states towards $H \rightarrow \{0, H_{\rm sat}\}$ and the typical commensurate Y, $\uparrow \uparrow \downarrow$ and V phases in the center, where order-by-disorder selection dominates.
	
	However, this option is ruled out by experiment:
	\dmnrc~{\it does} go through a structural transition at $T^\star \approx 118$~K, but our single-crystal x-ray diffraction data clearly establish that three-fold symmetry (i.e. $J' = J$) remains intact, at least down to $T_{\rm N}$ \cite{supplement}.
	Instead, we point out two alternative mechanisms which could account for the IC behavior, both of which rely on the strong spin-orbit effects.
	
	An obvious culprit would be the magnetoelastic coupling.
	The latter could drive a magneto-structural phase transition at $T_{\rm N2}$, resulting in a "dynamic" $J - J'$ isosceles distortion depending on both field and temperature.
	Such effects are often seen in pyrochlore systems like CdCr$_2$O$_4$ or HgCr$_2$O$_4$ (cf. "Spin-Peierls" effect \cite{pencHalfMagnetizationPlateauStabilized2004, tchernyshyovOrderDistortionString2002a}), where a {\it linear} gain in exchange energy upon distorting the lattice always leads to a small deformation below $T_{\rm N}$.
	As it turns out, this particular mechanism does not easily generalize to the 2D triangular lattice, since the leading correction to the classical energy $\Delta E \approx -\frac{9}{8}JS^2 \varepsilon_m^2$ for a monoclinic lattice strain $\varepsilon_m$ is only \textit{quadratic} \cite{supplement}.
	But analogous distortions may arise by invoking the pseudo Jahn-Teller effect in the presence of strong SOC, as recently proposed in $5d^5$ iridate perovskites to explain a small tetragonal distortion below $T_{\rm N}$ \cite{liuPseudoJahnTellerEffectMagnetoelastic2019}.
	Lattice deformations will modulate the spatial shape of the spin-orbital wavefunctions, which can generate new terms $\delta \mathcal{H}_{ij} \sim g_\gamma \varepsilon_\gamma Q_{ij}^\gamma$ in the Hamiltonian, describing the pseudospin-lattice coupling between two-site quadrupoles $Q_{ij}^\gamma \sim \hat{S}^\alpha_i \hat{S}^\beta_j$ of the appropriate symmetry and strains $\varepsilon_\gamma$.
	An extension of this effect to the triangular lattice geometry would be an interesting problem to tackle theoretically.
	
	However, there are several strong arguments to rule out any such magneto-elastic scenario.
	Firstly, if a lattice distortion is the primary driver for incommensurability, the $J - J'$ anisotropy and the IC splitting of the $\mathbf{q}$-vector must be directly proportional to the order parameter.
	This is in contradiction with our diffraction results.
	In the high-field "IC3" presaturation phase, $q_{\rm IC}$ moves \textit{away} from the commensurate position even as the magnetic Bragg intensity goes to zero [see Fig.~\ref{fig:DiffInt}(c,d)].
	Furthermore, the temperature evolution of the order parameter $|\mathbf{m}_\mathbf{q}|$ and the IC shift $\delta$ in the IC1 phase exhibit a completely different curvature [see Fig.~\ref{fig:DiffInt}(b)], clearly proving that the IC phases are not induced by magneto-elastic coupling.
	Another important clue comes from the low-field $\mathbf{q}^+$ satellites around the K-points [see Fig.~\ref{fig:Diffraction}(f,g)].
	Their presence as cross harmonics, coupling different arms of the propagation vector star, is direct proof that the IC1 phase exhibits multi-$\mathbf{q}$ order.
	This cannot be reconciled with a lattice distortion, which may only be stabilized through the energy gained by {\it breaking} three-fold symmetry.
	Finally, we have confirmed for all reflections in the family $\mathbf{Q} = (1/3,1/3,0) + \bm{\delta}$, that those with IC shift $\bm{\delta} \parallel \mathbf{Q}$ - that is $\mathbf{Q} \approx (0.35,0.35,0)$ and six-fold equivalents - carry no intensity, indicating an effect of the neutron polarization factor.
	This is inconsistent with all coplanar spiral configurations, favoring instead a multi-$\mathbf{q}$ order incorporating several longitudinal spin density wave contributions.

	Bond-dependent exchange anisotropy presents a simple alternative to stabilize an incommensurate structure.
	Specifically, the Heisenberg-Kitaev model on a triangular lattice has been predicted to realize the so called $\mathbb{Z}_2$ vortex crystal \cite{beckerSpinorbitPhysicsJ12015, rousochatzakisKitaevAnisotropyInduces2016, liCollectiveSpinDynamics2019}.
	Even for a tiny Kitaev-term $|K| \ll J$, the Fourier transformed exchange matrix is minimized by three different ordering vectors
	\begin{equation}
		\mathbf{q}^\gamma = \frac{\hat{\mathbf{e}}^\gamma}{\pi} \arccos \left( -\frac{J}{2(J+K)} \right),
	\end{equation}
	one for each spin component $\gamma \in (x,y,z)$.
	Locally, the magnetic structure would still mimic the typical $120^\circ$ order.
	But on larger length scales, one would see a superlattice of topological defects - $\mathbb{Z}_2$ vortices - which crystallize into a triangular arrangement (see Supplemental Material \hyperlinkcite{supplement}{S11}).
	This structure has three main $\mathbf{q}$-vectors, preserving three-fold symmetry, but with many higher harmonics to retain a uniform moment size - in agreement with the observation of $\mathbf{q}^+$ coupled harmonics.
	Since the bond-dependent terms break all continuous symmetries, the groundstate also acquires a finite spin gap $\Delta$ \cite{rousochatzakisKitaevAnisotropyInduces2016}, seemingly consistent with our heat capacity measurements indicating $\Delta \sim 0.4 J$.
	
	In practice, the presence of a small Kitaev term in \dmnrc should not be surprising.
	Bond dependent exchange anisotropies in the parallel-edge superexchange geometry have been proposed by several authors \cite{trebstKitaevMaterials2022, beckerSpinorbitPhysicsJ12015, catuneanuMagneticOrdersProximal2015}.
	Given the bond angles deviating from $90^\circ$ and the significant trigonal distortion $\Delta/\lambda \simeq -0.5$, we expect such terms to be sub-leading compared to $J$ \cite{winterModelsMaterialsGeneralized2017}.
	However, the observed propagation vector $q_{\rm IC} \approx 0.35$~r.l.u. can be generated by a modest ferromagnetic  Kitaev term $K \sim -0.1J$, which seems completely realistic.
	In fact, the contribution from $J_{\pm\pm}$ alone obtained through INS in the polarized regime would be significantly larger, with $K = -2J_{\pm\pm}$ amounting to $K/J \approx -0.4^{+0.3}_{-0.2}$.
	This makes \dmnrc a leading candidate to realize the $\mathbb{Z}_2$ vortex crystal phase.
	Given the large uncertainties on $J_{\pm\pm}$ and completely undetermined $J_{z\pm}$, more work should be put towards constraining these bond anisotropies.
	We note that the stability of this phase against finite inter-plane coupling $J_c$, anisotropies away from the Heisenberg-Kitaev limit $J_{z\pm} \neq 2\sqrt{2}J_{\pm\pm}$, or magnetic field $H$ have not been investigated theoretically, likely due to the inherent complexity of the vortex crystal state.
	All of these problems would be of imminent interest in understanding the groundstate of \dmnrc.
		
	Finally, we comment on the tunability of this new structural archetype MA$_2$ARuX$_6$.
	Powder samples of several different members in this family have been previously synthesized in \cite{vishnoiStructuralDiversityMagnetic2020}, confirming that at least the alkali A and halide X can easily be replaced.
	This already gives huge flexibility in tuning the magnetism of these exciting new quantum materials.
	Especially the hierarchy between the small perturbing terms like inter-plane interactions, bond anisotropies or magnetoelastic coupling may be easily modified by changing the anions participating in superexchange.

	In conclusion, the new quantum antiferromagnet \dmnrc poses a rare model system to combine the effects of strong spin-orbit coupling endemic to $4d/5d$ transition metals and the geometric frustration on a triangular lattice.
	Its magnetism is described by a quasi-2D nearest-neighbor Heisenberg Hamiltonian with the addition of inter-plane interactions, bond anisotropies and magnetoelastic coupling as small perturbing terms, begetting an extremely rich phase diagram including several incommenasurate states.
	Of particular interest is the multi-$\mathbf{q}$ ground state realized in zero magnetic field, posing a prime candidate to host the exotic $\mathbb{Z}_2$ vortex crystal phase.
	This calls for polarized diffraction studies to confirm the explicit spin configuration and spectroscopy experiments to investigate its dynamics.
	\dmnrc constitutes the first member in a large family of $4d/5d$-based quantum triangular lattice magnets, opening new pathways to study the interplay between geometric frustration and spin-orbit effects.

	\section{Methods}
	
	Single crystal samples of \dmnrc were grown using a hydrothermal method analogous to \cite{vishnoiStructuralDiversityMagnetic2020}.
	We obtain dark red crystals with highly symmetric hexagonal faces and typical sample mass of order $10 - 200$ mg.
	Fully deuterated ($\geq 98\%$ D) precursors were employed in the growth to limit the incoherent H-scattering in neutron experiments.
	The structure and sample quality of all experimental probes was confirmed using a Bruker APEX-II single-crystal x-ray diffractometer.
	Detailed refinements to characterize the low-temperature structural transition were carried out at the I19 synchrotron beamline (Diamond, UK).
	Datasets of 1-5 $\times 10^4$ reflections were collected at $\lambda = 0.6889$~\AA~on several $\sim 50 \mu$m single crystal samples between $T = 150$~K and 30~K.
	The DIALS \cite{winter_ACSD_DIALS2018} and shelx \cite{sheldrickShortHistorySHELX2008} software suites were employed for the data processing and structure solution, respectively.
	
	Magnetic susceptibility measurements on a $m = 20$~mg single crystal were performed in a commercial Magnetic Property Measurement System (MPMS) SQUID magnetometer in a $\mu_0 H = 0.1$~T probing field.
	High-$T$ magnetization curves up to $\mu_0 H = 7$~T were collected in the same configuration.
	The magnetization data at dilution temperatures were obtained on a 2~mg sample using a home-built Faraday balance setup \cite{blosserMiniatureCapacitiveFaraday2020} and calibrated to absolute units using the SQUID data at 2~K as reference.
	For the ESR experiments a tunable-frequency spectrometer (similar to the one described in Ref.~\cite{zvyagin_PhysB_2004_ESRinCuGeO3}) was employed, equipped with a $16$~T superconducting magnet. VDI microwave-chain sources (product of Virginia Diodes, Inc., USA) were used to generate radiation in the frequency range of $50-360$~GHz. A hot-electron n-InSb bolometer (product of QMC Instruments Ltd., UK), operated at $4.2$~K, was employed as a THz detector. For the frequencies below $50$~GHz a microwave vector network analyser (MVNA, production of AB Millimeter, France) was used. Sample holders in the Faraday and Voigt configuration were used for $\mathbf{H \parallel c}$ and $\mathbf{H \parallel a}$ measurements respectively. 2,2-diphenyl-1-picrylhydrazyl (DPPH) was used as a standard frequency-field marker.
	The heat capacity of a 0.4~mg single crystal sample was measured with the standard relaxation calorimetry technique on a 9~T Physical Property Measurement System (PPMS) with a $^3$He-$^4$He dilution refrigerator insert.
	Magnetostriction data were taken in the same configuration using a $L = 1.48$~mm long single crystal with natural faces in the (100) and (001) cleavage planes.
	The sample dilation was determined with a miniature capacitive dilatometer \cite{kuchlerNewApplicationsWorlds2023} in conjunction with an Andeen-Hagerling 2700A bridge, operated at 1.11~kHz.
	Sound velocity measurements of the longitudinal $c_{11}$ mode (polarization $\mathbf{u}$ and propagation $\mathbf{k}$ along the $\mathbf{a^\star}$-axis) were carried out at the Dresden High Magnetic Field Laboratory using a pulse-echo method with phase-sensitive detection \cite{luthiPhysicalAcousticsSolid1967, zherlitsyn_LTPhys_SpinLatticeEffects2014}. LiNbO$_3$ transducers were bonded with Thiokol to two parallel (100) faces of a $L = 2.8$~mm long single crystal sample.
	Experiments were carried out at $f = 40$ MHz using a $^3$He-$^4$He dilution refrigerator equipped with an 18~T magnet.
	
	Single-crystal neutron diffraction experiments were performed at WISH (ISIS, UK) \cite{wish, wish2} and ZEBRA (PSI, Switzerland) on the same $m = 55$~mg sample aligned in the $(hk0)$ scattering plane.
	The crystal was mounted in a $^3$He-$^4$He dilution refrigerator together with a 10~T (6~T) vertical cryomagnet. 
	A constant wavelength $\lambda = 1.383$~\AA~(Ge monochromator) was chosen for the latter experiment. 
	The incommensurate positions and intensities from WISH were extracted by fitting 2D Gaussian profiles with a fixed resolution $(\Delta q_\parallel, \Delta q_\perp)$ to the data.
	These data were analyzed using the mantid software suite \cite{arnoldMantidDataAnalysis2014}.
	
	Inelastic neutron scattering experiments were carried out on the cold triple-axis spectrometer THALES (ILL, France) \cite{thales} with Flatcone multi-analyzer setup.
	16 single crystals were coaligned in the $ab$ scattering plane for a total sample mass of $m \simeq 1.9$~g. The probe was installed with $(hk0)$ scattering geometry in a dilution refrigerator with 10~T vertical cryomagnet and cooled to $T\sim 70$~mK. All measurements were taken at a final energy $E_f = 4.06$~meV (FWHM resolution $\Delta E \approx 175$~$\mu$eV), fixed by the Flatcone setup. 
	35 constant-energy slices were collected in the field-polarized regime at $\mu_0 H = 7$~T, rotating 90$^\circ$ through the scattering plane in 0.5$^\circ$ steps.
	The triple-axis geometry allows for so called "Currat-Axe" spurions \cite{shiraneNeutronTripleAxis2002}, undesirable scattering events that occur when the sample angle accidentally matches a Bragg reflection, resulting in a spurious peak upon incoherent/thermal diffuse scattering in the analyzer.
	These features follow known paths in reciprocal space and are simply masked out (see white spots in Fig.~\ref{fig:INS}(a-c)).
	To reconstruct the spin-wave dispersion relation, regions affected by Currat-Axe spurions are discarded, a constant background was subtracted, the data were normalized by form factor and averaged over $> 20$ equivalent paths through several Brillouin zones.
	Further details about the sample and fitting procedures may be found in Supplemental Material \hyperlinkcite{supplement}{S8-9}.
	The SpinW software \cite{tothLinearSpinWave2015} was employed for the spin-wave intensity calculations.

	\section{Data Availability}
	
	All data are available upon reasonable request to the corresponding author. 
	The neutron scattering data collected at WISH (STFC ISIS Neutron and Muon Source) and THALES (Institut Laue-Langevin) are available at \url{https://doi.org/10.5286/ISIS.E.RB2420060} and \url{https://doi.ill.fr/10.5291/ILL-DATA.4-01-1857}, respectively.

	\section{Acknowledgments}
	
	We acknowledge the beam time allocation at ISIS (WISH id: RB2420060 \cite{wish} and RB2520045 \cite{wish2}), PSI (ZEBRA id: 20240853) and ILL (THALES id: 4-01-1857 \cite{thales}). 
	This work was carried out with the support of Diamond Light Source, instrument I19 (proposal CY39239).
	We also thank Sergei Zherlytsin (HZDR) for the assistance with ultrasound measurements.

	\section{Funding}
	
	This work is supported by a MINT grant of the Swiss National Science Foundation. 
	We acknowledge the support of the HLD at HZDR, member of the European Magnetic Field Laboratory (EMFL), and the W\"{u}rzburg-Dresden Cluster of Excellence on Complexity and Topology in Quantum Matter - ct.qmat (EXC 2147, project ID 390858490).
	D.K. was supported by the EPSRC under Grant No. EP/W00562X/1.

	\section{Author Contributions}
	
	Samples used in this study were grown by J.N and Z.Y.
	Magnetometry and dilatometry studies were carried out by J.N. and S.G.
	Heat capacity measurements were performed by J.N., B.D. and S.G.
	ESR measurements were performed by K.Y.P. and S.A.Z., while ultrasound data were collected by K.Y.P, J.S., B.V.S. and F.H.
	Neutron diffraction experiments at WISH and ZEBRA were carried out by J.N. and A.Z. with assistance of D.K., P.M. and F.O., and J.N., C.N. and A.Z. with help from O.Z., respectively.
	INS measurements at THALES were taken by J.N., S.G. and A.Z. with assistance from P.S. and A.H.
	Synchrotron x-ray experiments at I19 were performed by J.N., S.G. and A.Z. with assistance from D.A. and S.B.
	Analysis and modeling of experimental data was carried out by J.N.
	The manuscript was written by J.N. and A.Z. with input from all coauthors.

	\section{Competing Interests}
	
	The authors declare no competing interests.

	\bibliography{Bibliography_npj25}

\begin{thebibliography}{99}%
\makeatletter
\providecommand \@ifxundefined [1]{%
 \@ifx{#1\undefined}
}%
\providecommand \@ifnum [1]{%
 \ifnum #1\expandafter \@firstoftwo
 \else \expandafter \@secondoftwo
 \fi
}%
\providecommand \@ifx [1]{%
 \ifx #1\expandafter \@firstoftwo
 \else \expandafter \@secondoftwo
 \fi
}%
\providecommand \natexlab [1]{#1}%
\providecommand \enquote  [1]{``#1''}%
\providecommand \bibnamefont  [1]{#1}%
\providecommand \bibfnamefont [1]{#1}%
\providecommand \citenamefont [1]{#1}%
\providecommand \href@noop [0]{\@secondoftwo}%
\providecommand \href [0]{\begingroup \@sanitize@url \@href}%
\providecommand \@href[1]{\@@startlink{#1}\@@href}%
\providecommand \@@href[1]{\endgroup#1\@@endlink}%
\providecommand \@sanitize@url [0]{\catcode `\\12\catcode `\$12\catcode
  `\&12\catcode `\#12\catcode `\^12\catcode `\_12\catcode `\%12\relax}%
\providecommand \@@startlink[1]{}%
\providecommand \@@endlink[0]{}%
\providecommand \url  [0]{\begingroup\@sanitize@url \@url }%
\providecommand \@url [1]{\endgroup\@href {#1}{\urlprefix }}%
\providecommand \urlprefix  [0]{URL }%
\providecommand \Eprint [0]{\href }%
\providecommand \doibase [0]{https://doi.org/}%
\providecommand \selectlanguage [0]{\@gobble}%
\providecommand \bibinfo  [0]{\@secondoftwo}%
\providecommand \bibfield  [0]{\@secondoftwo}%
\providecommand \translation [1]{[#1]}%
\providecommand \BibitemOpen [0]{}%
\providecommand \bibitemStop [0]{}%
\providecommand \bibitemNoStop [0]{.\EOS\space}%
\providecommand \EOS [0]{\spacefactor3000\relax}%
\providecommand \BibitemShut  [1]{\csname bibitem#1\endcsname}%
\let\auto@bib@innerbib\@empty
\bibitem [{\citenamefont
  {Wannier}(1950)}]{wannierAntiferromagnetismTriangularIsing1950}%
  \BibitemOpen
  \bibfield  {author} {\bibinfo {author} {\bibfnamefont {G.~H.}\ \bibnamefont
  {Wannier}},\ }\bibfield  {title} {\bibinfo {title} {Antiferromagnetism. {{The
  Triangular Ising Net}}},\ }\href {https://doi.org/10.1103/PhysRev.79.357}
  {\bibfield  {journal} {\bibinfo  {journal} {Physical Review}\ }\textbf
  {\bibinfo {volume} {79}},\ \bibinfo {pages} {357} (\bibinfo {year}
  {1950})}\BibitemShut {NoStop}%
\bibitem [{\citenamefont {Balents}(2010)}]{balentsSpinLiquidsFrustrated2010}%
  \BibitemOpen
  \bibfield  {author} {\bibinfo {author} {\bibfnamefont {L.}~\bibnamefont
  {Balents}},\ }\bibfield  {title} {\bibinfo {title} {Spin liquids in
  frustrated magnets},\ }\href {https://doi.org/10.1038/nature08917} {\bibfield
   {journal} {\bibinfo  {journal} {Nature}\ }\textbf {\bibinfo {volume}
  {464}},\ \bibinfo {pages} {199} (\bibinfo {year} {2010})}\BibitemShut
  {NoStop}%
\bibitem [{\citenamefont {Lacroix}\ \emph {et~al.}(2011)\citenamefont
  {Lacroix}, \citenamefont {Mendels},\ and\ \citenamefont
  {Mila}}]{lacroixIntroductionFrustratedMagnetism2011}%
  \BibitemOpen
  \bibinfo {editor} {\bibfnamefont {C.}~\bibnamefont {Lacroix}}, \bibinfo
  {editor} {\bibfnamefont {P.}~\bibnamefont {Mendels}},\ and\ \bibinfo {editor}
  {\bibfnamefont {F.}~\bibnamefont {Mila}},\ eds.,\ \href
  {https://doi.org/10.1007/978-3-642-10589-0} {\emph {\bibinfo {title}
  {Introduction to {{Frustrated Magnetism}}: {{Materials}}, {{Experiments}},
  {{Theory}}}}},\ \bibinfo {edition} {1st}\ ed.,\ \bibinfo {series} {Springer
  {{Series}} in {{Solid-State Sciences}}}, Vol.\ \bibinfo {volume} {164}\
  (\bibinfo  {publisher} {Springer},\ \bibinfo {address} {Berlin},\ \bibinfo
  {year} {2011})\BibitemShut {NoStop}%
\bibitem [{\citenamefont {Chernyshev}\ and\ \citenamefont
  {Zhitomirsky}(2009)}]{chernyshevSpinWavesTriangular2009}%
  \BibitemOpen
  \bibfield  {author} {\bibinfo {author} {\bibfnamefont {A.~L.}\ \bibnamefont
  {Chernyshev}}\ and\ \bibinfo {author} {\bibfnamefont {M.~E.}\ \bibnamefont
  {Zhitomirsky}},\ }\bibfield  {title} {\bibinfo {title} {Spin waves in a
  triangular lattice antiferromagnet: {{Decays}}, spectrum renormalization, and
  singularities},\ }\href {https://doi.org/10.1103/PhysRevB.79.144416}
  {\bibfield  {journal} {\bibinfo  {journal} {Physical Review B}\ }\textbf
  {\bibinfo {volume} {79}},\ \bibinfo {pages} {144416} (\bibinfo {year}
  {2009})}\BibitemShut {NoStop}%
\bibitem [{\citenamefont {Ito}\ \emph {et~al.}(2017)\citenamefont {Ito},
  \citenamefont {Kurita}, \citenamefont {Tanaka}, \citenamefont
  {{Ohira-Kawamura}}, \citenamefont {Nakajima}, \citenamefont {Itoh},
  \citenamefont {Kuwahara},\ and\ \citenamefont
  {Kakurai}}]{itoStructureMagneticExcitations2017}%
  \BibitemOpen
  \bibfield  {author} {\bibinfo {author} {\bibfnamefont {S.}~\bibnamefont
  {Ito}}, \bibinfo {author} {\bibfnamefont {N.}~\bibnamefont {Kurita}},
  \bibinfo {author} {\bibfnamefont {H.}~\bibnamefont {Tanaka}}, \bibinfo
  {author} {\bibfnamefont {S.}~\bibnamefont {{Ohira-Kawamura}}}, \bibinfo
  {author} {\bibfnamefont {K.}~\bibnamefont {Nakajima}}, \bibinfo {author}
  {\bibfnamefont {S.}~\bibnamefont {Itoh}}, \bibinfo {author} {\bibfnamefont
  {K.}~\bibnamefont {Kuwahara}},\ and\ \bibinfo {author} {\bibfnamefont
  {K.}~\bibnamefont {Kakurai}},\ }\bibfield  {title} {\bibinfo {title}
  {Structure of the magnetic excitations in the spin-1/2 triangular-lattice
  {{Heisenberg}} antiferromagnet {{Ba$_3$CoSb2O$_9$}}},\ }\href
  {https://doi.org/10.1038/s41467-017-00316-x} {\bibfield  {journal} {\bibinfo
  {journal} {Nature Communications}\ }\textbf {\bibinfo {volume} {8}},\
  \bibinfo {pages} {235} (\bibinfo {year} {2017})}\BibitemShut {NoStop}%
\bibitem [{\citenamefont {Xie}\ \emph {et~al.}(2023)\citenamefont {Xie},
  \citenamefont {Eberharter}, \citenamefont {Xing}, \citenamefont {Nishimoto},
  \citenamefont {Brando}, \citenamefont {Khanenko}, \citenamefont
  {Sichelschmidt}, \citenamefont {Turrini}, \citenamefont {Mazzone},
  \citenamefont {Naumov}, \citenamefont {Sanjeewa}, \citenamefont {Harrison},
  \citenamefont {Sefat}, \citenamefont {Normand}, \citenamefont {L{\"a}uchli},
  \citenamefont {Podlesnyak},\ and\ \citenamefont
  {Nikitin}}]{xieCompleteFieldinducedSpectral2023a}%
  \BibitemOpen
  \bibfield  {author} {\bibinfo {author} {\bibfnamefont {T.}~\bibnamefont
  {Xie}}, \bibinfo {author} {\bibfnamefont {A.~A.}\ \bibnamefont {Eberharter}},
  \bibinfo {author} {\bibfnamefont {J.}~\bibnamefont {Xing}}, \bibinfo {author}
  {\bibfnamefont {S.}~\bibnamefont {Nishimoto}}, \bibinfo {author}
  {\bibfnamefont {M.}~\bibnamefont {Brando}}, \bibinfo {author} {\bibfnamefont
  {P.}~\bibnamefont {Khanenko}}, \bibinfo {author} {\bibfnamefont
  {J.}~\bibnamefont {Sichelschmidt}}, \bibinfo {author} {\bibfnamefont {A.~A.}\
  \bibnamefont {Turrini}}, \bibinfo {author} {\bibfnamefont {D.~G.}\
  \bibnamefont {Mazzone}}, \bibinfo {author} {\bibfnamefont {P.~G.}\
  \bibnamefont {Naumov}}, \bibinfo {author} {\bibfnamefont {L.~D.}\
  \bibnamefont {Sanjeewa}}, \bibinfo {author} {\bibfnamefont {N.}~\bibnamefont
  {Harrison}}, \bibinfo {author} {\bibfnamefont {A.~S.}\ \bibnamefont {Sefat}},
  \bibinfo {author} {\bibfnamefont {B.}~\bibnamefont {Normand}}, \bibinfo
  {author} {\bibfnamefont {A.~M.}\ \bibnamefont {L{\"a}uchli}}, \bibinfo
  {author} {\bibfnamefont {A.}~\bibnamefont {Podlesnyak}},\ and\ \bibinfo
  {author} {\bibfnamefont {S.~E.}\ \bibnamefont {Nikitin}},\ }\bibfield
  {title} {\bibinfo {title} {Complete field-induced spectral response of the
  spin-1/2 triangular-lattice antiferromagnet {{CsYbSe$_2$}}},\ }\href
  {https://doi.org/10.1038/s41535-023-00580-9} {\bibfield  {journal} {\bibinfo
  {journal} {npj Quantum Materials}\ }\textbf {\bibinfo {volume} {8}},\
  \bibinfo {pages} {48} (\bibinfo {year} {2023})}\BibitemShut {NoStop}%
\bibitem [{\citenamefont {Li}\ \emph {et~al.}(2020{\natexlab{a}})\citenamefont
  {Li}, \citenamefont {Gegenwart},\ and\ \citenamefont
  {Tsirlin}}]{liSpinLiquidsGeometrically2020}%
  \BibitemOpen
  \bibfield  {author} {\bibinfo {author} {\bibfnamefont {Y.}~\bibnamefont
  {Li}}, \bibinfo {author} {\bibfnamefont {P.}~\bibnamefont {Gegenwart}},\ and\
  \bibinfo {author} {\bibfnamefont {A.~A.}\ \bibnamefont {Tsirlin}},\
  }\bibfield  {title} {\bibinfo {title} {Spin liquids in geometrically perfect
  triangular antiferromagnets},\ }\href
  {https://doi.org/10.1088/1361-648X/ab724e} {\bibfield  {journal} {\bibinfo
  {journal} {Journal of Physics: Condensed Matter}\ }\textbf {\bibinfo {volume}
  {32}},\ \bibinfo {pages} {224004} (\bibinfo {year}
  {2020}{\natexlab{a}})}\BibitemShut {NoStop}%
\bibitem [{\citenamefont {Maksimov}\ \emph {et~al.}(2019)\citenamefont
  {Maksimov}, \citenamefont {Zhu}, \citenamefont {White},\ and\ \citenamefont
  {Chernyshev}}]{maksimovAnisotropicExchangeMagnetsTriangular2019}%
  \BibitemOpen
  \bibfield  {author} {\bibinfo {author} {\bibfnamefont {P.~A.}\ \bibnamefont
  {Maksimov}}, \bibinfo {author} {\bibfnamefont {Z.}~\bibnamefont {Zhu}},
  \bibinfo {author} {\bibfnamefont {S.~R.}\ \bibnamefont {White}},\ and\
  \bibinfo {author} {\bibfnamefont {A.~L.}\ \bibnamefont {Chernyshev}},\
  }\bibfield  {title} {\bibinfo {title} {Anisotropic-{{Exchange Magnets}} on a
  {{Triangular Lattice}}: {{Spin Waves}}, {{Accidental Degeneracies}}, and
  {{Dual Spin Liquids}}},\ }\href {https://doi.org/10.1103/PhysRevX.9.021017}
  {\bibfield  {journal} {\bibinfo  {journal} {Physical Review X}\ }\textbf
  {\bibinfo {volume} {9}},\ \bibinfo {pages} {021017} (\bibinfo {year}
  {2019})}\BibitemShut {NoStop}%
\bibitem [{\citenamefont {Zhu}\ \emph {et~al.}(2025)\citenamefont {Zhu},
  \citenamefont {Chinellato}, \citenamefont {Romerio}, \citenamefont {Murai},
  \citenamefont {{Ohira-Kawamura}}, \citenamefont {Balz}, \citenamefont {Yan},
  \citenamefont {Gvasaliya}, \citenamefont {Kato}, \citenamefont {Batista},\
  and\ \citenamefont {Zheludev}}]{zhuWannierStatesSpin2025}%
  \BibitemOpen
  \bibfield  {author} {\bibinfo {author} {\bibfnamefont {M.}~\bibnamefont
  {Zhu}}, \bibinfo {author} {\bibfnamefont {L.~M.}\ \bibnamefont {Chinellato}},
  \bibinfo {author} {\bibfnamefont {V.}~\bibnamefont {Romerio}}, \bibinfo
  {author} {\bibfnamefont {N.}~\bibnamefont {Murai}}, \bibinfo {author}
  {\bibfnamefont {S.}~\bibnamefont {{Ohira-Kawamura}}}, \bibinfo {author}
  {\bibfnamefont {C.}~\bibnamefont {Balz}}, \bibinfo {author} {\bibfnamefont
  {Z.}~\bibnamefont {Yan}}, \bibinfo {author} {\bibfnamefont {S.}~\bibnamefont
  {Gvasaliya}}, \bibinfo {author} {\bibfnamefont {Y.}~\bibnamefont {Kato}},
  \bibinfo {author} {\bibfnamefont {C.~D.}\ \bibnamefont {Batista}},\ and\
  \bibinfo {author} {\bibfnamefont {A.}~\bibnamefont {Zheludev}},\ }\bibfield
  {title} {\bibinfo {title} {Wannier states and spin supersolid physics in the
  triangular antiferromagnet {{K$_2$Co}}({{SeO$_3$}})$_2$},\ }\href
  {https://doi.org/10.1038/s41535-025-00791-2} {\bibfield  {journal} {\bibinfo
  {journal} {npj Quantum Materials}\ }\textbf {\bibinfo {volume} {10}},\
  \bibinfo {pages} {74} (\bibinfo {year} {2025})}\BibitemShut {NoStop}%
\bibitem [{\citenamefont {Woodland}\ \emph {et~al.}(2025)\citenamefont
  {Woodland}, \citenamefont {Okuma}, \citenamefont {Stewart}, \citenamefont
  {Balz},\ and\ \citenamefont
  {Coldea}}]{woodlandContinuumExcitationsSharp2025}%
  \BibitemOpen
  \bibfield  {author} {\bibinfo {author} {\bibfnamefont {L.}~\bibnamefont
  {Woodland}}, \bibinfo {author} {\bibfnamefont {R.}~\bibnamefont {Okuma}},
  \bibinfo {author} {\bibfnamefont {J.~R.}\ \bibnamefont {Stewart}}, \bibinfo
  {author} {\bibfnamefont {C.}~\bibnamefont {Balz}},\ and\ \bibinfo {author}
  {\bibfnamefont {R.}~\bibnamefont {Coldea}},\ }\bibfield  {title} {\bibinfo
  {title} {From continuum excitations to sharp magnons via transverse magnetic
  field in the spin-$\frac{1}{2}$ ising-like triangular lattice antiferromagnet
  ${{\mathrm{Na}}}_{2}\mathrm{BaCo}{({\mathrm{PO}}_{4})}_{2}$},\ }\href
  {https://doi.org/10.1103/1pvl-kzjm} {\bibfield  {journal} {\bibinfo
  {journal} {Phys. Rev. B}\ }\textbf {\bibinfo {volume} {112}},\ \bibinfo
  {pages} {104413} (\bibinfo {year} {2025})}\BibitemShut {NoStop}%
\bibitem [{\citenamefont {Scheie}\ \emph {et~al.}(2024)\citenamefont {Scheie},
  \citenamefont {Kamiya}, \citenamefont {Zhang}, \citenamefont {Lee},
  \citenamefont {Woods}, \citenamefont {Ajeesh}, \citenamefont {Gonzalez},
  \citenamefont {Bernu}, \citenamefont {Villanova}, \citenamefont {Xing},
  \citenamefont {Huang}, \citenamefont {Zhang}, \citenamefont {Ma},
  \citenamefont {Choi}, \citenamefont {Pajerowski}, \citenamefont {Zhou},
  \citenamefont {Sefat}, \citenamefont {Okamoto}, \citenamefont {Berlijn},
  \citenamefont {Messio}, \citenamefont {Movshovich}, \citenamefont {Batista},\
  and\ \citenamefont {Tennant}}]{scheieNonlinearMagnonsExchange2024}%
  \BibitemOpen
  \bibfield  {author} {\bibinfo {author} {\bibfnamefont {A.~O.}\ \bibnamefont
  {Scheie}}, \bibinfo {author} {\bibfnamefont {Y.}~\bibnamefont {Kamiya}},
  \bibinfo {author} {\bibfnamefont {H.}~\bibnamefont {Zhang}}, \bibinfo
  {author} {\bibfnamefont {S.}~\bibnamefont {Lee}}, \bibinfo {author}
  {\bibfnamefont {A.~J.}\ \bibnamefont {Woods}}, \bibinfo {author}
  {\bibfnamefont {M.~O.}\ \bibnamefont {Ajeesh}}, \bibinfo {author}
  {\bibfnamefont {M.~G.}\ \bibnamefont {Gonzalez}}, \bibinfo {author}
  {\bibfnamefont {B.}~\bibnamefont {Bernu}}, \bibinfo {author} {\bibfnamefont
  {J.~W.}\ \bibnamefont {Villanova}}, \bibinfo {author} {\bibfnamefont
  {J.}~\bibnamefont {Xing}}, \bibinfo {author} {\bibfnamefont {Q.}~\bibnamefont
  {Huang}}, \bibinfo {author} {\bibfnamefont {Q.}~\bibnamefont {Zhang}},
  \bibinfo {author} {\bibfnamefont {J.}~\bibnamefont {Ma}}, \bibinfo {author}
  {\bibfnamefont {E.~S.}\ \bibnamefont {Choi}}, \bibinfo {author}
  {\bibfnamefont {D.~M.}\ \bibnamefont {Pajerowski}}, \bibinfo {author}
  {\bibfnamefont {H.}~\bibnamefont {Zhou}}, \bibinfo {author} {\bibfnamefont
  {A.~S.}\ \bibnamefont {Sefat}}, \bibinfo {author} {\bibfnamefont
  {S.}~\bibnamefont {Okamoto}}, \bibinfo {author} {\bibfnamefont
  {T.}~\bibnamefont {Berlijn}}, \bibinfo {author} {\bibfnamefont
  {L.}~\bibnamefont {Messio}}, \bibinfo {author} {\bibfnamefont
  {R.}~\bibnamefont {Movshovich}}, \bibinfo {author} {\bibfnamefont {C.~D.}\
  \bibnamefont {Batista}},\ and\ \bibinfo {author} {\bibfnamefont {D.~A.}\
  \bibnamefont {Tennant}},\ }\bibfield  {title} {\bibinfo {title} {Nonlinear
  magnons and exchange {{Hamiltonians}} of the delafossite proximate quantum
  spin liquid candidates {{KYbSe$_2$}} and {{NaYbSe$_2$}}},\ }\href
  {https://doi.org/10.1103/PhysRevB.109.014425} {\bibfield  {journal} {\bibinfo
   {journal} {Physical Review B}\ }\textbf {\bibinfo {volume} {109}},\ \bibinfo
  {pages} {014425} (\bibinfo {year} {2024})}\BibitemShut {NoStop}%
\bibitem [{\citenamefont {Zhu}\ \emph {et~al.}(2017)\citenamefont {Zhu},
  \citenamefont {Maksimov}, \citenamefont {White},\ and\ \citenamefont
  {Chernyshev}}]{zhuDisorderInducedMimicrySpin2017}%
  \BibitemOpen
  \bibfield  {author} {\bibinfo {author} {\bibfnamefont {Z.}~\bibnamefont
  {Zhu}}, \bibinfo {author} {\bibfnamefont {P.~A.}\ \bibnamefont {Maksimov}},
  \bibinfo {author} {\bibfnamefont {S.~R.}\ \bibnamefont {White}},\ and\
  \bibinfo {author} {\bibfnamefont {A.~L.}\ \bibnamefont {Chernyshev}},\
  }\bibfield  {title} {\bibinfo {title} {Disorder-{{Induced Mimicry}} of a
  {{Spin Liquid}} in {{YbMgGaO$_4$}}},\ }\href
  {https://doi.org/10.1103/PhysRevLett.119.157201} {\bibfield  {journal}
  {\bibinfo  {journal} {Physical Review Letters}\ }\textbf {\bibinfo {volume}
  {119}},\ \bibinfo {pages} {157201} (\bibinfo {year} {2017})}\BibitemShut
  {NoStop}%
\bibitem [{\citenamefont {Jackeli}\ and\ \citenamefont
  {Khaliullin}(2009)}]{jackeliMottInsulatorsStrong2009}%
  \BibitemOpen
  \bibfield  {author} {\bibinfo {author} {\bibfnamefont {G.}~\bibnamefont
  {Jackeli}}\ and\ \bibinfo {author} {\bibfnamefont {G.}~\bibnamefont
  {Khaliullin}},\ }\bibfield  {title} {\bibinfo {title} {Mott {{Insulators}} in
  the {{Strong Spin-Orbit Coupling Limit}}: {{From Heisenberg}} to a {{Quantum
  Compass}} and {{Kitaev Models}}},\ }\href
  {https://doi.org/10.1103/PhysRevLett.102.017205} {\bibfield  {journal}
  {\bibinfo  {journal} {Physical Review Letters}\ }\textbf {\bibinfo {volume}
  {102}},\ \bibinfo {pages} {017205} (\bibinfo {year} {2009})}\BibitemShut
  {NoStop}%
\bibitem [{\citenamefont {Trebst}\ and\ \citenamefont
  {Hickey}(2022)}]{trebstKitaevMaterials2022}%
  \BibitemOpen
  \bibfield  {author} {\bibinfo {author} {\bibfnamefont {S.}~\bibnamefont
  {Trebst}}\ and\ \bibinfo {author} {\bibfnamefont {C.}~\bibnamefont
  {Hickey}},\ }\bibfield  {title} {\bibinfo {title} {Kitaev materials},\ }\href
  {https://doi.org/https://doi.org/10.1016/j.physrep.2021.11.003} {\bibfield
  {journal} {\bibinfo  {journal} {Physics Reports}\ }\textbf {\bibinfo {volume}
  {950}},\ \bibinfo {pages} {1} (\bibinfo {year} {2022})}\BibitemShut {NoStop}%
\bibitem [{\citenamefont {Zhou}\ \emph {et~al.}(2025)\citenamefont {Zhou},
  \citenamefont {Li}, \citenamefont {Liang},\ and\ \citenamefont
  {Zhou}}]{zhouTopologicalSpinTextures2025a}%
  \BibitemOpen
  \bibfield  {author} {\bibinfo {author} {\bibfnamefont {Y.}~\bibnamefont
  {Zhou}}, \bibinfo {author} {\bibfnamefont {S.}~\bibnamefont {Li}}, \bibinfo
  {author} {\bibfnamefont {X.}~\bibnamefont {Liang}},\ and\ \bibinfo {author}
  {\bibfnamefont {Y.}~\bibnamefont {Zhou}},\ }\bibfield  {title} {\bibinfo
  {title} {Topological {{Spin Textures}}: {{Basic Physics}} and {{Devices}}},\
  }\href {https://doi.org/10.1002/adma.202312935} {\bibfield  {journal}
  {\bibinfo  {journal} {Advanced Materials}\ }\textbf {\bibinfo {volume}
  {37}},\ \bibinfo {pages} {2312935} (\bibinfo {year} {2025})}\BibitemShut
  {NoStop}%
\bibitem [{\citenamefont {Kawamura}\ and\ \citenamefont
  {Miyashita}(1984)}]{kawamuraPhaseTransitionTwoDimensional1984}%
  \BibitemOpen
  \bibfield  {author} {\bibinfo {author} {\bibfnamefont {H.}~\bibnamefont
  {Kawamura}}\ and\ \bibinfo {author} {\bibfnamefont {S.}~\bibnamefont
  {Miyashita}},\ }\bibfield  {title} {\bibinfo {title} {Phase {{Transition}} of
  the {{Two-Dimensional Heisenberg Antiferromagnet}} on the {{Triangular
  Lattice}}},\ }\href {https://doi.org/10.1143/JPSJ.53.4138} {\bibfield
  {journal} {\bibinfo  {journal} {Journal of the Physical Society of Japan}\
  }\textbf {\bibinfo {volume} {53}},\ \bibinfo {pages} {4138} (\bibinfo {year}
  {1984})}\BibitemShut {NoStop}%
\bibitem [{\citenamefont {Rousochatzakis}\ \emph {et~al.}(2016)\citenamefont
  {Rousochatzakis}, \citenamefont {R{\"o}ssler}, \citenamefont {{van den
  Brink}},\ and\ \citenamefont
  {Daghofer}}]{rousochatzakisKitaevAnisotropyInduces2016}%
  \BibitemOpen
  \bibfield  {author} {\bibinfo {author} {\bibfnamefont {I.}~\bibnamefont
  {Rousochatzakis}}, \bibinfo {author} {\bibfnamefont {U.~K.}\ \bibnamefont
  {R{\"o}ssler}}, \bibinfo {author} {\bibfnamefont {J.}~\bibnamefont {{van den
  Brink}}},\ and\ \bibinfo {author} {\bibfnamefont {M.}~\bibnamefont
  {Daghofer}},\ }\bibfield  {title} {\bibinfo {title} {Kitaev anisotropy
  induces mesoscopic {{$\mathbb{Z}_2$}} vortex crystals in frustrated hexagonal
  antiferromagnets},\ }\href {https://doi.org/10.1103/PhysRevB.93.104417}
  {\bibfield  {journal} {\bibinfo  {journal} {Physical Review B}\ }\textbf
  {\bibinfo {volume} {93}},\ \bibinfo {pages} {104417} (\bibinfo {year}
  {2016})}\BibitemShut {NoStop}%
\bibitem [{\citenamefont {Skyrme}(1962)}]{skyrmeUnifiedFieldTheory1962}%
  \BibitemOpen
  \bibfield  {author} {\bibinfo {author} {\bibfnamefont {T.~H.~R.}\
  \bibnamefont {Skyrme}},\ }\bibfield  {title} {\bibinfo {title} {A unified
  field theory of mesons and baryons},\ }\href
  {https://doi.org/10.1016/0029-5582(62)90775-7} {\bibfield  {journal}
  {\bibinfo  {journal} {Nuclear Physics}\ }\textbf {\bibinfo {volume} {31}},\
  \bibinfo {pages} {556} (\bibinfo {year} {1962})}\BibitemShut {NoStop}%
\bibitem [{\citenamefont {Bogdanov}\ and\ \citenamefont
  {Yablonskii}(1989)}]{bogdanovThermodynamicallyStableVortices1989}%
  \BibitemOpen
  \bibfield  {author} {\bibinfo {author} {\bibfnamefont {A.~N.}\ \bibnamefont
  {Bogdanov}}\ and\ \bibinfo {author} {\bibfnamefont {D.~A.}\ \bibnamefont
  {Yablonskii}},\ }\bibfield  {title} {\bibinfo {title} {Thermodynamically
  stable ``vortices'' in magnetically ordered crystals. {{The}} mixed state of
  magnets},\ }\href {http://www.jetp.ras.ru/cgi-bin/e/index/r/95/1/p178?a=list}
  {\bibfield  {journal} {\bibinfo  {journal} {Zh. Eksp. Teor. Fiz}\ }\textbf
  {\bibinfo {volume} {95}},\ \bibinfo {pages} {178} (\bibinfo {year}
  {1989})}\BibitemShut {NoStop}%
\bibitem [{\citenamefont {M{\"u}hlbauer}\ \emph {et~al.}(2009)\citenamefont
  {M{\"u}hlbauer}, \citenamefont {Binz}, \citenamefont {Jonietz}, \citenamefont
  {Pfleiderer}, \citenamefont {Rosch}, \citenamefont {Neubauer}, \citenamefont
  {Georgii},\ and\ \citenamefont
  {B{\"o}ni}}]{muhlbauerSkyrmionLatticeChiral2009a}%
  \BibitemOpen
  \bibfield  {author} {\bibinfo {author} {\bibfnamefont {S.}~\bibnamefont
  {M{\"u}hlbauer}}, \bibinfo {author} {\bibfnamefont {B.}~\bibnamefont {Binz}},
  \bibinfo {author} {\bibfnamefont {F.}~\bibnamefont {Jonietz}}, \bibinfo
  {author} {\bibfnamefont {C.}~\bibnamefont {Pfleiderer}}, \bibinfo {author}
  {\bibfnamefont {A.}~\bibnamefont {Rosch}}, \bibinfo {author} {\bibfnamefont
  {A.}~\bibnamefont {Neubauer}}, \bibinfo {author} {\bibfnamefont
  {R.}~\bibnamefont {Georgii}},\ and\ \bibinfo {author} {\bibfnamefont
  {P.}~\bibnamefont {B{\"o}ni}},\ }\bibfield  {title} {\bibinfo {title}
  {Skyrmion {{Lattice}} in a {{Chiral Magnet}}},\ }\href
  {https://doi.org/10.1126/science.1166767} {\bibfield  {journal} {\bibinfo
  {journal} {Science}\ }\textbf {\bibinfo {volume} {323}},\ \bibinfo {pages}
  {915} (\bibinfo {year} {2009})}\BibitemShut {NoStop}%
\bibitem [{\citenamefont {Seki}\ \emph {et~al.}(2012)\citenamefont {Seki},
  \citenamefont {Yu}, \citenamefont {Ishiwata},\ and\ \citenamefont
  {Tokura}}]{sekiObservationSkyrmionsMultiferroic2012}%
  \BibitemOpen
  \bibfield  {author} {\bibinfo {author} {\bibfnamefont {S.}~\bibnamefont
  {Seki}}, \bibinfo {author} {\bibfnamefont {X.~Z.}\ \bibnamefont {Yu}},
  \bibinfo {author} {\bibfnamefont {S.}~\bibnamefont {Ishiwata}},\ and\
  \bibinfo {author} {\bibfnamefont {Y.}~\bibnamefont {Tokura}},\ }\bibfield
  {title} {\bibinfo {title} {Observation of {{Skyrmions}} in a {{Multiferroic
  Material}}},\ }\href {https://doi.org/10.1126/science.1214143} {\bibfield
  {journal} {\bibinfo  {journal} {Science}\ }\textbf {\bibinfo {volume}
  {336}},\ \bibinfo {pages} {198} (\bibinfo {year} {2012})}\BibitemShut
  {NoStop}%
\bibitem [{\citenamefont
  {Abrikosov}(1957)}]{abrikosovMagneticPropertiesSuperconductors1957}%
  \BibitemOpen
  \bibfield  {author} {\bibinfo {author} {\bibfnamefont {A.~A.}\ \bibnamefont
  {Abrikosov}},\ }\bibfield  {title} {\bibinfo {title} {On the {{Magnetic
  Properties}} of {{Superconductors}} of the {{Second Group}}},\ }\href
  {http://jetp.ras.ru/cgi-bin/e/index/e/5/6/p1174?a=list} {\bibfield  {journal}
  {\bibinfo  {journal} {Sov. Phys. JETP}\ }\textbf {\bibinfo {volume} {5}},\
  \bibinfo {pages} {1174} (\bibinfo {year} {1957})}\BibitemShut {NoStop}%
\bibitem [{\citenamefont {Essmann}\ and\ \citenamefont
  {Tr{\"a}uble}(1967)}]{essmannDirectObservationIndividual1967}%
  \BibitemOpen
  \bibfield  {author} {\bibinfo {author} {\bibfnamefont {U.}~\bibnamefont
  {Essmann}}\ and\ \bibinfo {author} {\bibfnamefont {H.}~\bibnamefont
  {Tr{\"a}uble}},\ }\bibfield  {title} {\bibinfo {title} {The direct
  observation of individual flux lines in type {{II}} superconductors},\ }\href
  {https://doi.org/10.1016/0375-9601(67)90819-5} {\bibfield  {journal}
  {\bibinfo  {journal} {Physics Letters A}\ }\textbf {\bibinfo {volume} {24}},\
  \bibinfo {pages} {526} (\bibinfo {year} {1967})}\BibitemShut {NoStop}%
\bibitem [{\citenamefont {Hess}\ \emph {et~al.}(1989)\citenamefont {Hess},
  \citenamefont {Robinson}, \citenamefont {Dynes}, \citenamefont {Valles},\
  and\ \citenamefont
  {Waszczak}}]{hessScanningTunnelingMicroscopeObservationAbrikosov1989}%
  \BibitemOpen
  \bibfield  {author} {\bibinfo {author} {\bibfnamefont {H.~F.}\ \bibnamefont
  {Hess}}, \bibinfo {author} {\bibfnamefont {R.~B.}\ \bibnamefont {Robinson}},
  \bibinfo {author} {\bibfnamefont {R.~C.}\ \bibnamefont {Dynes}}, \bibinfo
  {author} {\bibfnamefont {J.~M.}\ \bibnamefont {Valles}},\ and\ \bibinfo
  {author} {\bibfnamefont {J.~V.}\ \bibnamefont {Waszczak}},\ }\bibfield
  {title} {\bibinfo {title} {Scanning-{{Tunneling-Microscope Observation}} of
  the {{Abrikosov Flux Lattice}} and the {{Density}} of {{States}} near and
  inside a {{Fluxoid}}},\ }\href {https://doi.org/10.1103/PhysRevLett.62.214}
  {\bibfield  {journal} {\bibinfo  {journal} {Physical Review Letters}\
  }\textbf {\bibinfo {volume} {62}},\ \bibinfo {pages} {214} (\bibinfo {year}
  {1989})}\BibitemShut {NoStop}%
\bibitem [{\citenamefont {Becker}\ \emph {et~al.}(2015)\citenamefont {Becker},
  \citenamefont {Hermanns}, \citenamefont {Bauer}, \citenamefont {Garst},\ and\
  \citenamefont {Trebst}}]{beckerSpinorbitPhysicsJ12015}%
  \BibitemOpen
  \bibfield  {author} {\bibinfo {author} {\bibfnamefont {M.}~\bibnamefont
  {Becker}}, \bibinfo {author} {\bibfnamefont {M.}~\bibnamefont {Hermanns}},
  \bibinfo {author} {\bibfnamefont {B.}~\bibnamefont {Bauer}}, \bibinfo
  {author} {\bibfnamefont {M.}~\bibnamefont {Garst}},\ and\ \bibinfo {author}
  {\bibfnamefont {S.}~\bibnamefont {Trebst}},\ }\bibfield  {title} {\bibinfo
  {title} {Spin-orbit physics of {{$j=\frac{1}{2}$}} {{Mott}} insulators on the
  triangular lattice},\ }\href {https://doi.org/10.1103/PhysRevB.91.155135}
  {\bibfield  {journal} {\bibinfo  {journal} {Physical Review B}\ }\textbf
  {\bibinfo {volume} {91}},\ \bibinfo {pages} {155135} (\bibinfo {year}
  {2015})}\BibitemShut {NoStop}%
\bibitem [{\citenamefont {Catuneanu}\ \emph {et~al.}(2015)\citenamefont
  {Catuneanu}, \citenamefont {Rau}, \citenamefont {Kim},\ and\ \citenamefont
  {Kee}}]{catuneanuMagneticOrdersProximal2015}%
  \BibitemOpen
  \bibfield  {author} {\bibinfo {author} {\bibfnamefont {A.}~\bibnamefont
  {Catuneanu}}, \bibinfo {author} {\bibfnamefont {J.~G.}\ \bibnamefont {Rau}},
  \bibinfo {author} {\bibfnamefont {H.-S.}\ \bibnamefont {Kim}},\ and\ \bibinfo
  {author} {\bibfnamefont {H.-Y.}\ \bibnamefont {Kee}},\ }\bibfield  {title}
  {\bibinfo {title} {Magnetic orders proximal to the {{Kitaev}} limit in
  frustrated triangular systems: {{Application}} to {{Ba$_3$IrTi$_2$O$_9$}}},\
  }\href {https://doi.org/10.1103/PhysRevB.92.165108} {\bibfield  {journal}
  {\bibinfo  {journal} {Physical Review B}\ }\textbf {\bibinfo {volume} {92}},\
  \bibinfo {pages} {165108} (\bibinfo {year} {2015})}\BibitemShut {NoStop}%
\bibitem [{\citenamefont {Wang}\ \emph {et~al.}(2015)\citenamefont {Wang},
  \citenamefont {Kamiya}, \citenamefont {Nevidomskyy},\ and\ \citenamefont
  {Batista}}]{wangThreeDimensionalCrystallizationVortex2015}%
  \BibitemOpen
  \bibfield  {author} {\bibinfo {author} {\bibfnamefont {Z.}~\bibnamefont
  {Wang}}, \bibinfo {author} {\bibfnamefont {Y.}~\bibnamefont {Kamiya}},
  \bibinfo {author} {\bibfnamefont {A.~H.}\ \bibnamefont {Nevidomskyy}},\ and\
  \bibinfo {author} {\bibfnamefont {C.~D.}\ \bibnamefont {Batista}},\
  }\bibfield  {title} {\bibinfo {title} {Three-{{Dimensional Crystallization}}
  of {{Vortex Strings}} in {{Frustrated Quantum Magnets}}},\ }\href
  {https://doi.org/10.1103/PhysRevLett.115.107201} {\bibfield  {journal}
  {\bibinfo  {journal} {Physical Review Letters}\ }\textbf {\bibinfo {volume}
  {115}},\ \bibinfo {pages} {107201} (\bibinfo {year} {2015})}\BibitemShut
  {NoStop}%
\bibitem [{\citenamefont {Yao}\ and\ \citenamefont
  {Dong}(2018)}]{yaoStabilizationModulationTopological2018}%
  \BibitemOpen
  \bibfield  {author} {\bibinfo {author} {\bibfnamefont {X.}~\bibnamefont
  {Yao}}\ and\ \bibinfo {author} {\bibfnamefont {S.}~\bibnamefont {Dong}},\
  }\bibfield  {title} {\bibinfo {title} {Stabilization and modulation of the
  topological magnetic phase with a {{$\mathbb{Z}_2$}}-vortex lattice in the
  {{Kitaev-Heisenberg}} honeycomb model: {{The}} key role of the
  third-nearest-neighbor interaction},\ }\href
  {https://doi.org/10.1103/PhysRevB.98.054413} {\bibfield  {journal} {\bibinfo
  {journal} {Physical Review B}\ }\textbf {\bibinfo {volume} {98}},\ \bibinfo
  {pages} {054413} (\bibinfo {year} {2018})}\BibitemShut {NoStop}%
\bibitem [{\citenamefont {Kishimoto}\ \emph {et~al.}(2018)\citenamefont
  {Kishimoto}, \citenamefont {Morita}, \citenamefont {Matsubayashi},
  \citenamefont {Sota}, \citenamefont {Yunoki},\ and\ \citenamefont
  {Tohyama}}]{kishimotoGroundStatePhase2018}%
  \BibitemOpen
  \bibfield  {author} {\bibinfo {author} {\bibfnamefont {M.}~\bibnamefont
  {Kishimoto}}, \bibinfo {author} {\bibfnamefont {K.}~\bibnamefont {Morita}},
  \bibinfo {author} {\bibfnamefont {Y.}~\bibnamefont {Matsubayashi}}, \bibinfo
  {author} {\bibfnamefont {S.}~\bibnamefont {Sota}}, \bibinfo {author}
  {\bibfnamefont {S.}~\bibnamefont {Yunoki}},\ and\ \bibinfo {author}
  {\bibfnamefont {T.}~\bibnamefont {Tohyama}},\ }\bibfield  {title} {\bibinfo
  {title} {Ground state phase diagram of the {{Kitaev-Heisenberg}} model on a
  honeycomb-triangular lattice},\ }\href
  {https://doi.org/10.1103/PhysRevB.98.054411} {\bibfield  {journal} {\bibinfo
  {journal} {Physical Review B}\ }\textbf {\bibinfo {volume} {98}},\ \bibinfo
  {pages} {054411} (\bibinfo {year} {2018})}\BibitemShut {NoStop}%
\bibitem [{\citenamefont {Li}\ \emph {et~al.}(2019{\natexlab{a}})\citenamefont
  {Li}, \citenamefont {Perkins},\ and\ \citenamefont
  {Rousochatzakis}}]{liCollectiveSpinDynamics2019}%
  \BibitemOpen
  \bibfield  {author} {\bibinfo {author} {\bibfnamefont {M.}~\bibnamefont
  {Li}}, \bibinfo {author} {\bibfnamefont {N.~B.}\ \bibnamefont {Perkins}},\
  and\ \bibinfo {author} {\bibfnamefont {I.}~\bibnamefont {Rousochatzakis}},\
  }\bibfield  {title} {\bibinfo {title} {Collective spin dynamics of
  {{$\mathbb{Z}_2$}} vortex crystals in triangular {{Kitaev-Heisenberg}}
  antiferromagnets},\ }\href {https://doi.org/10.1103/PhysRevResearch.1.013002}
  {\bibfield  {journal} {\bibinfo  {journal} {Physical Review Research}\
  }\textbf {\bibinfo {volume} {1}},\ \bibinfo {pages} {013002} (\bibinfo {year}
  {2019}{\natexlab{a}})}\BibitemShut {NoStop}%
\bibitem [{\citenamefont {Osorio}\ \emph {et~al.}(2019)\citenamefont {Osorio},
  \citenamefont {Sturla}, \citenamefont {Rosales},\ and\ \citenamefont
  {Cabra}}]{osorioSkyrmions$mathbbZ_2$Vortices2019}%
  \BibitemOpen
  \bibfield  {author} {\bibinfo {author} {\bibfnamefont {S.~A.}\ \bibnamefont
  {Osorio}}, \bibinfo {author} {\bibfnamefont {M.~B.}\ \bibnamefont {Sturla}},
  \bibinfo {author} {\bibfnamefont {H.~D.}\ \bibnamefont {Rosales}},\ and\
  \bibinfo {author} {\bibfnamefont {D.~C.}\ \bibnamefont {Cabra}},\ }\bibfield
  {title} {\bibinfo {title} {From skyrmions to {{$\mathbb{Z}_2$}} vortices in
  distorted chiral antiferromagnets},\ }\href
  {https://doi.org/10.1103/PhysRevB.100.220404} {\bibfield  {journal} {\bibinfo
   {journal} {Physical Review B}\ }\textbf {\bibinfo {volume} {100}},\ \bibinfo
  {pages} {220404} (\bibinfo {year} {2019})}\BibitemShut {NoStop}%
\bibitem [{\citenamefont {Seabrook}\ \emph {et~al.}(2020)\citenamefont
  {Seabrook}, \citenamefont {Baez},\ and\ \citenamefont
  {Reuther}}]{seabrook$mathbbZ_2$VorticesGround2020}%
  \BibitemOpen
  \bibfield  {author} {\bibinfo {author} {\bibfnamefont {E.}~\bibnamefont
  {Seabrook}}, \bibinfo {author} {\bibfnamefont {M.~L.}\ \bibnamefont {Baez}},\
  and\ \bibinfo {author} {\bibfnamefont {J.}~\bibnamefont {Reuther}},\
  }\bibfield  {title} {\bibinfo {title} {{{$\mathbb{Z}_2$}} vortices in the
  ground states of classical {{Kitaev-Heisenberg}} models},\ }\href
  {https://doi.org/10.1103/PhysRevB.101.174443} {\bibfield  {journal} {\bibinfo
   {journal} {Physical Review B}\ }\textbf {\bibinfo {volume} {101}},\ \bibinfo
  {pages} {174443} (\bibinfo {year} {2020})}\BibitemShut {NoStop}%
\bibitem [{\citenamefont {Shinjo}\ \emph {et~al.}(2016)\citenamefont {Shinjo},
  \citenamefont {Sota}, \citenamefont {Yunoki}, \citenamefont {Totsuka},\ and\
  \citenamefont {Tohyama}}]{shinjoDensityMatrixRenormalizationGroup2016}%
  \BibitemOpen
  \bibfield  {author} {\bibinfo {author} {\bibfnamefont {K.}~\bibnamefont
  {Shinjo}}, \bibinfo {author} {\bibfnamefont {S.}~\bibnamefont {Sota}},
  \bibinfo {author} {\bibfnamefont {S.}~\bibnamefont {Yunoki}}, \bibinfo
  {author} {\bibfnamefont {K.}~\bibnamefont {Totsuka}},\ and\ \bibinfo {author}
  {\bibfnamefont {T.}~\bibnamefont {Tohyama}},\ }\bibfield  {title} {\bibinfo
  {title} {Density-{{Matrix Renormalization Group Study}} of
  {{Kitaev}}--{{Heisenberg Model}} on a {{Triangular Lattice}}},\ }\href
  {https://doi.org/10.7566/JPSJ.85.114710} {\bibfield  {journal} {\bibinfo
  {journal} {Journal of the Physical Society of Japan}\ }\textbf {\bibinfo
  {volume} {85}},\ \bibinfo {pages} {114710} (\bibinfo {year}
  {2016})}\BibitemShut {NoStop}%
\bibitem [{\citenamefont {Bhattacharyya}\ \emph {et~al.}(2023)\citenamefont
  {Bhattacharyya}, \citenamefont {Bogdanov}, \citenamefont {Nishimoto},
  \citenamefont {Wilson},\ and\ \citenamefont
  {Hozoi}}]{bhattacharyyaNaRuO2KitaevHeisenbergExchange2023}%
  \BibitemOpen
  \bibfield  {author} {\bibinfo {author} {\bibfnamefont {P.}~\bibnamefont
  {Bhattacharyya}}, \bibinfo {author} {\bibfnamefont {N.~A.}\ \bibnamefont
  {Bogdanov}}, \bibinfo {author} {\bibfnamefont {S.}~\bibnamefont {Nishimoto}},
  \bibinfo {author} {\bibfnamefont {S.~D.}\ \bibnamefont {Wilson}},\ and\
  \bibinfo {author} {\bibfnamefont {L.}~\bibnamefont {Hozoi}},\ }\bibfield
  {title} {\bibinfo {title} {{{NaRuO$_2$}}: {{Kitaev-Heisenberg}} exchange in
  triangular-lattice setting},\ }\href
  {https://doi.org/10.1038/s41535-023-00582-7} {\bibfield  {journal} {\bibinfo
  {journal} {npj Quantum Materials}\ }\textbf {\bibinfo {volume} {8}},\
  \bibinfo {pages} {52} (\bibinfo {year} {2023})}\BibitemShut {NoStop}%
\bibitem [{\citenamefont {Vishnoi}\ \emph {et~al.}(2020)\citenamefont
  {Vishnoi}, \citenamefont {Zuo}, \citenamefont {Strom}, \citenamefont {Wu},
  \citenamefont {Wilson}, \citenamefont {Seshadri},\ and\ \citenamefont
  {Cheetham}}]{vishnoiStructuralDiversityMagnetic2020}%
  \BibitemOpen
  \bibfield  {author} {\bibinfo {author} {\bibfnamefont {P.}~\bibnamefont
  {Vishnoi}}, \bibinfo {author} {\bibfnamefont {J.~L.}\ \bibnamefont {Zuo}},
  \bibinfo {author} {\bibfnamefont {T.~A.}\ \bibnamefont {Strom}}, \bibinfo
  {author} {\bibfnamefont {G.}~\bibnamefont {Wu}}, \bibinfo {author}
  {\bibfnamefont {S.~D.}\ \bibnamefont {Wilson}}, \bibinfo {author}
  {\bibfnamefont {R.}~\bibnamefont {Seshadri}},\ and\ \bibinfo {author}
  {\bibfnamefont {A.~K.}\ \bibnamefont {Cheetham}},\ }\bibfield  {title}
  {\bibinfo {title} {Structural {{Diversity}} and {{Magnetic Properties}} of
  {{Hybrid Ruthenium Halide Perovskites}} and {{Related Compounds}}},\ }\href
  {https://doi.org/10.1002/anie.202003095} {\bibfield  {journal} {\bibinfo
  {journal} {Angewandte Chemie International Edition}\ }\textbf {\bibinfo
  {volume} {59}},\ \bibinfo {pages} {8974} (\bibinfo {year}
  {2020})}\BibitemShut {NoStop}%
\bibitem [{sup()}]{supplement}%
  \BibitemOpen
  \href@noop {} {\bibinfo {title} {See {{Supplemental Material}} for further
  details about the low-temperature crystal structure, the single-ion fitting
  procedures \& {{ESR}} data, the magnetoelastic coupling, our inelastic
  neutron spectroscopy experiment and the proposed magnetic structures,
  including {{Supplemental Figures}} \hyperlinkcite{supplement}{S1-11},
  {{Tables}} \hyperlinkcite{supplement}{T1-5}, as well as {{Refs.}}
  \cite{waskowskaTemperaturePressuredependentLattice2010a,
  kajitaFerroaxialTransitionsGlaseriteType2024,
  hearmonElectricFieldControl2012, zelenskiyMagneticPhasesS52021,
  kubotaSuccessiveMagneticPhase2015, bergmanModelsDegeneracyBreaking2006,
  capriottiLongRangeNeelOrder1999, zhengExcitationSpectraSpin$frac12$2006,
  whiteNeelOrderSquare2007a, mouhatNecessarySufficientElastic2014,
  villainOrderEffectDisorder1980, heinilaSelectionGroundState1993c,
  pelissettoCriticalPhenomena2002}.}}\BibitemShut {Stop}%
\bibitem [{\citenamefont {Ponomaryov}\ \emph {et~al.}(2020)\citenamefont
  {Ponomaryov}, \citenamefont {Zviagina}, \citenamefont {Wosnitza},
  \citenamefont {Lampen-Kelley}, \citenamefont {Banerjee}, \citenamefont {Yan},
  \citenamefont {Bridges}, \citenamefont {Mandrus}, \citenamefont {Nagler},\
  and\ \citenamefont {Zvyagin}}]{ponomaryov_PRL_NatureMagExc2013}%
  \BibitemOpen
  \bibfield  {author} {\bibinfo {author} {\bibfnamefont {A.~N.}\ \bibnamefont
  {Ponomaryov}}, \bibinfo {author} {\bibfnamefont {L.}~\bibnamefont
  {Zviagina}}, \bibinfo {author} {\bibfnamefont {J.}~\bibnamefont {Wosnitza}},
  \bibinfo {author} {\bibfnamefont {P.}~\bibnamefont {Lampen-Kelley}}, \bibinfo
  {author} {\bibfnamefont {A.}~\bibnamefont {Banerjee}}, \bibinfo {author}
  {\bibfnamefont {J.-Q.}\ \bibnamefont {Yan}}, \bibinfo {author} {\bibfnamefont
  {C.~A.}\ \bibnamefont {Bridges}}, \bibinfo {author} {\bibfnamefont {D.~G.}\
  \bibnamefont {Mandrus}}, \bibinfo {author} {\bibfnamefont {S.~E.}\
  \bibnamefont {Nagler}},\ and\ \bibinfo {author} {\bibfnamefont {S.~A.}\
  \bibnamefont {Zvyagin}},\ }\bibfield  {title} {\bibinfo {title} {Nature of
  magnetic excitations in the high-field phase of
  {{$\ensuremath{\alpha}\text{\ensuremath{-}}{\mathrm{RuCl}}_{3}$}}},\ }\href
  {https://doi.org/10.1103/PhysRevLett.125.037202} {\bibfield  {journal}
  {\bibinfo  {journal} {Phys. Rev. Lett.}\ }\textbf {\bibinfo {volume} {125}},\
  \bibinfo {pages} {037202} (\bibinfo {year} {2020})}\BibitemShut {NoStop}%
\bibitem [{\citenamefont {Geschwind}\ and\ \citenamefont
  {Remeika}(1962)}]{geschwindSpinResonanceTransition1962}%
  \BibitemOpen
  \bibfield  {author} {\bibinfo {author} {\bibfnamefont {S.}~\bibnamefont
  {Geschwind}}\ and\ \bibinfo {author} {\bibfnamefont {J.~P.}\ \bibnamefont
  {Remeika}},\ }\bibfield  {title} {\bibinfo {title} {Spin {{Resonance}} of
  {{Transition Metal Ions}} in {{Corundum}}},\ }\href
  {https://doi.org/10.1063/1.1777126} {\bibfield  {journal} {\bibinfo
  {journal} {Journal of Applied Physics}\ }\textbf {\bibinfo {volume} {33}},\
  \bibinfo {pages} {370} (\bibinfo {year} {1962})}\BibitemShut {NoStop}%
\bibitem [{\citenamefont {Abragam}\ and\ \citenamefont
  {Bleaney}(1970)}]{abragamElectronParamagneticResonance}%
  \BibitemOpen
  \bibfield  {author} {\bibinfo {author} {\bibfnamefont {A.}~\bibnamefont
  {Abragam}}\ and\ \bibinfo {author} {\bibfnamefont {B.}~\bibnamefont
  {Bleaney}},\ }\href
  {https://global.oup.com/academic/product/electron-paramagnetic-resonance-of-transition-ions-9780199651528?cc=ch&lang=en&}
  {\emph {\bibinfo {title} {Electron {{Paramagnetic Resonance}} of {{Transition
  Ions}}}}},\ \bibinfo {edition} {1st}\ ed.,\ Oxford classic texts in the
  physical sciences\ (\bibinfo  {publisher} {Oxford University Press},\
  \bibinfo {year} {1970})\BibitemShut {NoStop}%
\bibitem [{\citenamefont {Liu}\ \emph {et~al.}(2022)\citenamefont {Liu},
  \citenamefont {Chaloupka},\ and\ \citenamefont
  {Khaliullin}}]{liuExchangeInteractionsD52022}%
  \BibitemOpen
  \bibfield  {author} {\bibinfo {author} {\bibfnamefont {H.}~\bibnamefont
  {Liu}}, \bibinfo {author} {\bibfnamefont {J.}~\bibnamefont {Chaloupka}},\
  and\ \bibinfo {author} {\bibfnamefont {G.}~\bibnamefont {Khaliullin}},\
  }\bibfield  {title} {\bibinfo {title} {Exchange interactions in $d^5$
  {{Kitaev}} materials: {{From Na$_2$IrO$_3$}} to {{$\alpha$-RuCl$_3$}}},\
  }\href {https://doi.org/10.1103/PhysRevB.105.214411} {\bibfield  {journal}
  {\bibinfo  {journal} {Physical Review B}\ }\textbf {\bibinfo {volume}
  {105}},\ \bibinfo {pages} {214411} (\bibinfo {year} {2022})}\BibitemShut
  {NoStop}%
\bibitem [{\citenamefont {Winter}\ \emph {et~al.}(2017)\citenamefont {Winter},
  \citenamefont {Tsirlin}, \citenamefont {Daghofer}, \citenamefont {van~den
  Brink}, \citenamefont {Singh}, \citenamefont {Gegenwart},\ and\ \citenamefont
  {Valent{\'i}}}]{winterModelsMaterialsGeneralized2017}%
  \BibitemOpen
  \bibfield  {author} {\bibinfo {author} {\bibfnamefont {S.~M.}\ \bibnamefont
  {Winter}}, \bibinfo {author} {\bibfnamefont {A.~A.}\ \bibnamefont {Tsirlin}},
  \bibinfo {author} {\bibfnamefont {M.}~\bibnamefont {Daghofer}}, \bibinfo
  {author} {\bibfnamefont {J.}~\bibnamefont {van~den Brink}}, \bibinfo {author}
  {\bibfnamefont {Y.}~\bibnamefont {Singh}}, \bibinfo {author} {\bibfnamefont
  {P.}~\bibnamefont {Gegenwart}},\ and\ \bibinfo {author} {\bibfnamefont
  {R.}~\bibnamefont {Valent{\'i}}},\ }\bibfield  {title} {\bibinfo {title}
  {Models and materials for generalized {{Kitaev}} magnetism},\ }\href
  {https://doi.org/10.1088/1361-648X/aa8cf5} {\bibfield  {journal} {\bibinfo
  {journal} {Journal of Physics: Condensed Matter}\ }\textbf {\bibinfo {volume}
  {29}},\ \bibinfo {pages} {493002} (\bibinfo {year} {2017})}\BibitemShut
  {NoStop}%
\bibitem [{\citenamefont {Chubukov}\ and\ \citenamefont
  {Golosov}(1991)}]{chubukovQuantumTheoryAntiferromagnet1991a}%
  \BibitemOpen
  \bibfield  {author} {\bibinfo {author} {\bibfnamefont {A.~V.}\ \bibnamefont
  {Chubukov}}\ and\ \bibinfo {author} {\bibfnamefont {D.~I.}\ \bibnamefont
  {Golosov}},\ }\bibfield  {title} {\bibinfo {title} {Quantum theory of an
  antiferromagnet on a triangular lattice in a magnetic field},\ }\href
  {https://doi.org/10.1088/0953-8984/3/1/005} {\bibfield  {journal} {\bibinfo
  {journal} {Journal of Physics: Condensed Matter}\ }\textbf {\bibinfo {volume}
  {3}},\ \bibinfo {pages} {69} (\bibinfo {year} {1991})}\BibitemShut {NoStop}%
\bibitem [{\citenamefont {Seabra}\ \emph {et~al.}(2011)\citenamefont {Seabra},
  \citenamefont {Momoi}, \citenamefont {Sindzingre},\ and\ \citenamefont
  {Shannon}}]{seabraPhaseDiagramClassical2011}%
  \BibitemOpen
  \bibfield  {author} {\bibinfo {author} {\bibfnamefont {L.}~\bibnamefont
  {Seabra}}, \bibinfo {author} {\bibfnamefont {T.}~\bibnamefont {Momoi}},
  \bibinfo {author} {\bibfnamefont {P.}~\bibnamefont {Sindzingre}},\ and\
  \bibinfo {author} {\bibfnamefont {N.}~\bibnamefont {Shannon}},\ }\bibfield
  {title} {\bibinfo {title} {Phase diagram of the classical {{Heisenberg}}
  antiferromagnet on a triangular lattice in an applied magnetic field},\
  }\href {https://doi.org/10.1103/PhysRevB.84.214418} {\bibfield  {journal}
  {\bibinfo  {journal} {Physical Review B}\ }\textbf {\bibinfo {volume} {84}},\
  \bibinfo {pages} {214418} (\bibinfo {year} {2011})}\BibitemShut {NoStop}%
\bibitem [{\citenamefont {Yamamoto}\ \emph {et~al.}(2014)\citenamefont
  {Yamamoto}, \citenamefont {Marmorini},\ and\ \citenamefont
  {Danshita}}]{yamamotoQuantumPhaseDiagram2014}%
  \BibitemOpen
  \bibfield  {author} {\bibinfo {author} {\bibfnamefont {D.}~\bibnamefont
  {Yamamoto}}, \bibinfo {author} {\bibfnamefont {G.}~\bibnamefont
  {Marmorini}},\ and\ \bibinfo {author} {\bibfnamefont {I.}~\bibnamefont
  {Danshita}},\ }\bibfield  {title} {\bibinfo {title} {Quantum {{Phase
  Diagram}} of the {{Triangular-Lattice}} {{XXZ}} {{Model}} in a {{Magnetic
  Field}}},\ }\href {https://doi.org/10.1103/PhysRevLett.112.127203} {\bibfield
   {journal} {\bibinfo  {journal} {Physical Review Letters}\ }\textbf {\bibinfo
  {volume} {112}},\ \bibinfo {pages} {127203} (\bibinfo {year}
  {2014})}\BibitemShut {NoStop}%
\bibitem [{\citenamefont {Lines}(1979)}]{linesElasticPropertiesMagnetic1979}%
  \BibitemOpen
  \bibfield  {author} {\bibinfo {author} {\bibfnamefont {M.~E.}\ \bibnamefont
  {Lines}},\ }\bibfield  {title} {\bibinfo {title} {Elastic properties of
  magnetic materials},\ }\href {https://doi.org/10.1016/0370-1573(79)90039-5}
  {\bibfield  {journal} {\bibinfo  {journal} {Physics Reports}\ }\textbf
  {\bibinfo {volume} {55}},\ \bibinfo {pages} {133} (\bibinfo {year}
  {1979})}\BibitemShut {NoStop}%
\bibitem [{\citenamefont {Johannsen}\ \emph {et~al.}(2005)\citenamefont
  {Johannsen}, \citenamefont {Vasiliev}, \citenamefont {Oosawa}, \citenamefont
  {Tanaka},\ and\ \citenamefont
  {Lorenz}}]{johannsenMagnetoelasticCoupling2005}%
  \BibitemOpen
  \bibfield  {author} {\bibinfo {author} {\bibfnamefont {N.}~\bibnamefont
  {Johannsen}}, \bibinfo {author} {\bibfnamefont {A.}~\bibnamefont {Vasiliev}},
  \bibinfo {author} {\bibfnamefont {A.}~\bibnamefont {Oosawa}}, \bibinfo
  {author} {\bibfnamefont {H.}~\bibnamefont {Tanaka}},\ and\ \bibinfo {author}
  {\bibfnamefont {T.}~\bibnamefont {Lorenz}},\ }\bibfield  {title} {\bibinfo
  {title} {Magnetoelastic coupling in the spin-dimer system {{TlCuCl$_3$}}},\
  }\href {https://doi.org/10.1103/PhysRevLett.95.017205} {\bibfield  {journal}
  {\bibinfo  {journal} {Phys. Rev. Lett.}\ }\textbf {\bibinfo {volume} {95}},\
  \bibinfo {pages} {017205} (\bibinfo {year} {2005})}\BibitemShut {NoStop}%
\bibitem [{\citenamefont {Kocsis}\ \emph {et~al.}(2022)\citenamefont {Kocsis},
  \citenamefont {Kaib}, \citenamefont {Riedl}, \citenamefont {Gass},
  \citenamefont {Lampen-Kelley}, \citenamefont {Mandrus}, \citenamefont
  {Nagler}, \citenamefont {P\'erez}, \citenamefont {Nielsch}, \citenamefont
  {B\"uchner}, \citenamefont {Wolter},\ and\ \citenamefont
  {Valent\'{\i}}}]{kocsisMagnetoelasticCoupling2022}%
  \BibitemOpen
  \bibfield  {author} {\bibinfo {author} {\bibfnamefont {V.}~\bibnamefont
  {Kocsis}}, \bibinfo {author} {\bibfnamefont {D.~A.~S.}\ \bibnamefont {Kaib}},
  \bibinfo {author} {\bibfnamefont {K.}~\bibnamefont {Riedl}}, \bibinfo
  {author} {\bibfnamefont {S.}~\bibnamefont {Gass}}, \bibinfo {author}
  {\bibfnamefont {P.}~\bibnamefont {Lampen-Kelley}}, \bibinfo {author}
  {\bibfnamefont {D.~G.}\ \bibnamefont {Mandrus}}, \bibinfo {author}
  {\bibfnamefont {S.~E.}\ \bibnamefont {Nagler}}, \bibinfo {author}
  {\bibfnamefont {N.}~\bibnamefont {P\'erez}}, \bibinfo {author} {\bibfnamefont
  {K.}~\bibnamefont {Nielsch}}, \bibinfo {author} {\bibfnamefont
  {B.}~\bibnamefont {B\"uchner}}, \bibinfo {author} {\bibfnamefont {A.~U.~B.}\
  \bibnamefont {Wolter}},\ and\ \bibinfo {author} {\bibfnamefont
  {R.}~\bibnamefont {Valent\'{\i}}},\ }\bibfield  {title} {\bibinfo {title}
  {Magnetoelastic coupling anisotropy in the kitaev material
  {{$\alpha$-RuCl$_3$}}},\ }\href {https://doi.org/10.1103/PhysRevB.105.094410}
  {\bibfield  {journal} {\bibinfo  {journal} {Phys. Rev. B}\ }\textbf {\bibinfo
  {volume} {105}},\ \bibinfo {pages} {094410} (\bibinfo {year}
  {2022})}\BibitemShut {NoStop}%
\bibitem [{\citenamefont {Hohenberg}\ and\ \citenamefont
  {Halperin}(1977)}]{hohenbergTheoryDynamicCritical1977}%
  \BibitemOpen
  \bibfield  {author} {\bibinfo {author} {\bibfnamefont {P.~C.}\ \bibnamefont
  {Hohenberg}}\ and\ \bibinfo {author} {\bibfnamefont {B.~I.}\ \bibnamefont
  {Halperin}},\ }\bibfield  {title} {\bibinfo {title} {Theory of dynamic
  critical phenomena},\ }\href {https://doi.org/10.1103/RevModPhys.49.435}
  {\bibfield  {journal} {\bibinfo  {journal} {Reviews of Modern Physics}\
  }\textbf {\bibinfo {volume} {49}},\ \bibinfo {pages} {435} (\bibinfo {year}
  {1977})}\BibitemShut {NoStop}%
\bibitem [{\citenamefont {Li}\ \emph {et~al.}(2019{\natexlab{b}})\citenamefont
  {Li}, \citenamefont {Zelenskiy}, \citenamefont {Quilliam}, \citenamefont
  {Dun}, \citenamefont {Zhou}, \citenamefont {Plumer},\ and\ \citenamefont
  {Quirion}}]{liMagnetoelasticCouplingMagnetization2019a}%
  \BibitemOpen
  \bibfield  {author} {\bibinfo {author} {\bibfnamefont {M.}~\bibnamefont
  {Li}}, \bibinfo {author} {\bibfnamefont {A.}~\bibnamefont {Zelenskiy}},
  \bibinfo {author} {\bibfnamefont {J.~A.}\ \bibnamefont {Quilliam}}, \bibinfo
  {author} {\bibfnamefont {Z.~L.}\ \bibnamefont {Dun}}, \bibinfo {author}
  {\bibfnamefont {H.~D.}\ \bibnamefont {Zhou}}, \bibinfo {author}
  {\bibfnamefont {M.~L.}\ \bibnamefont {Plumer}},\ and\ \bibinfo {author}
  {\bibfnamefont {G.}~\bibnamefont {Quirion}},\ }\bibfield  {title} {\bibinfo
  {title} {Magnetoelastic coupling and the magnetization plateau in
  {{Ba$_3$CoSb$_2$O$_9$}}},\ }\href
  {https://doi.org/10.1103/PhysRevB.99.094408} {\bibfield  {journal} {\bibinfo
  {journal} {Physical Review B}\ }\textbf {\bibinfo {volume} {99}},\ \bibinfo
  {pages} {094408} (\bibinfo {year} {2019}{\natexlab{b}})}\BibitemShut
  {NoStop}%
\bibitem [{\citenamefont {Cong}\ \emph {et~al.}(2016)\citenamefont {Cong},
  \citenamefont {Postulka}, \citenamefont {Wolf}, \citenamefont {Van~Well},
  \citenamefont {Ritter}, \citenamefont {Assmus}, \citenamefont {Krellner},\
  and\ \citenamefont {Lang}}]{congMagnetoacousticStudyQuantum2016}%
  \BibitemOpen
  \bibfield  {author} {\bibinfo {author} {\bibfnamefont {P.~T.}\ \bibnamefont
  {Cong}}, \bibinfo {author} {\bibfnamefont {L.}~\bibnamefont {Postulka}},
  \bibinfo {author} {\bibfnamefont {B.}~\bibnamefont {Wolf}}, \bibinfo {author}
  {\bibfnamefont {N.}~\bibnamefont {Van~Well}}, \bibinfo {author}
  {\bibfnamefont {F.}~\bibnamefont {Ritter}}, \bibinfo {author} {\bibfnamefont
  {W.}~\bibnamefont {Assmus}}, \bibinfo {author} {\bibfnamefont
  {C.}~\bibnamefont {Krellner}},\ and\ \bibinfo {author} {\bibfnamefont
  {M.}~\bibnamefont {Lang}},\ }\bibfield  {title} {\bibinfo {title}
  {Magneto-acoustic study near the quantum critical point of the frustrated
  quantum antiferromagnet {{Cs$_2$CuCl$_4$}}},\ }\href
  {https://doi.org/10.1063/1.4961710} {\bibfield  {journal} {\bibinfo
  {journal} {Journal of Applied Physics}\ }\textbf {\bibinfo {volume} {120}},\
  \bibinfo {pages} {142113} (\bibinfo {year} {2016})}\BibitemShut {NoStop}%
\bibitem [{\citenamefont {Thallapaka}\ \emph {et~al.}(2019)\citenamefont
  {Thallapaka}, \citenamefont {Wolf}, \citenamefont {Gati}, \citenamefont
  {Postulka}, \citenamefont {Tutsch}, \citenamefont {Schmidt}, \citenamefont
  {Thalmeier}, \citenamefont {Ritter}, \citenamefont {Krellner}, \citenamefont
  {Li}, \citenamefont {Borisov}, \citenamefont {Valent{\'i}},\ and\
  \citenamefont {Lang}}]{thallapakaMagnetoStructuralPropertiesLayered2019a}%
  \BibitemOpen
  \bibfield  {author} {\bibinfo {author} {\bibfnamefont {S.~K.}\ \bibnamefont
  {Thallapaka}}, \bibinfo {author} {\bibfnamefont {B.}~\bibnamefont {Wolf}},
  \bibinfo {author} {\bibfnamefont {E.}~\bibnamefont {Gati}}, \bibinfo {author}
  {\bibfnamefont {L.}~\bibnamefont {Postulka}}, \bibinfo {author}
  {\bibfnamefont {U.}~\bibnamefont {Tutsch}}, \bibinfo {author} {\bibfnamefont
  {B.}~\bibnamefont {Schmidt}}, \bibinfo {author} {\bibfnamefont
  {P.}~\bibnamefont {Thalmeier}}, \bibinfo {author} {\bibfnamefont
  {F.}~\bibnamefont {Ritter}}, \bibinfo {author} {\bibfnamefont
  {C.}~\bibnamefont {Krellner}}, \bibinfo {author} {\bibfnamefont
  {Y.}~\bibnamefont {Li}}, \bibinfo {author} {\bibfnamefont {V.}~\bibnamefont
  {Borisov}}, \bibinfo {author} {\bibfnamefont {R.}~\bibnamefont
  {Valent{\'i}}},\ and\ \bibinfo {author} {\bibfnamefont {M.}~\bibnamefont
  {Lang}},\ }\bibfield  {title} {\bibinfo {title} {Magneto-{{Structural
  Properties}} of the {{Layered Quasi-2D Triangular-Lattice Antiferromagnets
  Cs$_2$CuCl$_{4-x}$Br$_x$}} for x = 0, 1, 2 and 4},\ }\href
  {https://doi.org/10.1002/pssb.201900044} {\bibfield  {journal} {\bibinfo
  {journal} {Physica Status Solidi (B)}\ }\textbf {\bibinfo {volume} {256}},\
  \bibinfo {pages} {1900044} (\bibinfo {year} {2019})}\BibitemShut {NoStop}%
\bibitem [{\citenamefont {Blosser}\ \emph {et~al.}(2020)\citenamefont
  {Blosser}, \citenamefont {Facheris},\ and\ \citenamefont
  {Zheludev}}]{blosserMiniatureCapacitiveFaraday2020}%
  \BibitemOpen
  \bibfield  {author} {\bibinfo {author} {\bibfnamefont {D.}~\bibnamefont
  {Blosser}}, \bibinfo {author} {\bibfnamefont {L.}~\bibnamefont {Facheris}},\
  and\ \bibinfo {author} {\bibfnamefont {A.}~\bibnamefont {Zheludev}},\
  }\bibfield  {title} {\bibinfo {title} {Miniature capacitive {{Faraday}} force
  magnetometer for magnetization measurements at low temperatures and high
  magnetic fields},\ }\href {https://doi.org/10.1063/5.0005850} {\bibfield
  {journal} {\bibinfo  {journal} {Review of Scientific Instruments}\ }\textbf
  {\bibinfo {volume} {91}},\ \bibinfo {pages} {073905} (\bibinfo {year}
  {2020})}\BibitemShut {NoStop}%
\bibitem [{\citenamefont {Janssen}\ and\ \citenamefont
  {Vojta}(2019)}]{janssen_JPhysCondMat_HKMag2019}%
  \BibitemOpen
  \bibfield  {author} {\bibinfo {author} {\bibfnamefont {L.}~\bibnamefont
  {Janssen}}\ and\ \bibinfo {author} {\bibfnamefont {M.}~\bibnamefont
  {Vojta}},\ }\bibfield  {title} {\bibinfo {title} {Heisenberg–kitaev physics
  in magnetic fields},\ }\href {https://doi.org/10.1088/1361-648X/ab283e}
  {\bibfield  {journal} {\bibinfo  {journal} {Journal of Physics: Condensed
  Matter}\ }\textbf {\bibinfo {volume} {31}},\ \bibinfo {pages} {423002}
  (\bibinfo {year} {2019})}\BibitemShut {NoStop}%
\bibitem [{\citenamefont {Pelissetto}\ and\ \citenamefont
  {Vicari}(2002)}]{pelissettoCriticalPhenomena2002}%
  \BibitemOpen
  \bibfield  {author} {\bibinfo {author} {\bibfnamefont {A.}~\bibnamefont
  {Pelissetto}}\ and\ \bibinfo {author} {\bibfnamefont {E.}~\bibnamefont
  {Vicari}},\ }\bibfield  {title} {\bibinfo {title} {Critical phenomena and
  renormalization-group theory},\ }\href
  {https://doi.org/https://doi.org/10.1016/S0370-1573(02)00219-3} {\bibfield
  {journal} {\bibinfo  {journal} {Physics Reports}\ }\textbf {\bibinfo {volume}
  {368}},\ \bibinfo {pages} {549} (\bibinfo {year} {2002})}\BibitemShut
  {NoStop}%
\bibitem [{\citenamefont {Coldea}\ \emph {et~al.}(2002)\citenamefont {Coldea},
  \citenamefont {Tennant}, \citenamefont {Habicht}, \citenamefont {Smeibidl},
  \citenamefont {Wolters},\ and\ \citenamefont
  {Tylczynski}}]{coldeaDirectMeasurementSpin2002}%
  \BibitemOpen
  \bibfield  {author} {\bibinfo {author} {\bibfnamefont {R.}~\bibnamefont
  {Coldea}}, \bibinfo {author} {\bibfnamefont {D.~A.}\ \bibnamefont {Tennant}},
  \bibinfo {author} {\bibfnamefont {K.}~\bibnamefont {Habicht}}, \bibinfo
  {author} {\bibfnamefont {P.}~\bibnamefont {Smeibidl}}, \bibinfo {author}
  {\bibfnamefont {C.}~\bibnamefont {Wolters}},\ and\ \bibinfo {author}
  {\bibfnamefont {Z.}~\bibnamefont {Tylczynski}},\ }\bibfield  {title}
  {\bibinfo {title} {Direct {{Measurement}} of the {{Spin Hamiltonian}} and
  {{Observation}} of {{Condensation}} of {{Magnons}} in the {{2D Frustrated
  Quantum Magnet Cs$_2$CuCl$_4$}}},\ }\href
  {https://doi.org/10.1103/PhysRevLett.88.137203} {\bibfield  {journal}
  {\bibinfo  {journal} {Physical Review Letters}\ }\textbf {\bibinfo {volume}
  {88}},\ \bibinfo {pages} {137203} (\bibinfo {year} {2002})}\BibitemShut
  {NoStop}%
\bibitem [{\citenamefont
  {Zhitomirsky}(2015)}]{zhitomirskyRealspacePerturbationTheory2015}%
  \BibitemOpen
  \bibfield  {author} {\bibinfo {author} {\bibfnamefont {M.~E.}\ \bibnamefont
  {Zhitomirsky}},\ }\bibfield  {title} {\bibinfo {title} {Real-space
  perturbation theory for frustrated magnets: Application to magnetization
  plateaus},\ }\href {https://doi.org/10.1088/1742-6596/592/1/012110}
  {\bibfield  {journal} {\bibinfo  {journal} {Journal of Physics: Conference
  Series}\ }\textbf {\bibinfo {volume} {592}},\ \bibinfo {pages} {012110}
  (\bibinfo {year} {2015})}\BibitemShut {NoStop}%
\bibitem [{\citenamefont {Yamamoto}\ \emph {et~al.}(2015)\citenamefont
  {Yamamoto}, \citenamefont {Marmorini},\ and\ \citenamefont
  {Danshita}}]{yamamotoMicroscopicModelCalculations2015}%
  \BibitemOpen
  \bibfield  {author} {\bibinfo {author} {\bibfnamefont {D.}~\bibnamefont
  {Yamamoto}}, \bibinfo {author} {\bibfnamefont {G.}~\bibnamefont
  {Marmorini}},\ and\ \bibinfo {author} {\bibfnamefont {I.}~\bibnamefont
  {Danshita}},\ }\bibfield  {title} {\bibinfo {title} {Microscopic {{Model
  Calculations}} for the {{Magnetization Process}} of {{Layered
  Triangular-Lattice Quantum Antiferromagnets}}},\ }\href
  {https://doi.org/10.1103/PhysRevLett.114.027201} {\bibfield  {journal}
  {\bibinfo  {journal} {Physical Review Letters}\ }\textbf {\bibinfo {volume}
  {114}},\ \bibinfo {pages} {027201} (\bibinfo {year} {2015})}\BibitemShut
  {NoStop}%
\bibitem [{\citenamefont {Li}\ \emph {et~al.}(2020{\natexlab{b}})\citenamefont
  {Li}, \citenamefont {Plumer},\ and\ \citenamefont
  {Quirion}}]{liEffectsInterlayerBiquadratic2020}%
  \BibitemOpen
  \bibfield  {author} {\bibinfo {author} {\bibfnamefont {M.}~\bibnamefont
  {Li}}, \bibinfo {author} {\bibfnamefont {M.~L.}\ \bibnamefont {Plumer}},\
  and\ \bibinfo {author} {\bibfnamefont {G.}~\bibnamefont {Quirion}},\
  }\bibfield  {title} {\bibinfo {title} {Effects of interlayer and bi-quadratic
  exchange coupling on layered triangular lattice antiferromagnets},\ }\href
  {https://doi.org/10.1088/1361-648X/ab5ea6} {\bibfield  {journal} {\bibinfo
  {journal} {Journal of Physics: Condensed Matter}\ }\textbf {\bibinfo {volume}
  {32}},\ \bibinfo {pages} {135803} (\bibinfo {year}
  {2020}{\natexlab{b}})}\BibitemShut {NoStop}%
\bibitem [{\citenamefont {Okada}\ \emph {et~al.}(2022)\citenamefont {Okada},
  \citenamefont {Tanaka}, \citenamefont {Kurita}, \citenamefont {Yamamoto},
  \citenamefont {Matsuo},\ and\ \citenamefont
  {Kindo}}]{okadaFieldorientationDependenceQuantum2022}%
  \BibitemOpen
  \bibfield  {author} {\bibinfo {author} {\bibfnamefont {K.}~\bibnamefont
  {Okada}}, \bibinfo {author} {\bibfnamefont {H.}~\bibnamefont {Tanaka}},
  \bibinfo {author} {\bibfnamefont {N.}~\bibnamefont {Kurita}}, \bibinfo
  {author} {\bibfnamefont {D.}~\bibnamefont {Yamamoto}}, \bibinfo {author}
  {\bibfnamefont {A.}~\bibnamefont {Matsuo}},\ and\ \bibinfo {author}
  {\bibfnamefont {K.}~\bibnamefont {Kindo}},\ }\bibfield  {title} {\bibinfo
  {title} {Field-orientation dependence of quantum phase transitions in the
  {{S}}=1/2 triangular-lattice antiferromagnet {{Ba$_3$CoSb2O$_9$}}},\ }\href
  {https://doi.org/10.1103/PhysRevB.106.104415} {\bibfield  {journal} {\bibinfo
   {journal} {Physical Review B}\ }\textbf {\bibinfo {volume} {106}},\ \bibinfo
  {pages} {104415} (\bibinfo {year} {2022})}\BibitemShut {NoStop}%
\bibitem [{\citenamefont {Koutroulakis}\ \emph {et~al.}(2015)\citenamefont
  {Koutroulakis}, \citenamefont {Zhou}, \citenamefont {Kamiya}, \citenamefont
  {Thompson}, \citenamefont {Zhou}, \citenamefont {Batista},\ and\
  \citenamefont {Brown}}]{koutroulakisQuantumPhaseDiagram2015}%
  \BibitemOpen
  \bibfield  {author} {\bibinfo {author} {\bibfnamefont {G.}~\bibnamefont
  {Koutroulakis}}, \bibinfo {author} {\bibfnamefont {T.}~\bibnamefont {Zhou}},
  \bibinfo {author} {\bibfnamefont {Y.}~\bibnamefont {Kamiya}}, \bibinfo
  {author} {\bibfnamefont {J.~D.}\ \bibnamefont {Thompson}}, \bibinfo {author}
  {\bibfnamefont {H.~D.}\ \bibnamefont {Zhou}}, \bibinfo {author}
  {\bibfnamefont {C.~D.}\ \bibnamefont {Batista}},\ and\ \bibinfo {author}
  {\bibfnamefont {S.~E.}\ \bibnamefont {Brown}},\ }\bibfield  {title} {\bibinfo
  {title} {Quantum phase diagram of the {{$S = \frac{1}{2}$}}
  triangular-lattice antiferromagnet {{Ba$_3$CoSb$_2$O$_9$}}},\ }\href
  {https://doi.org/10.1103/PhysRevB.91.024410} {\bibfield  {journal} {\bibinfo
  {journal} {Physical Review B}\ }\textbf {\bibinfo {volume} {91}},\ \bibinfo
  {pages} {024410} (\bibinfo {year} {2015})}\BibitemShut {NoStop}%
\bibitem [{\citenamefont {Chen}\ \emph {et~al.}(2013)\citenamefont {Chen},
  \citenamefont {Ju}, \citenamefont {Jiang}, \citenamefont {Starykh},\ and\
  \citenamefont {Balents}}]{chenGroundStatesSpin12013}%
  \BibitemOpen
  \bibfield  {author} {\bibinfo {author} {\bibfnamefont {R.}~\bibnamefont
  {Chen}}, \bibinfo {author} {\bibfnamefont {H.}~\bibnamefont {Ju}}, \bibinfo
  {author} {\bibfnamefont {H.-C.}\ \bibnamefont {Jiang}}, \bibinfo {author}
  {\bibfnamefont {O.~A.}\ \bibnamefont {Starykh}},\ and\ \bibinfo {author}
  {\bibfnamefont {L.}~\bibnamefont {Balents}},\ }\bibfield  {title} {\bibinfo
  {title} {Ground states of spin-1/2 triangular antiferromagnets in a magnetic
  field},\ }\href {https://doi.org/10.1103/PhysRevB.87.165123} {\bibfield
  {journal} {\bibinfo  {journal} {Physical Review B}\ }\textbf {\bibinfo
  {volume} {87}},\ \bibinfo {pages} {165123} (\bibinfo {year}
  {2013})}\BibitemShut {NoStop}%
\bibitem [{\citenamefont {Griset}\ \emph {et~al.}(2011)\citenamefont {Griset},
  \citenamefont {Head}, \citenamefont {Alicea},\ and\ \citenamefont
  {Starykh}}]{grisetDeformedTriangularLattice2011}%
  \BibitemOpen
  \bibfield  {author} {\bibinfo {author} {\bibfnamefont {C.}~\bibnamefont
  {Griset}}, \bibinfo {author} {\bibfnamefont {S.}~\bibnamefont {Head}},
  \bibinfo {author} {\bibfnamefont {J.}~\bibnamefont {Alicea}},\ and\ \bibinfo
  {author} {\bibfnamefont {O.~A.}\ \bibnamefont {Starykh}},\ }\bibfield
  {title} {\bibinfo {title} {Deformed triangular lattice antiferromagnets in a
  magnetic field: {{Role}} of spatial anisotropy and {{Dzyaloshinskii-Moriya}}
  interactions},\ }\href {https://doi.org/10.1103/PhysRevB.84.245108}
  {\bibfield  {journal} {\bibinfo  {journal} {Physical Review B}\ }\textbf
  {\bibinfo {volume} {84}},\ \bibinfo {pages} {245108} (\bibinfo {year}
  {2011})}\BibitemShut {NoStop}%
\bibitem [{\citenamefont {Tay}\ and\ \citenamefont
  {Motrunich}(2010)}]{tayVariationalStudiesTriangular2010}%
  \BibitemOpen
  \bibfield  {author} {\bibinfo {author} {\bibfnamefont {T.}~\bibnamefont
  {Tay}}\ and\ \bibinfo {author} {\bibfnamefont {O.~I.}\ \bibnamefont
  {Motrunich}},\ }\bibfield  {title} {\bibinfo {title} {Variational studies of
  triangular {{Heisenberg}} antiferromagnet in magnetic field},\ }\href
  {https://doi.org/10.1103/PhysRevB.81.165116} {\bibfield  {journal} {\bibinfo
  {journal} {Physical Review B}\ }\textbf {\bibinfo {volume} {81}},\ \bibinfo
  {pages} {165116} (\bibinfo {year} {2010})}\BibitemShut {NoStop}%
\bibitem [{\citenamefont {Coldea}\ \emph {et~al.}(2001)\citenamefont {Coldea},
  \citenamefont {Tennant}, \citenamefont {Tsvelik},\ and\ \citenamefont
  {Tylczynski}}]{coldeaExperimentalRealization2D2001}%
  \BibitemOpen
  \bibfield  {author} {\bibinfo {author} {\bibfnamefont {R.}~\bibnamefont
  {Coldea}}, \bibinfo {author} {\bibfnamefont {D.~A.}\ \bibnamefont {Tennant}},
  \bibinfo {author} {\bibfnamefont {A.~M.}\ \bibnamefont {Tsvelik}},\ and\
  \bibinfo {author} {\bibfnamefont {Z.}~\bibnamefont {Tylczynski}},\ }\bibfield
   {title} {\bibinfo {title} {Experimental {{Realization}} of a {{2D Fractional
  Quantum Spin Liquid}}},\ }\href {https://doi.org/10.1103/PhysRevLett.86.1335}
  {\bibfield  {journal} {\bibinfo  {journal} {Physical Review Letters}\
  }\textbf {\bibinfo {volume} {86}},\ \bibinfo {pages} {1335} (\bibinfo {year}
  {2001})}\BibitemShut {NoStop}%
\bibitem [{\citenamefont {Tokiwa}\ \emph {et~al.}(2006)\citenamefont {Tokiwa},
  \citenamefont {Radu}, \citenamefont {Coldea}, \citenamefont {Wilhelm},
  \citenamefont {Tylczynski},\ and\ \citenamefont
  {Steglich}}]{tokiwaMagneticPhaseTransitions2006}%
  \BibitemOpen
  \bibfield  {author} {\bibinfo {author} {\bibfnamefont {Y.}~\bibnamefont
  {Tokiwa}}, \bibinfo {author} {\bibfnamefont {T.}~\bibnamefont {Radu}},
  \bibinfo {author} {\bibfnamefont {R.}~\bibnamefont {Coldea}}, \bibinfo
  {author} {\bibfnamefont {H.}~\bibnamefont {Wilhelm}}, \bibinfo {author}
  {\bibfnamefont {Z.}~\bibnamefont {Tylczynski}},\ and\ \bibinfo {author}
  {\bibfnamefont {F.}~\bibnamefont {Steglich}},\ }\bibfield  {title} {\bibinfo
  {title} {Magnetic phase transitions in the two-dimensional frustrated quantum
  antiferromagnet {{Cs$_2$CuCl$_4$}}},\ }\href
  {https://doi.org/10.1103/PhysRevB.73.134414} {\bibfield  {journal} {\bibinfo
  {journal} {Physical Review B}\ }\textbf {\bibinfo {volume} {73}},\ \bibinfo
  {pages} {134414} (\bibinfo {year} {2006})}\BibitemShut {NoStop}%
\bibitem [{\citenamefont {Ono}\ \emph {et~al.}(2004)\citenamefont {Ono},
  \citenamefont {Tanaka}, \citenamefont {Kolomiyets}, \citenamefont {Mitamura},
  \citenamefont {Goto}, \citenamefont {Nakajima}, \citenamefont {Oosawa},
  \citenamefont {Koike}, \citenamefont {Kakurai}, \citenamefont {Klenke},
  \citenamefont {Smeibidle},\ and\ \citenamefont
  {Mei{\ss}ner}}]{onoMagnetizationPlateaux12004}%
  \BibitemOpen
  \bibfield  {author} {\bibinfo {author} {\bibfnamefont {T.}~\bibnamefont
  {Ono}}, \bibinfo {author} {\bibfnamefont {H.}~\bibnamefont {Tanaka}},
  \bibinfo {author} {\bibfnamefont {O.}~\bibnamefont {Kolomiyets}}, \bibinfo
  {author} {\bibfnamefont {H.}~\bibnamefont {Mitamura}}, \bibinfo {author}
  {\bibfnamefont {T.}~\bibnamefont {Goto}}, \bibinfo {author} {\bibfnamefont
  {K.}~\bibnamefont {Nakajima}}, \bibinfo {author} {\bibfnamefont
  {A.}~\bibnamefont {Oosawa}}, \bibinfo {author} {\bibfnamefont
  {Y.}~\bibnamefont {Koike}}, \bibinfo {author} {\bibfnamefont
  {K.}~\bibnamefont {Kakurai}}, \bibinfo {author} {\bibfnamefont
  {J.}~\bibnamefont {Klenke}}, \bibinfo {author} {\bibfnamefont
  {P.}~\bibnamefont {Smeibidle}},\ and\ \bibinfo {author} {\bibfnamefont
  {M.}~\bibnamefont {Mei{\ss}ner}},\ }\bibfield  {title} {\bibinfo {title}
  {Magnetization plateaux of the {{S}} = 1/2 two-dimensional frustrated
  antiferromagnet {{Cs$_2$CuBr$_4$}}},\ }\href
  {https://doi.org/10.1088/0953-8984/16/11/028} {\bibfield  {journal} {\bibinfo
   {journal} {Journal of Physics: Condensed Matter}\ }\textbf {\bibinfo
  {volume} {16}},\ \bibinfo {pages} {S773} (\bibinfo {year}
  {2004})}\BibitemShut {NoStop}%
\bibitem [{\citenamefont {Fortune}\ \emph {et~al.}(2009)\citenamefont
  {Fortune}, \citenamefont {Hannahs}, \citenamefont {Yoshida}, \citenamefont
  {Sherline}, \citenamefont {Ono}, \citenamefont {Tanaka},\ and\ \citenamefont
  {Takano}}]{fortuneCascadeMagneticFieldInducedQuantum2009}%
  \BibitemOpen
  \bibfield  {author} {\bibinfo {author} {\bibfnamefont {N.~A.}\ \bibnamefont
  {Fortune}}, \bibinfo {author} {\bibfnamefont {S.~T.}\ \bibnamefont
  {Hannahs}}, \bibinfo {author} {\bibfnamefont {Y.}~\bibnamefont {Yoshida}},
  \bibinfo {author} {\bibfnamefont {T.~E.}\ \bibnamefont {Sherline}}, \bibinfo
  {author} {\bibfnamefont {T.}~\bibnamefont {Ono}}, \bibinfo {author}
  {\bibfnamefont {H.}~\bibnamefont {Tanaka}},\ and\ \bibinfo {author}
  {\bibfnamefont {Y.}~\bibnamefont {Takano}},\ }\bibfield  {title} {\bibinfo
  {title} {Cascade of {{Magnetic-Field-Induced Quantum Phase Transitions}} in a
  {{Spin-}} 1/2 {{Triangular-Lattice Antiferromagnet}}},\ }\href
  {https://doi.org/10.1103/PhysRevLett.102.257201} {\bibfield  {journal}
  {\bibinfo  {journal} {Physical Review Letters}\ }\textbf {\bibinfo {volume}
  {102}},\ \bibinfo {pages} {257201} (\bibinfo {year} {2009})}\BibitemShut
  {NoStop}%
\bibitem [{\citenamefont {Hirai}\ \emph {et~al.}(2020)\citenamefont {Hirai},
  \citenamefont {Yajima}, \citenamefont {Nawa}, \citenamefont {Kawamura},\ and\
  \citenamefont {Hiroi}}]{hirai_ACS_AnisotropicTriangular2020}%
  \BibitemOpen
  \bibfield  {author} {\bibinfo {author} {\bibfnamefont {D.}~\bibnamefont
  {Hirai}}, \bibinfo {author} {\bibfnamefont {T.}~\bibnamefont {Yajima}},
  \bibinfo {author} {\bibfnamefont {K.}~\bibnamefont {Nawa}}, \bibinfo {author}
  {\bibfnamefont {M.}~\bibnamefont {Kawamura}},\ and\ \bibinfo {author}
  {\bibfnamefont {Z.}~\bibnamefont {Hiroi}},\ }\bibfield  {title} {\bibinfo
  {title} {Anisotropic triangular lattice realized in rhenium oxychlorides
  {{A$_3$ReO$_5$Cl$_2$ (A = Ba, Sr)}}},\ }\href
  {https://doi.org/10.1021/acs.inorgchem.0c01187} {\bibfield  {journal}
  {\bibinfo  {journal} {Inorganic Chemistry}\ }\textbf {\bibinfo {volume}
  {59}},\ \bibinfo {pages} {10025} (\bibinfo {year} {2020})}\BibitemShut
  {NoStop}%
\bibitem [{\citenamefont {Nawa}\ \emph {et~al.}(2020)\citenamefont {Nawa},
  \citenamefont {Hirai}, \citenamefont {Kofu}, \citenamefont {Nakajima},
  \citenamefont {Murasaki}, \citenamefont {Kogane}, \citenamefont {Kimata},
  \citenamefont {Nojiri}, \citenamefont {Hiroi},\ and\ \citenamefont
  {Sato}}]{nawa_PRR_BoundSpinon2020}%
  \BibitemOpen
  \bibfield  {author} {\bibinfo {author} {\bibfnamefont {K.}~\bibnamefont
  {Nawa}}, \bibinfo {author} {\bibfnamefont {D.}~\bibnamefont {Hirai}},
  \bibinfo {author} {\bibfnamefont {M.}~\bibnamefont {Kofu}}, \bibinfo {author}
  {\bibfnamefont {K.}~\bibnamefont {Nakajima}}, \bibinfo {author}
  {\bibfnamefont {R.}~\bibnamefont {Murasaki}}, \bibinfo {author}
  {\bibfnamefont {S.}~\bibnamefont {Kogane}}, \bibinfo {author} {\bibfnamefont
  {M.}~\bibnamefont {Kimata}}, \bibinfo {author} {\bibfnamefont
  {H.}~\bibnamefont {Nojiri}}, \bibinfo {author} {\bibfnamefont
  {Z.}~\bibnamefont {Hiroi}},\ and\ \bibinfo {author} {\bibfnamefont {T.~J.}\
  \bibnamefont {Sato}},\ }\bibfield  {title} {\bibinfo {title} {Bound spinon
  excitations in the spin-$\frac{1}{2}$ anisotropic triangular antiferromagnet
  {{${\mathrm{Ca}}_{3}\mathrm{Re}{\mathrm{O}}_{5}{\mathrm{Cl}}_{2}$}}},\ }\href
  {https://doi.org/10.1103/PhysRevResearch.2.043121} {\bibfield  {journal}
  {\bibinfo  {journal} {Phys. Rev. Res.}\ }\textbf {\bibinfo {volume} {2}},\
  \bibinfo {pages} {043121} (\bibinfo {year} {2020})}\BibitemShut {NoStop}%
\bibitem [{\citenamefont {Weihong}\ \emph {et~al.}(1999)\citenamefont
  {Weihong}, \citenamefont {McKenzie},\ and\ \citenamefont
  {Singh}}]{weihongPhaseDiagramClass1999}%
  \BibitemOpen
  \bibfield  {author} {\bibinfo {author} {\bibfnamefont {Z.}~\bibnamefont
  {Weihong}}, \bibinfo {author} {\bibfnamefont {R.~H.}\ \bibnamefont
  {McKenzie}},\ and\ \bibinfo {author} {\bibfnamefont {R.~R.~P.}\ \bibnamefont
  {Singh}},\ }\bibfield  {title} {\bibinfo {title} {Phase diagram for a class
  of spin-1/2 {{Heisenberg}} models interpolating between the square-lattice,
  the triangular-lattice, and the linear-chain limits},\ }\href
  {https://doi.org/10.1103/PhysRevB.59.14367} {\bibfield  {journal} {\bibinfo
  {journal} {Physical Review B}\ }\textbf {\bibinfo {volume} {59}},\ \bibinfo
  {pages} {14367} (\bibinfo {year} {1999})}\BibitemShut {NoStop}%
\bibitem [{\citenamefont {Thesberg}\ and\ \citenamefont
  {S{\o}rensen}(2014)}]{thesbergExactDiagonalizationStudy2014}%
  \BibitemOpen
  \bibfield  {author} {\bibinfo {author} {\bibfnamefont {M.}~\bibnamefont
  {Thesberg}}\ and\ \bibinfo {author} {\bibfnamefont {E.~S.}\ \bibnamefont
  {S{\o}rensen}},\ }\bibfield  {title} {\bibinfo {title} {Exact diagonalization
  study of the anisotropic triangular lattice {{Heisenberg}} model using
  twisted boundary conditions},\ }\href
  {https://doi.org/10.1103/PhysRevB.90.115117} {\bibfield  {journal} {\bibinfo
  {journal} {Physical Review B}\ }\textbf {\bibinfo {volume} {90}},\ \bibinfo
  {pages} {115117} (\bibinfo {year} {2014})}\BibitemShut {NoStop}%
\bibitem [{\citenamefont {Hasik}\ and\ \citenamefont
  {Corboz}(2024)}]{hasikIncommensurateOrderTranslationally2024}%
  \BibitemOpen
  \bibfield  {author} {\bibinfo {author} {\bibfnamefont {J.}~\bibnamefont
  {Hasik}}\ and\ \bibinfo {author} {\bibfnamefont {P.}~\bibnamefont {Corboz}},\
  }\bibfield  {title} {\bibinfo {title} {Incommensurate {{Order}} with
  {{Translationally Invariant Projected Entangled-Pair States}}: {{Spiral
  States}} and {{Quantum Spin Liquid}} on the {{Anisotropic Triangular
  Lattice}}},\ }\href {https://doi.org/10.1103/PhysRevLett.133.176502}
  {\bibfield  {journal} {\bibinfo  {journal} {Physical Review Letters}\
  }\textbf {\bibinfo {volume} {133}},\ \bibinfo {pages} {176502} (\bibinfo
  {year} {2024})}\BibitemShut {NoStop}%
\bibitem [{\citenamefont {Penc}\ \emph {et~al.}(2004)\citenamefont {Penc},
  \citenamefont {Shannon},\ and\ \citenamefont
  {Shiba}}]{pencHalfMagnetizationPlateauStabilized2004}%
  \BibitemOpen
  \bibfield  {author} {\bibinfo {author} {\bibfnamefont {K.}~\bibnamefont
  {Penc}}, \bibinfo {author} {\bibfnamefont {N.}~\bibnamefont {Shannon}},\ and\
  \bibinfo {author} {\bibfnamefont {H.}~\bibnamefont {Shiba}},\ }\bibfield
  {title} {\bibinfo {title} {Half-{{Magnetization Plateau Stabilized}} by
  {{Structural Distortion}} in the {{Antiferromagnetic Heisenberg Model}} on a
  {{Pyrochlore Lattice}}},\ }\href
  {https://doi.org/10.1103/PhysRevLett.93.197203} {\bibfield  {journal}
  {\bibinfo  {journal} {Physical Review Letters}\ }\textbf {\bibinfo {volume}
  {93}},\ \bibinfo {pages} {197203} (\bibinfo {year} {2004})}\BibitemShut
  {NoStop}%
\bibitem [{\citenamefont {Tchernyshyov}\ \emph {et~al.}(2002)\citenamefont
  {Tchernyshyov}, \citenamefont {Moessner},\ and\ \citenamefont
  {Sondhi}}]{tchernyshyovOrderDistortionString2002a}%
  \BibitemOpen
  \bibfield  {author} {\bibinfo {author} {\bibfnamefont {O.}~\bibnamefont
  {Tchernyshyov}}, \bibinfo {author} {\bibfnamefont {R.}~\bibnamefont
  {Moessner}},\ and\ \bibinfo {author} {\bibfnamefont {S.~L.}\ \bibnamefont
  {Sondhi}},\ }\bibfield  {title} {\bibinfo {title} {Order by {{Distortion}}
  and {{String Modes}} in {{Pyrochlore Antiferromagnets}}},\ }\href
  {https://doi.org/10.1103/PhysRevLett.88.067203} {\bibfield  {journal}
  {\bibinfo  {journal} {Physical Review Letters}\ }\textbf {\bibinfo {volume}
  {88}},\ \bibinfo {pages} {067203} (\bibinfo {year} {2002})}\BibitemShut
  {NoStop}%
\bibitem [{\citenamefont {Liu}\ and\ \citenamefont
  {Khaliullin}(2019)}]{liuPseudoJahnTellerEffectMagnetoelastic2019}%
  \BibitemOpen
  \bibfield  {author} {\bibinfo {author} {\bibfnamefont {H.}~\bibnamefont
  {Liu}}\ and\ \bibinfo {author} {\bibfnamefont {G.}~\bibnamefont
  {Khaliullin}},\ }\bibfield  {title} {\bibinfo {title} {Pseudo-{{Jahn-Teller
  Effect}} and {{Magnetoelastic Coupling}} in {{Spin-Orbit Mott Insulators}}},\
  }\href {https://doi.org/10.1103/PhysRevLett.122.057203} {\bibfield  {journal}
  {\bibinfo  {journal} {Physical Review Letters}\ }\textbf {\bibinfo {volume}
  {122}},\ \bibinfo {pages} {057203} (\bibinfo {year} {2019})}\BibitemShut
  {NoStop}%
\bibitem [{\citenamefont {Winter}\ \emph {et~al.}(2018)\citenamefont {Winter},
  \citenamefont {Waterman}, \citenamefont {Parkhurst}, \citenamefont
  {Brewster}, \citenamefont {Gildea}, \citenamefont {Gerstel}, \citenamefont
  {Fuentes-Montero}, \citenamefont {Vollmar}, \citenamefont {Michels-Clark},
  \citenamefont {Young}, \citenamefont {Sauter},\ and\ \citenamefont
  {Evans}}]{winter_ACSD_DIALS2018}%
  \BibitemOpen
  \bibfield  {author} {\bibinfo {author} {\bibfnamefont {G.}~\bibnamefont
  {Winter}}, \bibinfo {author} {\bibfnamefont {D.~G.}\ \bibnamefont
  {Waterman}}, \bibinfo {author} {\bibfnamefont {J.~M.}\ \bibnamefont
  {Parkhurst}}, \bibinfo {author} {\bibfnamefont {A.~S.}\ \bibnamefont
  {Brewster}}, \bibinfo {author} {\bibfnamefont {R.~J.}\ \bibnamefont
  {Gildea}}, \bibinfo {author} {\bibfnamefont {M.}~\bibnamefont {Gerstel}},
  \bibinfo {author} {\bibfnamefont {L.}~\bibnamefont {Fuentes-Montero}},
  \bibinfo {author} {\bibfnamefont {M.}~\bibnamefont {Vollmar}}, \bibinfo
  {author} {\bibfnamefont {T.}~\bibnamefont {Michels-Clark}}, \bibinfo {author}
  {\bibfnamefont {I.~D.}\ \bibnamefont {Young}}, \bibinfo {author}
  {\bibfnamefont {N.~K.}\ \bibnamefont {Sauter}},\ and\ \bibinfo {author}
  {\bibfnamefont {G.}~\bibnamefont {Evans}},\ }\bibfield  {title} {\bibinfo
  {title} {{{\it DIALS}: implementation and evaluation of a new integration
  package}},\ }\href {https://doi.org/10.1107/S2059798317017235} {\bibfield
  {journal} {\bibinfo  {journal} {Acta Crystallographica Section D}\ }\textbf
  {\bibinfo {volume} {74}},\ \bibinfo {pages} {85} (\bibinfo {year}
  {2018})}\BibitemShut {NoStop}%
\bibitem [{\citenamefont {Sheldrick}(2008)}]{sheldrickShortHistorySHELX2008}%
  \BibitemOpen
  \bibfield  {author} {\bibinfo {author} {\bibfnamefont {G.~M.}\ \bibnamefont
  {Sheldrick}},\ }\bibfield  {title} {\bibinfo {title} {A short history of
  {{SHELX}}},\ }\href {https://doi.org/10.1107/S0108767307043930} {\bibfield
  {journal} {\bibinfo  {journal} {Acta Crystallographica Section A}\ }\textbf
  {\bibinfo {volume} {64}},\ \bibinfo {pages} {112} (\bibinfo {year}
  {2008})}\BibitemShut {NoStop}%
\bibitem [{\citenamefont {Zvyagin}\ \emph {et~al.}(2004)\citenamefont
  {Zvyagin}, \citenamefont {Krzystek}, \citenamefont {{van Loosdrecht}},
  \citenamefont {Dhalenne},\ and\ \citenamefont
  {Revcolevschi}}]{zvyagin_PhysB_2004_ESRinCuGeO3}%
  \BibitemOpen
  \bibfield  {author} {\bibinfo {author} {\bibfnamefont {S.~A.}\ \bibnamefont
  {Zvyagin}}, \bibinfo {author} {\bibfnamefont {J.}~\bibnamefont {Krzystek}},
  \bibinfo {author} {\bibfnamefont {P.~H.~M.}\ \bibnamefont {{van
  Loosdrecht}}}, \bibinfo {author} {\bibfnamefont {G.}~\bibnamefont
  {Dhalenne}},\ and\ \bibinfo {author} {\bibfnamefont {A.}~\bibnamefont
  {Revcolevschi}},\ }\bibfield  {title} {\bibinfo {title} {{High-field ESR
  study of the dimerized-incommensurate phase transition in the spin-Peierls
  compound CuGeO$_3$}},\ }\href
  {https://doi.org/https://doi.org/10.1016/j.physb.2004.01.009} {\bibfield
  {journal} {\bibinfo  {journal} {Physica B}\ }\textbf {\bibinfo {volume}
  {346-347}},\ \bibinfo {pages} {1} (\bibinfo {year} {2004})}\BibitemShut
  {NoStop}%
\bibitem [{\citenamefont {K{\"u}chler}\ \emph {et~al.}(2023)\citenamefont
  {K{\"u}chler}, \citenamefont {Wawrzy{\'n}czak}, \citenamefont {{Dawczak-D{\k
  e}bicki}}, \citenamefont {Gooth},\ and\ \citenamefont
  {Galeski}}]{kuchlerNewApplicationsWorlds2023}%
  \BibitemOpen
  \bibfield  {author} {\bibinfo {author} {\bibfnamefont {R.}~\bibnamefont
  {K{\"u}chler}}, \bibinfo {author} {\bibfnamefont {R.}~\bibnamefont
  {Wawrzy{\'n}czak}}, \bibinfo {author} {\bibfnamefont {H.}~\bibnamefont
  {{Dawczak-D{\k e}bicki}}}, \bibinfo {author} {\bibfnamefont {J.}~\bibnamefont
  {Gooth}},\ and\ \bibinfo {author} {\bibfnamefont {S.}~\bibnamefont
  {Galeski}},\ }\bibfield  {title} {\bibinfo {title} {New applications for the
  world's smallest high-precision capacitance dilatometer and its
  stress-implementing counterpart},\ }\href {https://doi.org/10.1063/5.0141974}
  {\bibfield  {journal} {\bibinfo  {journal} {Review of Scientific
  Instruments}\ }\textbf {\bibinfo {volume} {94}},\ \bibinfo {pages} {045108}
  (\bibinfo {year} {2023})}\BibitemShut {NoStop}%
\bibitem [{\citenamefont {L{\"u}thi}(2005)}]{luthiPhysicalAcousticsSolid1967}%
  \BibitemOpen
  \bibfield  {author} {\bibinfo {author} {\bibfnamefont {B.}~\bibnamefont
  {L{\"u}thi}},\ }\href {https://doi.org/10.1007/b138867} {\emph {\bibinfo
  {title} {Physical {{Acoustics}} in the {{Solid State}}}}},\ \bibinfo
  {edition} {1st}\ ed.\ (\bibinfo  {publisher} {Springer},\ \bibinfo {address}
  {Berlin},\ \bibinfo {year} {2005})\BibitemShut {NoStop}%
\bibitem [{\citenamefont {Zherlitsyn}\ \emph {et~al.}(2014)\citenamefont
  {Zherlitsyn}, \citenamefont {Yasin}, \citenamefont {Wosnitza}, \citenamefont
  {Zvyagin}, \citenamefont {Andreev},\ and\ \citenamefont
  {Tsurkan}}]{zherlitsyn_LTPhys_SpinLatticeEffects2014}%
  \BibitemOpen
  \bibfield  {author} {\bibinfo {author} {\bibfnamefont {S.}~\bibnamefont
  {Zherlitsyn}}, \bibinfo {author} {\bibfnamefont {S.}~\bibnamefont {Yasin}},
  \bibinfo {author} {\bibfnamefont {J.}~\bibnamefont {Wosnitza}}, \bibinfo
  {author} {\bibfnamefont {A.~A.}\ \bibnamefont {Zvyagin}}, \bibinfo {author}
  {\bibfnamefont {A.~V.}\ \bibnamefont {Andreev}},\ and\ \bibinfo {author}
  {\bibfnamefont {V.}~\bibnamefont {Tsurkan}},\ }\bibfield  {title} {\bibinfo
  {title} {Spin-lattice effects in selected antiferromagnetic materials (review
  article)},\ }\href {https://doi.org/10.1063/1.4865559} {\bibfield  {journal}
  {\bibinfo  {journal} {Low Temperature Physics}\ }\textbf {\bibinfo {volume}
  {40}},\ \bibinfo {pages} {123} (\bibinfo {year} {2014})}\BibitemShut
  {NoStop}%
\bibitem [{\citenamefont {Nagl}\ \emph {et~al.}(2024)\citenamefont {Nagl},
  \citenamefont {Zheludev},\ and\ \citenamefont {Khalyavin}}]{wish}%
  \BibitemOpen
  \bibfield  {author} {\bibinfo {author} {\bibfnamefont {J.}~\bibnamefont
  {Nagl}}, \bibinfo {author} {\bibfnamefont {A.}~\bibnamefont {Zheludev}},\
  and\ \bibinfo {author} {\bibfnamefont {D.}~\bibnamefont {Khalyavin}},\
  }\bibfield  {title} {\bibinfo {title} {Field-dependent magnetic structure of
  the triangular-lattice antiferromagnet {{(CD$_3$ND$_3$)$_2$NaRuCl$_6$}}},\
  }\bibfield  {journal} {\bibinfo  {journal} {STFC ISIS Neutron and Muon
  Source}\ }\href {https://doi.org/10.5286/ISIS.E.RB2420060}
  {10.5286/ISIS.E.RB2420060} (\bibinfo {year} {2024})\BibitemShut {NoStop}%
\bibitem [{\citenamefont {Nagl}\ \emph
  {et~al.}(2025{\natexlab{a}})\citenamefont {Nagl}, \citenamefont {Zheludev},\
  and\ \citenamefont {Manuel}}]{wish2}%
  \BibitemOpen
  \bibfield  {author} {\bibinfo {author} {\bibfnamefont {J.}~\bibnamefont
  {Nagl}}, \bibinfo {author} {\bibfnamefont {A.}~\bibnamefont {Zheludev}},\
  and\ \bibinfo {author} {\bibfnamefont {P.}~\bibnamefont {Manuel}},\
  }\bibfield  {title} {\bibinfo {title} {Field-induced incommensurate magnetic
  structures of the triangular-lattice antiferromagnet
  {{(CD$_3$ND$_3$)$_2$NaRuCl$_6$}}},\ }\bibfield  {journal} {\bibinfo
  {journal} {STFC ISIS Neutron and Muon Source}\ }\href
  {https://doi.org/10.5286/ISIS.E.RB2520045} {10.5286/ISIS.E.RB2520045}
  (\bibinfo {year} {2025}{\natexlab{a}})\BibitemShut {NoStop}%
\bibitem [{\citenamefont {Arnold}\ \emph {et~al.}(2014)\citenamefont {Arnold},
  \citenamefont {Bilheux}, \citenamefont {Borreguero}, \citenamefont {Buts},
  \citenamefont {Campbell}, \citenamefont {Chapon}, \citenamefont {Doucet},
  \citenamefont {Draper}, \citenamefont {Ferraz~Leal}, \citenamefont {Gigg},
  \citenamefont {Lynch}, \citenamefont {Markvardsen}, \citenamefont
  {Mikkelson}, \citenamefont {Mikkelson}, \citenamefont {Miller}, \citenamefont
  {Palmen}, \citenamefont {Parker}, \citenamefont {Passos}, \citenamefont
  {Perring}, \citenamefont {Peterson}, \citenamefont {Ren}, \citenamefont
  {Reuter}, \citenamefont {Savici}, \citenamefont {Taylor}, \citenamefont
  {Taylor}, \citenamefont {Tolchenov}, \citenamefont {Zhou},\ and\
  \citenamefont {Zikovsky}}]{arnoldMantidDataAnalysis2014}%
  \BibitemOpen
  \bibfield  {author} {\bibinfo {author} {\bibfnamefont {O.}~\bibnamefont
  {Arnold}}, \bibinfo {author} {\bibfnamefont {J.~C.}\ \bibnamefont {Bilheux}},
  \bibinfo {author} {\bibfnamefont {J.~M.}\ \bibnamefont {Borreguero}},
  \bibinfo {author} {\bibfnamefont {A.}~\bibnamefont {Buts}}, \bibinfo {author}
  {\bibfnamefont {S.~I.}\ \bibnamefont {Campbell}}, \bibinfo {author}
  {\bibfnamefont {L.}~\bibnamefont {Chapon}}, \bibinfo {author} {\bibfnamefont
  {M.}~\bibnamefont {Doucet}}, \bibinfo {author} {\bibfnamefont
  {N.}~\bibnamefont {Draper}}, \bibinfo {author} {\bibfnamefont
  {R.}~\bibnamefont {Ferraz~Leal}}, \bibinfo {author} {\bibfnamefont {M.~A.}\
  \bibnamefont {Gigg}}, \bibinfo {author} {\bibfnamefont {V.~E.}\ \bibnamefont
  {Lynch}}, \bibinfo {author} {\bibfnamefont {A.}~\bibnamefont {Markvardsen}},
  \bibinfo {author} {\bibfnamefont {D.~J.}\ \bibnamefont {Mikkelson}}, \bibinfo
  {author} {\bibfnamefont {R.~L.}\ \bibnamefont {Mikkelson}}, \bibinfo {author}
  {\bibfnamefont {R.}~\bibnamefont {Miller}}, \bibinfo {author} {\bibfnamefont
  {K.}~\bibnamefont {Palmen}}, \bibinfo {author} {\bibfnamefont
  {P.}~\bibnamefont {Parker}}, \bibinfo {author} {\bibfnamefont
  {G.}~\bibnamefont {Passos}}, \bibinfo {author} {\bibfnamefont {T.~G.}\
  \bibnamefont {Perring}}, \bibinfo {author} {\bibfnamefont {P.~F.}\
  \bibnamefont {Peterson}}, \bibinfo {author} {\bibfnamefont {S.}~\bibnamefont
  {Ren}}, \bibinfo {author} {\bibfnamefont {M.~A.}\ \bibnamefont {Reuter}},
  \bibinfo {author} {\bibfnamefont {A.~T.}\ \bibnamefont {Savici}}, \bibinfo
  {author} {\bibfnamefont {J.~W.}\ \bibnamefont {Taylor}}, \bibinfo {author}
  {\bibfnamefont {R.~J.}\ \bibnamefont {Taylor}}, \bibinfo {author}
  {\bibfnamefont {R.}~\bibnamefont {Tolchenov}}, \bibinfo {author}
  {\bibfnamefont {W.}~\bibnamefont {Zhou}},\ and\ \bibinfo {author}
  {\bibfnamefont {J.}~\bibnamefont {Zikovsky}},\ }\bibfield  {title} {\bibinfo
  {title} {Mantid---{{Data}} analysis and visualization package for neutron
  scattering and {{$\mu$SR}} experiments},\ }\href
  {https://doi.org/10.1016/j.nima.2014.07.029} {\bibfield  {journal} {\bibinfo
  {journal} {Nuclear Instruments and Methods in Physics Research Section A:
  Accelerators, Spectrometers, Detectors and Associated Equipment}\ }\textbf
  {\bibinfo {volume} {764}},\ \bibinfo {pages} {156} (\bibinfo {year}
  {2014})}\BibitemShut {NoStop}%
\bibitem [{\citenamefont {Nagl}\ \emph
  {et~al.}(2025{\natexlab{b}})\citenamefont {Nagl}, \citenamefont {Gvasaliya},
  \citenamefont {Zheludev}, \citenamefont {Hiess},\ and\ \citenamefont
  {Steffens}}]{thales}%
  \BibitemOpen
  \bibfield  {author} {\bibinfo {author} {\bibfnamefont {J.}~\bibnamefont
  {Nagl}}, \bibinfo {author} {\bibfnamefont {S.}~\bibnamefont {Gvasaliya}},
  \bibinfo {author} {\bibfnamefont {A.}~\bibnamefont {Zheludev}}, \bibinfo
  {author} {\bibfnamefont {A.}~\bibnamefont {Hiess}},\ and\ \bibinfo {author}
  {\bibfnamefont {P.}~\bibnamefont {Steffens}},\ }\bibfield  {title} {\bibinfo
  {title} {Excitation spectrum of the triangular-lattice antiferromagnet
  {{(CD$_3$ND$_3$)$_2$NaRuCl$_6$}}},\ }\bibfield  {journal} {\bibinfo
  {journal} {Institut Laue-Langevin (ILL)}\ }\href
  {https://doi.org/10.5291/ILL-DATA.4-01-1857} {10.5291/ILL-DATA.4-01-1857}
  (\bibinfo {year} {2025}{\natexlab{b}})\BibitemShut {NoStop}%
\bibitem [{\citenamefont {Shirane}\ \emph {et~al.}(2002)\citenamefont
  {Shirane}, \citenamefont {Shapiro},\ and\ \citenamefont
  {Tranquada}}]{shiraneNeutronTripleAxis2002}%
  \BibitemOpen
  \bibfield  {author} {\bibinfo {author} {\bibfnamefont {G.}~\bibnamefont
  {Shirane}}, \bibinfo {author} {\bibfnamefont {S.~M.}\ \bibnamefont
  {Shapiro}},\ and\ \bibinfo {author} {\bibfnamefont {J.~M.}\ \bibnamefont
  {Tranquada}},\ }\href
  {https://doi.org/https://doi.org/10.1017/CBO9780511534881} {\emph {\bibinfo
  {title} {Neutron Scattering with a Triple-Axis Spectrometer: Basic
  Techniques}}}\ (\bibinfo  {publisher} {Cambridge University Press},\ \bibinfo
  {address} {Cambridge},\ \bibinfo {year} {2002})\BibitemShut {NoStop}%
\bibitem [{\citenamefont {Toth}\ and\ \citenamefont
  {Lake}(2015)}]{tothLinearSpinWave2015}%
  \BibitemOpen
  \bibfield  {author} {\bibinfo {author} {\bibfnamefont {S.}~\bibnamefont
  {Toth}}\ and\ \bibinfo {author} {\bibfnamefont {B.}~\bibnamefont {Lake}},\
  }\bibfield  {title} {\bibinfo {title} {Linear spin wave theory for
  single-{{Q}} incommensurate magnetic structures},\ }\href
  {https://doi.org/10.1088/0953-8984/27/16/166002} {\bibfield  {journal}
  {\bibinfo  {journal} {Journal of Physics: Condensed Matter}\ }\textbf
  {\bibinfo {volume} {27}},\ \bibinfo {pages} {166002} (\bibinfo {year}
  {2015})}\BibitemShut {NoStop}%
\bibitem [{\citenamefont {Wa{\'s}kowska}\ \emph {et~al.}(2010)\citenamefont
  {Wa{\'s}kowska}, \citenamefont {Gerward}, \citenamefont {Staun~Olsen},
  \citenamefont {Morgenroth}, \citenamefont {M{\k a}czka},\ and\ \citenamefont
  {Hermanowicz}}]{waskowskaTemperaturePressuredependentLattice2010a}%
  \BibitemOpen
  \bibfield  {author} {\bibinfo {author} {\bibfnamefont {A.}~\bibnamefont
  {Wa{\'s}kowska}}, \bibinfo {author} {\bibfnamefont {L.}~\bibnamefont
  {Gerward}}, \bibinfo {author} {\bibfnamefont {J.}~\bibnamefont
  {Staun~Olsen}}, \bibinfo {author} {\bibfnamefont {W.}~\bibnamefont
  {Morgenroth}}, \bibinfo {author} {\bibfnamefont {M.}~\bibnamefont {M{\k
  a}czka}},\ and\ \bibinfo {author} {\bibfnamefont {K.}~\bibnamefont
  {Hermanowicz}},\ }\bibfield  {title} {\bibinfo {title} {Temperature- and
  pressure-dependent lattice behaviour of {{RbFe}}({{MoO$_4$}})$_2$},\ }\href
  {https://doi.org/10.1088/0953-8984/22/5/055406} {\bibfield  {journal}
  {\bibinfo  {journal} {Journal of Physics: Condensed Matter}\ }\textbf
  {\bibinfo {volume} {22}},\ \bibinfo {pages} {055406} (\bibinfo {year}
  {2010})}\BibitemShut {NoStop}%
\bibitem [{\citenamefont {Kajita}\ \emph {et~al.}(2024)\citenamefont {Kajita},
  \citenamefont {Nagai}, \citenamefont {Yamagishi}, \citenamefont {Kimura},
  \citenamefont {Hagihala},\ and\ \citenamefont
  {Kimura}}]{kajitaFerroaxialTransitionsGlaseriteType2024}%
  \BibitemOpen
  \bibfield  {author} {\bibinfo {author} {\bibfnamefont {Y.}~\bibnamefont
  {Kajita}}, \bibinfo {author} {\bibfnamefont {T.}~\bibnamefont {Nagai}},
  \bibinfo {author} {\bibfnamefont {S.}~\bibnamefont {Yamagishi}}, \bibinfo
  {author} {\bibfnamefont {K.}~\bibnamefont {Kimura}}, \bibinfo {author}
  {\bibfnamefont {M.}~\bibnamefont {Hagihala}},\ and\ \bibinfo {author}
  {\bibfnamefont {T.}~\bibnamefont {Kimura}},\ }\bibfield  {title} {\bibinfo
  {title} {Ferroaxial {{Transitions}} in {{Glaserite-Type
  Na$_2$BaM}}({{PO$_4$}})$_2$ ({{M}} = {{Mg}}, {{Mn}}, {{Co}}, and {{Ni}})},\
  }\href {https://doi.org/10.1021/acs.chemmater.4c01406} {\bibfield  {journal}
  {\bibinfo  {journal} {Chemistry of Materials}\ }\textbf {\bibinfo {volume}
  {36}},\ \bibinfo {pages} {7451} (\bibinfo {year} {2024})}\BibitemShut
  {NoStop}%
\bibitem [{\citenamefont {Hearmon}\ \emph {et~al.}(2012)\citenamefont
  {Hearmon}, \citenamefont {Fabrizi}, \citenamefont {Chapon}, \citenamefont
  {Johnson}, \citenamefont {Prabhakaran}, \citenamefont {Streltsov},
  \citenamefont {Brown},\ and\ \citenamefont
  {Radaelli}}]{hearmonElectricFieldControl2012}%
  \BibitemOpen
  \bibfield  {author} {\bibinfo {author} {\bibfnamefont {A.~J.}\ \bibnamefont
  {Hearmon}}, \bibinfo {author} {\bibfnamefont {F.}~\bibnamefont {Fabrizi}},
  \bibinfo {author} {\bibfnamefont {L.~C.}\ \bibnamefont {Chapon}}, \bibinfo
  {author} {\bibfnamefont {R.~D.}\ \bibnamefont {Johnson}}, \bibinfo {author}
  {\bibfnamefont {D.}~\bibnamefont {Prabhakaran}}, \bibinfo {author}
  {\bibfnamefont {S.~V.}\ \bibnamefont {Streltsov}}, \bibinfo {author}
  {\bibfnamefont {P.~J.}\ \bibnamefont {Brown}},\ and\ \bibinfo {author}
  {\bibfnamefont {P.~G.}\ \bibnamefont {Radaelli}},\ }\bibfield  {title}
  {\bibinfo {title} {Electric {{Field Control}} of the {{Magnetic Chiralities}}
  in {{Ferroaxial Multiferroic}} {{RbFe(MoO$_4$)$_2$}}},\ }\href
  {https://doi.org/10.1103/PhysRevLett.108.237201} {\bibfield  {journal}
  {\bibinfo  {journal} {Physical Review Letters}\ }\textbf {\bibinfo {volume}
  {108}},\ \bibinfo {pages} {237201} (\bibinfo {year} {2012})}\BibitemShut
  {NoStop}%
\bibitem [{\citenamefont {Zelenskiy}\ \emph {et~al.}(2021)\citenamefont
  {Zelenskiy}, \citenamefont {Quilliam}, \citenamefont {Shapiro},\ and\
  \citenamefont {Quirion}}]{zelenskiyMagneticPhasesS52021}%
  \BibitemOpen
  \bibfield  {author} {\bibinfo {author} {\bibfnamefont {A.}~\bibnamefont
  {Zelenskiy}}, \bibinfo {author} {\bibfnamefont {J.~A.}\ \bibnamefont
  {Quilliam}}, \bibinfo {author} {\bibfnamefont {A.~{\relax Ya}.}\ \bibnamefont
  {Shapiro}},\ and\ \bibinfo {author} {\bibfnamefont {G.}~\bibnamefont
  {Quirion}},\ }\bibfield  {title} {\bibinfo {title} {Magnetic phases of the
  {{S}}=5/2 triangular-lattice antiferromagnet {{RbFe}}({{MO$_4$}})$_2$ as
  determined by ultrasound velocity measurements},\ }\href
  {https://doi.org/10.1103/PhysRevB.103.224422} {\bibfield  {journal} {\bibinfo
   {journal} {Physical Review B}\ }\textbf {\bibinfo {volume} {103}},\ \bibinfo
  {pages} {224422} (\bibinfo {year} {2021})}\BibitemShut {NoStop}%
\bibitem [{\citenamefont {Kubota}\ \emph {et~al.}(2015)\citenamefont {Kubota},
  \citenamefont {Tanaka}, \citenamefont {Ono}, \citenamefont {Narumi},\ and\
  \citenamefont {Kindo}}]{kubotaSuccessiveMagneticPhase2015}%
  \BibitemOpen
  \bibfield  {author} {\bibinfo {author} {\bibfnamefont {Y.}~\bibnamefont
  {Kubota}}, \bibinfo {author} {\bibfnamefont {H.}~\bibnamefont {Tanaka}},
  \bibinfo {author} {\bibfnamefont {T.}~\bibnamefont {Ono}}, \bibinfo {author}
  {\bibfnamefont {Y.}~\bibnamefont {Narumi}},\ and\ \bibinfo {author}
  {\bibfnamefont {K.}~\bibnamefont {Kindo}},\ }\bibfield  {title} {\bibinfo
  {title} {Successive magnetic phase transitions in {{RuCl$_3$}}: {{XY-like}}
  frustrated magnet on the honeycomb lattice},\ }\href
  {https://doi.org/10.1103/PhysRevB.91.094422} {\bibfield  {journal} {\bibinfo
  {journal} {Physical Review B}\ }\textbf {\bibinfo {volume} {91}},\ \bibinfo
  {pages} {094422} (\bibinfo {year} {2015})}\BibitemShut {NoStop}%
\bibitem [{\citenamefont {Bergman}\ \emph {et~al.}(2006)\citenamefont
  {Bergman}, \citenamefont {Shindou}, \citenamefont {Fiete},\ and\
  \citenamefont {Balents}}]{bergmanModelsDegeneracyBreaking2006}%
  \BibitemOpen
  \bibfield  {author} {\bibinfo {author} {\bibfnamefont {D.~L.}\ \bibnamefont
  {Bergman}}, \bibinfo {author} {\bibfnamefont {R.}~\bibnamefont {Shindou}},
  \bibinfo {author} {\bibfnamefont {G.~A.}\ \bibnamefont {Fiete}},\ and\
  \bibinfo {author} {\bibfnamefont {L.}~\bibnamefont {Balents}},\ }\bibfield
  {title} {\bibinfo {title} {Models of degeneracy breaking in pyrochlore
  antiferromagnets},\ }\href {https://doi.org/10.1103/PhysRevB.74.134409}
  {\bibfield  {journal} {\bibinfo  {journal} {Physical Review B}\ }\textbf
  {\bibinfo {volume} {74}},\ \bibinfo {pages} {134409} (\bibinfo {year}
  {2006})}\BibitemShut {NoStop}%
\bibitem [{\citenamefont {Capriotti}\ \emph {et~al.}(1999)\citenamefont
  {Capriotti}, \citenamefont {Trumper},\ and\ \citenamefont
  {Sorella}}]{capriottiLongRangeNeelOrder1999}%
  \BibitemOpen
  \bibfield  {author} {\bibinfo {author} {\bibfnamefont {L.}~\bibnamefont
  {Capriotti}}, \bibinfo {author} {\bibfnamefont {A.~E.}\ \bibnamefont
  {Trumper}},\ and\ \bibinfo {author} {\bibfnamefont {S.}~\bibnamefont
  {Sorella}},\ }\bibfield  {title} {\bibinfo {title} {{{N{\'e}el}} {{Order}} in
  the {{Triangular Heisenberg Model}}},\ }\href
  {https://doi.org/10.1103/PhysRevLett.82.3899} {\bibfield  {journal} {\bibinfo
   {journal} {Physical Review Letters}\ }\textbf {\bibinfo {volume} {82}},\
  \bibinfo {pages} {3899} (\bibinfo {year} {1999})}\BibitemShut {NoStop}%
\bibitem [{\citenamefont {Zheng}\ \emph {et~al.}(2006)\citenamefont {Zheng},
  \citenamefont {Fj{\ae}restad}, \citenamefont {Singh}, \citenamefont
  {McKenzie},\ and\ \citenamefont
  {Coldea}}]{zhengExcitationSpectraSpin$frac12$2006}%
  \BibitemOpen
  \bibfield  {author} {\bibinfo {author} {\bibfnamefont {W.}~\bibnamefont
  {Zheng}}, \bibinfo {author} {\bibfnamefont {J.~O.}\ \bibnamefont
  {Fj{\ae}restad}}, \bibinfo {author} {\bibfnamefont {R.~R.~P.}\ \bibnamefont
  {Singh}}, \bibinfo {author} {\bibfnamefont {R.~H.}\ \bibnamefont
  {McKenzie}},\ and\ \bibinfo {author} {\bibfnamefont {R.}~\bibnamefont
  {Coldea}},\ }\bibfield  {title} {\bibinfo {title} {Excitation spectra of the
  spin-{{$\frac{1}{2}$}} triangular-lattice {{Heisenberg}} antiferromagnet},\
  }\href {https://doi.org/10.1103/PhysRevB.74.224420} {\bibfield  {journal}
  {\bibinfo  {journal} {Physical Review B}\ }\textbf {\bibinfo {volume} {74}},\
  \bibinfo {pages} {224420} (\bibinfo {year} {2006})}\BibitemShut {NoStop}%
\bibitem [{\citenamefont {White}\ and\ \citenamefont
  {Chernyshev}(2007)}]{whiteNeelOrderSquare2007a}%
  \BibitemOpen
  \bibfield  {author} {\bibinfo {author} {\bibfnamefont {S.~R.}\ \bibnamefont
  {White}}\ and\ \bibinfo {author} {\bibfnamefont {A.~L.}\ \bibnamefont
  {Chernyshev}},\ }\bibfield  {title} {\bibinfo {title} {{{N{\'e}el}} {{Order}}
  in {{Square}} and {{Triangular Lattice Heisenberg Models}}},\ }\href
  {https://doi.org/10.1103/PhysRevLett.99.127004} {\bibfield  {journal}
  {\bibinfo  {journal} {Physical Review Letters}\ }\textbf {\bibinfo {volume}
  {99}},\ \bibinfo {pages} {127004} (\bibinfo {year} {2007})}\BibitemShut
  {NoStop}%
\bibitem [{\citenamefont {Mouhat}\ and\ \citenamefont
  {Coudert}(2014)}]{mouhatNecessarySufficientElastic2014}%
  \BibitemOpen
  \bibfield  {author} {\bibinfo {author} {\bibfnamefont {F.}~\bibnamefont
  {Mouhat}}\ and\ \bibinfo {author} {\bibfnamefont {F.-X.}\ \bibnamefont
  {Coudert}},\ }\bibfield  {title} {\bibinfo {title} {Necessary and sufficient
  elastic stability conditions in various crystal systems},\ }\href
  {https://doi.org/10.1103/PhysRevB.90.224104} {\bibfield  {journal} {\bibinfo
  {journal} {Physical Review B}\ }\textbf {\bibinfo {volume} {90}},\ \bibinfo
  {pages} {224104} (\bibinfo {year} {2014})}\BibitemShut {NoStop}%
\bibitem [{\citenamefont {Villain}\ \emph {et~al.}(1980)\citenamefont
  {Villain}, \citenamefont {Bidaux}, \citenamefont {Carton},\ and\
  \citenamefont {Conte}}]{villainOrderEffectDisorder1980}%
  \BibitemOpen
  \bibfield  {author} {\bibinfo {author} {\bibfnamefont {J.}~\bibnamefont
  {Villain}}, \bibinfo {author} {\bibfnamefont {R.}~\bibnamefont {Bidaux}},
  \bibinfo {author} {\bibfnamefont {J.-P.}\ \bibnamefont {Carton}},\ and\
  \bibinfo {author} {\bibfnamefont {R.}~\bibnamefont {Conte}},\ }\bibfield
  {title} {\bibinfo {title} {Order as an effect of disorder},\ }\href
  {https://doi.org/10.1051/jphys:0198000410110126300} {\bibfield  {journal}
  {\bibinfo  {journal} {Journal de Physique}\ }\textbf {\bibinfo {volume}
  {41}},\ \bibinfo {pages} {1263} (\bibinfo {year} {1980})}\BibitemShut
  {NoStop}%
\bibitem [{\citenamefont {Heinil{\"a}}\ and\ \citenamefont
  {Oja}(1993)}]{heinilaSelectionGroundState1993c}%
  \BibitemOpen
  \bibfield  {author} {\bibinfo {author} {\bibfnamefont {M.~T.}\ \bibnamefont
  {Heinil{\"a}}}\ and\ \bibinfo {author} {\bibfnamefont {A.~S.}\ \bibnamefont
  {Oja}},\ }\bibfield  {title} {\bibinfo {title} {Selection of the ground state
  in type-{{I}} fcc antiferromagnets in an external magnetic field},\ }\href
  {https://doi.org/10.1103/PhysRevB.48.7227} {\bibfield  {journal} {\bibinfo
  {journal} {Physical Review B}\ }\textbf {\bibinfo {volume} {48}},\ \bibinfo
  {pages} {7227} (\bibinfo {year} {1993})}\BibitemShut {NoStop}%
\end{thebibliography}%


\begin{thebibliography}{29}%
\makeatletter
\providecommand \@ifxundefined [1]{%
 \@ifx{#1\undefined}
}%
\providecommand \@ifnum [1]{%
 \ifnum #1\expandafter \@firstoftwo
 \else \expandafter \@secondoftwo
 \fi
}%
\providecommand \@ifx [1]{%
 \ifx #1\expandafter \@firstoftwo
 \else \expandafter \@secondoftwo
 \fi
}%
\providecommand \natexlab [1]{#1}%
\providecommand \enquote  [1]{``#1''}%
\providecommand \bibnamefont  [1]{#1}%
\providecommand \bibfnamefont [1]{#1}%
\providecommand \citenamefont [1]{#1}%
\providecommand \href@noop [0]{\@secondoftwo}%
\providecommand \href [0]{\begingroup \@sanitize@url \@href}%
\providecommand \@href[1]{\@@startlink{#1}\@@href}%
\providecommand \@@href[1]{\endgroup#1\@@endlink}%
\providecommand \@sanitize@url [0]{\catcode `\\12\catcode `\$12\catcode
  `\&12\catcode `\#12\catcode `\^12\catcode `\_12\catcode `\%12\relax}%
\providecommand \@@startlink[1]{}%
\providecommand \@@endlink[0]{}%
\providecommand \url  [0]{\begingroup\@sanitize@url \@url }%
\providecommand \@url [1]{\endgroup\@href {#1}{\urlprefix }}%
\providecommand \urlprefix  [0]{URL }%
\providecommand \Eprint [0]{\href }%
\providecommand \doibase [0]{https://doi.org/}%
\providecommand \selectlanguage [0]{\@gobble}%
\providecommand \bibinfo  [0]{\@secondoftwo}%
\providecommand \bibfield  [0]{\@secondoftwo}%
\providecommand \translation [1]{[#1]}%
\providecommand \BibitemOpen [0]{}%
\providecommand \bibitemStop [0]{}%
\providecommand \bibitemNoStop [0]{.\EOS\space}%
\providecommand \EOS [0]{\spacefactor3000\relax}%
\providecommand \BibitemShut  [1]{\csname bibitem#1\endcsname}%
\let\auto@bib@innerbib\@empty
\bibitem [{\citenamefont {Vishnoi}\ \emph {et~al.}(2020)\citenamefont
  {Vishnoi}, \citenamefont {Zuo}, \citenamefont {Strom}, \citenamefont {Wu},
  \citenamefont {Wilson}, \citenamefont {Seshadri},\ and\ \citenamefont
  {Cheetham}}]{vishnoiStructuralDiversityMagnetic2020}%
  \BibitemOpen
  \bibfield  {author} {\bibinfo {author} {\bibfnamefont {P.}~\bibnamefont
  {Vishnoi}}, \bibinfo {author} {\bibfnamefont {J.~L.}\ \bibnamefont {Zuo}},
  \bibinfo {author} {\bibfnamefont {T.~A.}\ \bibnamefont {Strom}}, \bibinfo
  {author} {\bibfnamefont {G.}~\bibnamefont {Wu}}, \bibinfo {author}
  {\bibfnamefont {S.~D.}\ \bibnamefont {Wilson}}, \bibinfo {author}
  {\bibfnamefont {R.}~\bibnamefont {Seshadri}},\ and\ \bibinfo {author}
  {\bibfnamefont {A.~K.}\ \bibnamefont {Cheetham}},\ }\bibfield  {title}
  {\bibinfo {title} {Structural {{Diversity}} and {{Magnetic Properties}} of
  {{Hybrid Ruthenium Halide Perovskites}} and {{Related Compounds}}},\ }\href
  {https://doi.org/10.1002/anie.202003095} {\bibfield  {journal} {\bibinfo
  {journal} {Angewandte Chemie International Edition}\ }\textbf {\bibinfo
  {volume} {59}},\ \bibinfo {pages} {8974} (\bibinfo {year}
  {2020})}\BibitemShut {NoStop}%
\bibitem [{\citenamefont {Wa{\'s}kowska}\ \emph {et~al.}(2010)\citenamefont
  {Wa{\'s}kowska}, \citenamefont {Gerward}, \citenamefont {Staun~Olsen},
  \citenamefont {Morgenroth}, \citenamefont {M{\k a}czka},\ and\ \citenamefont
  {Hermanowicz}}]{waskowskaTemperaturePressuredependentLattice2010a}%
  \BibitemOpen
  \bibfield  {author} {\bibinfo {author} {\bibfnamefont {A.}~\bibnamefont
  {Wa{\'s}kowska}}, \bibinfo {author} {\bibfnamefont {L.}~\bibnamefont
  {Gerward}}, \bibinfo {author} {\bibfnamefont {J.}~\bibnamefont
  {Staun~Olsen}}, \bibinfo {author} {\bibfnamefont {W.}~\bibnamefont
  {Morgenroth}}, \bibinfo {author} {\bibfnamefont {M.}~\bibnamefont {M{\k
  a}czka}},\ and\ \bibinfo {author} {\bibfnamefont {K.}~\bibnamefont
  {Hermanowicz}},\ }\bibfield  {title} {\bibinfo {title} {Temperature- and
  pressure-dependent lattice behaviour of {{RbFe}}({{MoO$_4$}})$_2$},\ }\href
  {https://doi.org/10.1088/0953-8984/22/5/055406} {\bibfield  {journal}
  {\bibinfo  {journal} {Journal of Physics: Condensed Matter}\ }\textbf
  {\bibinfo {volume} {22}},\ \bibinfo {pages} {055406} (\bibinfo {year}
  {2010})}\BibitemShut {NoStop}%
\bibitem [{\citenamefont {Kajita}\ \emph {et~al.}(2024)\citenamefont {Kajita},
  \citenamefont {Nagai}, \citenamefont {Yamagishi}, \citenamefont {Kimura},
  \citenamefont {Hagihala},\ and\ \citenamefont
  {Kimura}}]{kajitaFerroaxialTransitionsGlaseriteType2024}%
  \BibitemOpen
  \bibfield  {author} {\bibinfo {author} {\bibfnamefont {Y.}~\bibnamefont
  {Kajita}}, \bibinfo {author} {\bibfnamefont {T.}~\bibnamefont {Nagai}},
  \bibinfo {author} {\bibfnamefont {S.}~\bibnamefont {Yamagishi}}, \bibinfo
  {author} {\bibfnamefont {K.}~\bibnamefont {Kimura}}, \bibinfo {author}
  {\bibfnamefont {M.}~\bibnamefont {Hagihala}},\ and\ \bibinfo {author}
  {\bibfnamefont {T.}~\bibnamefont {Kimura}},\ }\bibfield  {title} {\bibinfo
  {title} {Ferroaxial {{Transitions}} in {{Glaserite-Type
  Na$_2$BaM}}({{PO$_4$}})$_2$ ({{M}} = {{Mg}}, {{Mn}}, {{Co}}, and {{Ni}})},\
  }\href {https://doi.org/10.1021/acs.chemmater.4c01406} {\bibfield  {journal}
  {\bibinfo  {journal} {Chemistry of Materials}\ }\textbf {\bibinfo {volume}
  {36}},\ \bibinfo {pages} {7451} (\bibinfo {year} {2024})}\BibitemShut
  {NoStop}%
\bibitem [{\citenamefont {Woodland}\ \emph {et~al.}(2025)\citenamefont
  {Woodland}, \citenamefont {Okuma}, \citenamefont {Stewart}, \citenamefont
  {Balz},\ and\ \citenamefont
  {Coldea}}]{woodlandContinuumExcitationsSharp2025}%
  \BibitemOpen
  \bibfield  {author} {\bibinfo {author} {\bibfnamefont {L.}~\bibnamefont
  {Woodland}}, \bibinfo {author} {\bibfnamefont {R.}~\bibnamefont {Okuma}},
  \bibinfo {author} {\bibfnamefont {J.~R.}\ \bibnamefont {Stewart}}, \bibinfo
  {author} {\bibfnamefont {C.}~\bibnamefont {Balz}},\ and\ \bibinfo {author}
  {\bibfnamefont {R.}~\bibnamefont {Coldea}},\ }\bibfield  {title} {\bibinfo
  {title} {From continuum excitations to sharp magnons via transverse magnetic
  field in the spin-$\frac{1}{2}$ ising-like triangular lattice antiferromagnet
  ${{\mathrm{Na}}}_{2}\mathrm{BaCo}{({\mathrm{PO}}_{4})}_{2}$},\ }\href
  {https://doi.org/10.1103/1pvl-kzjm} {\bibfield  {journal} {\bibinfo
  {journal} {Phys. Rev. B}\ }\textbf {\bibinfo {volume} {112}},\ \bibinfo
  {pages} {104413} (\bibinfo {year} {2025})}\BibitemShut {NoStop}%
\bibitem [{\citenamefont {Hearmon}\ \emph {et~al.}(2012)\citenamefont
  {Hearmon}, \citenamefont {Fabrizi}, \citenamefont {Chapon}, \citenamefont
  {Johnson}, \citenamefont {Prabhakaran}, \citenamefont {Streltsov},
  \citenamefont {Brown},\ and\ \citenamefont
  {Radaelli}}]{hearmonElectricFieldControl2012}%
  \BibitemOpen
  \bibfield  {author} {\bibinfo {author} {\bibfnamefont {A.~J.}\ \bibnamefont
  {Hearmon}}, \bibinfo {author} {\bibfnamefont {F.}~\bibnamefont {Fabrizi}},
  \bibinfo {author} {\bibfnamefont {L.~C.}\ \bibnamefont {Chapon}}, \bibinfo
  {author} {\bibfnamefont {R.~D.}\ \bibnamefont {Johnson}}, \bibinfo {author}
  {\bibfnamefont {D.}~\bibnamefont {Prabhakaran}}, \bibinfo {author}
  {\bibfnamefont {S.~V.}\ \bibnamefont {Streltsov}}, \bibinfo {author}
  {\bibfnamefont {P.~J.}\ \bibnamefont {Brown}},\ and\ \bibinfo {author}
  {\bibfnamefont {P.~G.}\ \bibnamefont {Radaelli}},\ }\bibfield  {title}
  {\bibinfo {title} {Electric {{Field Control}} of the {{Magnetic Chiralities}}
  in {{Ferroaxial Multiferroic}} {{RbFe(MoO$_4$)$_2$}}},\ }\href
  {https://doi.org/10.1103/PhysRevLett.108.237201} {\bibfield  {journal}
  {\bibinfo  {journal} {Physical Review Letters}\ }\textbf {\bibinfo {volume}
  {108}},\ \bibinfo {pages} {237201} (\bibinfo {year} {2012})}\BibitemShut
  {NoStop}%
\bibitem [{\citenamefont {Zelenskiy}\ \emph {et~al.}(2021)\citenamefont
  {Zelenskiy}, \citenamefont {Quilliam}, \citenamefont {Shapiro},\ and\
  \citenamefont {Quirion}}]{zelenskiyMagneticPhasesS52021}%
  \BibitemOpen
  \bibfield  {author} {\bibinfo {author} {\bibfnamefont {A.}~\bibnamefont
  {Zelenskiy}}, \bibinfo {author} {\bibfnamefont {J.~A.}\ \bibnamefont
  {Quilliam}}, \bibinfo {author} {\bibfnamefont {A.~{\relax Ya}.}\ \bibnamefont
  {Shapiro}},\ and\ \bibinfo {author} {\bibfnamefont {G.}~\bibnamefont
  {Quirion}},\ }\bibfield  {title} {\bibinfo {title} {Magnetic phases of the
  {{S}}=5/2 triangular-lattice antiferromagnet {{RbFe}}({{MO$_4$}})$_2$ as
  determined by ultrasound velocity measurements},\ }\href
  {https://doi.org/10.1103/PhysRevB.103.224422} {\bibfield  {journal} {\bibinfo
   {journal} {Physical Review B}\ }\textbf {\bibinfo {volume} {103}},\ \bibinfo
  {pages} {224422} (\bibinfo {year} {2021})}\BibitemShut {NoStop}%
\bibitem [{\citenamefont {Abragam}\ and\ \citenamefont
  {Bleaney}(1970)}]{abragamElectronParamagneticResonance}%
  \BibitemOpen
  \bibfield  {author} {\bibinfo {author} {\bibfnamefont {A.}~\bibnamefont
  {Abragam}}\ and\ \bibinfo {author} {\bibfnamefont {B.}~\bibnamefont
  {Bleaney}},\ }\href
  {https://global.oup.com/academic/product/electron-paramagnetic-resonance-of-transition-ions-9780199651528?cc=ch&lang=en&}
  {\emph {\bibinfo {title} {Electron {{Paramagnetic Resonance}} of {{Transition
  Ions}}}}},\ \bibinfo {edition} {1st}\ ed.,\ Oxford classic texts in the
  physical sciences\ (\bibinfo  {publisher} {Oxford University Press},\
  \bibinfo {year} {1970})\BibitemShut {NoStop}%
\bibitem [{\citenamefont {Kubota}\ \emph {et~al.}(2015)\citenamefont {Kubota},
  \citenamefont {Tanaka}, \citenamefont {Ono}, \citenamefont {Narumi},\ and\
  \citenamefont {Kindo}}]{kubotaSuccessiveMagneticPhase2015}%
  \BibitemOpen
  \bibfield  {author} {\bibinfo {author} {\bibfnamefont {Y.}~\bibnamefont
  {Kubota}}, \bibinfo {author} {\bibfnamefont {H.}~\bibnamefont {Tanaka}},
  \bibinfo {author} {\bibfnamefont {T.}~\bibnamefont {Ono}}, \bibinfo {author}
  {\bibfnamefont {Y.}~\bibnamefont {Narumi}},\ and\ \bibinfo {author}
  {\bibfnamefont {K.}~\bibnamefont {Kindo}},\ }\bibfield  {title} {\bibinfo
  {title} {Successive magnetic phase transitions in {{RuCl$_3$}}: {{XY-like}}
  frustrated magnet on the honeycomb lattice},\ }\href
  {https://doi.org/10.1103/PhysRevB.91.094422} {\bibfield  {journal} {\bibinfo
  {journal} {Physical Review B}\ }\textbf {\bibinfo {volume} {91}},\ \bibinfo
  {pages} {094422} (\bibinfo {year} {2015})}\BibitemShut {NoStop}%
\bibitem [{\citenamefont {Lines}(1979)}]{linesElasticPropertiesMagnetic1979}%
  \BibitemOpen
  \bibfield  {author} {\bibinfo {author} {\bibfnamefont {M.~E.}\ \bibnamefont
  {Lines}},\ }\bibfield  {title} {\bibinfo {title} {Elastic properties of
  magnetic materials},\ }\href {https://doi.org/10.1016/0370-1573(79)90039-5}
  {\bibfield  {journal} {\bibinfo  {journal} {Physics Reports}\ }\textbf
  {\bibinfo {volume} {55}},\ \bibinfo {pages} {133} (\bibinfo {year}
  {1979})}\BibitemShut {NoStop}%
\bibitem [{\citenamefont {Penc}\ \emph {et~al.}(2004)\citenamefont {Penc},
  \citenamefont {Shannon},\ and\ \citenamefont
  {Shiba}}]{pencHalfMagnetizationPlateauStabilized2004}%
  \BibitemOpen
  \bibfield  {author} {\bibinfo {author} {\bibfnamefont {K.}~\bibnamefont
  {Penc}}, \bibinfo {author} {\bibfnamefont {N.}~\bibnamefont {Shannon}},\ and\
  \bibinfo {author} {\bibfnamefont {H.}~\bibnamefont {Shiba}},\ }\bibfield
  {title} {\bibinfo {title} {Half-{{Magnetization Plateau Stabilized}} by
  {{Structural Distortion}} in the {{Antiferromagnetic Heisenberg Model}} on a
  {{Pyrochlore Lattice}}},\ }\href
  {https://doi.org/10.1103/PhysRevLett.93.197203} {\bibfield  {journal}
  {\bibinfo  {journal} {Physical Review Letters}\ }\textbf {\bibinfo {volume}
  {93}},\ \bibinfo {pages} {197203} (\bibinfo {year} {2004})}\BibitemShut
  {NoStop}%
\bibitem [{\citenamefont {Bergman}\ \emph {et~al.}(2006)\citenamefont
  {Bergman}, \citenamefont {Shindou}, \citenamefont {Fiete},\ and\
  \citenamefont {Balents}}]{bergmanModelsDegeneracyBreaking2006}%
  \BibitemOpen
  \bibfield  {author} {\bibinfo {author} {\bibfnamefont {D.~L.}\ \bibnamefont
  {Bergman}}, \bibinfo {author} {\bibfnamefont {R.}~\bibnamefont {Shindou}},
  \bibinfo {author} {\bibfnamefont {G.~A.}\ \bibnamefont {Fiete}},\ and\
  \bibinfo {author} {\bibfnamefont {L.}~\bibnamefont {Balents}},\ }\bibfield
  {title} {\bibinfo {title} {Models of degeneracy breaking in pyrochlore
  antiferromagnets},\ }\href {https://doi.org/10.1103/PhysRevB.74.134409}
  {\bibfield  {journal} {\bibinfo  {journal} {Physical Review B}\ }\textbf
  {\bibinfo {volume} {74}},\ \bibinfo {pages} {134409} (\bibinfo {year}
  {2006})}\BibitemShut {NoStop}%
\bibitem [{\citenamefont {Tchernyshyov}\ \emph {et~al.}(2002)\citenamefont
  {Tchernyshyov}, \citenamefont {Moessner},\ and\ \citenamefont
  {Sondhi}}]{tchernyshyovOrderDistortionString2002a}%
  \BibitemOpen
  \bibfield  {author} {\bibinfo {author} {\bibfnamefont {O.}~\bibnamefont
  {Tchernyshyov}}, \bibinfo {author} {\bibfnamefont {R.}~\bibnamefont
  {Moessner}},\ and\ \bibinfo {author} {\bibfnamefont {S.~L.}\ \bibnamefont
  {Sondhi}},\ }\bibfield  {title} {\bibinfo {title} {Order by {{Distortion}}
  and {{String Modes}} in {{Pyrochlore Antiferromagnets}}},\ }\href
  {https://doi.org/10.1103/PhysRevLett.88.067203} {\bibfield  {journal}
  {\bibinfo  {journal} {Physical Review Letters}\ }\textbf {\bibinfo {volume}
  {88}},\ \bibinfo {pages} {067203} (\bibinfo {year} {2002})}\BibitemShut
  {NoStop}%
\bibitem [{\citenamefont {Capriotti}\ \emph {et~al.}(1999)\citenamefont
  {Capriotti}, \citenamefont {Trumper},\ and\ \citenamefont
  {Sorella}}]{capriottiLongRangeNeelOrder1999}%
  \BibitemOpen
  \bibfield  {author} {\bibinfo {author} {\bibfnamefont {L.}~\bibnamefont
  {Capriotti}}, \bibinfo {author} {\bibfnamefont {A.~E.}\ \bibnamefont
  {Trumper}},\ and\ \bibinfo {author} {\bibfnamefont {S.}~\bibnamefont
  {Sorella}},\ }\bibfield  {title} {\bibinfo {title} {{{N{\'e}el}} {{Order}} in
  the {{Triangular Heisenberg Model}}},\ }\href
  {https://doi.org/10.1103/PhysRevLett.82.3899} {\bibfield  {journal} {\bibinfo
   {journal} {Physical Review Letters}\ }\textbf {\bibinfo {volume} {82}},\
  \bibinfo {pages} {3899} (\bibinfo {year} {1999})}\BibitemShut {NoStop}%
\bibitem [{\citenamefont {Zheng}\ \emph {et~al.}(2006)\citenamefont {Zheng},
  \citenamefont {Fj{\ae}restad}, \citenamefont {Singh}, \citenamefont
  {McKenzie},\ and\ \citenamefont
  {Coldea}}]{zhengExcitationSpectraSpin$frac12$2006}%
  \BibitemOpen
  \bibfield  {author} {\bibinfo {author} {\bibfnamefont {W.}~\bibnamefont
  {Zheng}}, \bibinfo {author} {\bibfnamefont {J.~O.}\ \bibnamefont
  {Fj{\ae}restad}}, \bibinfo {author} {\bibfnamefont {R.~R.~P.}\ \bibnamefont
  {Singh}}, \bibinfo {author} {\bibfnamefont {R.~H.}\ \bibnamefont
  {McKenzie}},\ and\ \bibinfo {author} {\bibfnamefont {R.}~\bibnamefont
  {Coldea}},\ }\bibfield  {title} {\bibinfo {title} {Excitation spectra of the
  spin-{{$\frac{1}{2}$}} triangular-lattice {{Heisenberg}} antiferromagnet},\
  }\href {https://doi.org/10.1103/PhysRevB.74.224420} {\bibfield  {journal}
  {\bibinfo  {journal} {Physical Review B}\ }\textbf {\bibinfo {volume} {74}},\
  \bibinfo {pages} {224420} (\bibinfo {year} {2006})}\BibitemShut {NoStop}%
\bibitem [{\citenamefont {White}\ and\ \citenamefont
  {Chernyshev}(2007)}]{whiteNeelOrderSquare2007a}%
  \BibitemOpen
  \bibfield  {author} {\bibinfo {author} {\bibfnamefont {S.~R.}\ \bibnamefont
  {White}}\ and\ \bibinfo {author} {\bibfnamefont {A.~L.}\ \bibnamefont
  {Chernyshev}},\ }\bibfield  {title} {\bibinfo {title} {{{N{\'e}el}} {{Order}}
  in {{Square}} and {{Triangular Lattice Heisenberg Models}}},\ }\href
  {https://doi.org/10.1103/PhysRevLett.99.127004} {\bibfield  {journal}
  {\bibinfo  {journal} {Physical Review Letters}\ }\textbf {\bibinfo {volume}
  {99}},\ \bibinfo {pages} {127004} (\bibinfo {year} {2007})}\BibitemShut
  {NoStop}%
\bibitem [{\citenamefont {Mouhat}\ and\ \citenamefont
  {Coudert}(2014)}]{mouhatNecessarySufficientElastic2014}%
  \BibitemOpen
  \bibfield  {author} {\bibinfo {author} {\bibfnamefont {F.}~\bibnamefont
  {Mouhat}}\ and\ \bibinfo {author} {\bibfnamefont {F.-X.}\ \bibnamefont
  {Coudert}},\ }\bibfield  {title} {\bibinfo {title} {Necessary and sufficient
  elastic stability conditions in various crystal systems},\ }\href
  {https://doi.org/10.1103/PhysRevB.90.224104} {\bibfield  {journal} {\bibinfo
  {journal} {Physical Review B}\ }\textbf {\bibinfo {volume} {90}},\ \bibinfo
  {pages} {224104} (\bibinfo {year} {2014})}\BibitemShut {NoStop}%
\bibitem [{\citenamefont {Coldea}\ \emph {et~al.}(2002)\citenamefont {Coldea},
  \citenamefont {Tennant}, \citenamefont {Habicht}, \citenamefont {Smeibidl},
  \citenamefont {Wolters},\ and\ \citenamefont
  {Tylczynski}}]{coldeaDirectMeasurementSpin2002}%
  \BibitemOpen
  \bibfield  {author} {\bibinfo {author} {\bibfnamefont {R.}~\bibnamefont
  {Coldea}}, \bibinfo {author} {\bibfnamefont {D.~A.}\ \bibnamefont {Tennant}},
  \bibinfo {author} {\bibfnamefont {K.}~\bibnamefont {Habicht}}, \bibinfo
  {author} {\bibfnamefont {P.}~\bibnamefont {Smeibidl}}, \bibinfo {author}
  {\bibfnamefont {C.}~\bibnamefont {Wolters}},\ and\ \bibinfo {author}
  {\bibfnamefont {Z.}~\bibnamefont {Tylczynski}},\ }\bibfield  {title}
  {\bibinfo {title} {Direct {{Measurement}} of the {{Spin Hamiltonian}} and
  {{Observation}} of {{Condensation}} of {{Magnons}} in the {{2D Frustrated
  Quantum Magnet Cs$_2$CuCl$_4$}}},\ }\href
  {https://doi.org/10.1103/PhysRevLett.88.137203} {\bibfield  {journal}
  {\bibinfo  {journal} {Physical Review Letters}\ }\textbf {\bibinfo {volume}
  {88}},\ \bibinfo {pages} {137203} (\bibinfo {year} {2002})}\BibitemShut
  {NoStop}%
\bibitem [{\citenamefont {Yamamoto}\ \emph {et~al.}(2015)\citenamefont
  {Yamamoto}, \citenamefont {Marmorini},\ and\ \citenamefont
  {Danshita}}]{yamamotoMicroscopicModelCalculations2015}%
  \BibitemOpen
  \bibfield  {author} {\bibinfo {author} {\bibfnamefont {D.}~\bibnamefont
  {Yamamoto}}, \bibinfo {author} {\bibfnamefont {G.}~\bibnamefont
  {Marmorini}},\ and\ \bibinfo {author} {\bibfnamefont {I.}~\bibnamefont
  {Danshita}},\ }\bibfield  {title} {\bibinfo {title} {Microscopic {{Model
  Calculations}} for the {{Magnetization Process}} of {{Layered
  Triangular-Lattice Quantum Antiferromagnets}}},\ }\href
  {https://doi.org/10.1103/PhysRevLett.114.027201} {\bibfield  {journal}
  {\bibinfo  {journal} {Physical Review Letters}\ }\textbf {\bibinfo {volume}
  {114}},\ \bibinfo {pages} {027201} (\bibinfo {year} {2015})}\BibitemShut
  {NoStop}%
\bibitem [{\citenamefont {Villain}\ \emph {et~al.}(1980)\citenamefont
  {Villain}, \citenamefont {Bidaux}, \citenamefont {Carton},\ and\
  \citenamefont {Conte}}]{villainOrderEffectDisorder1980}%
  \BibitemOpen
  \bibfield  {author} {\bibinfo {author} {\bibfnamefont {J.}~\bibnamefont
  {Villain}}, \bibinfo {author} {\bibfnamefont {R.}~\bibnamefont {Bidaux}},
  \bibinfo {author} {\bibfnamefont {J.-P.}\ \bibnamefont {Carton}},\ and\
  \bibinfo {author} {\bibfnamefont {R.}~\bibnamefont {Conte}},\ }\bibfield
  {title} {\bibinfo {title} {Order as an effect of disorder},\ }\href
  {https://doi.org/10.1051/jphys:0198000410110126300} {\bibfield  {journal}
  {\bibinfo  {journal} {Journal de Physique}\ }\textbf {\bibinfo {volume}
  {41}},\ \bibinfo {pages} {1263} (\bibinfo {year} {1980})}\BibitemShut
  {NoStop}%
\bibitem [{\citenamefont {Chubukov}\ and\ \citenamefont
  {Golosov}(1991)}]{chubukovQuantumTheoryAntiferromagnet1991a}%
  \BibitemOpen
  \bibfield  {author} {\bibinfo {author} {\bibfnamefont {A.~V.}\ \bibnamefont
  {Chubukov}}\ and\ \bibinfo {author} {\bibfnamefont {D.~I.}\ \bibnamefont
  {Golosov}},\ }\bibfield  {title} {\bibinfo {title} {Quantum theory of an
  antiferromagnet on a triangular lattice in a magnetic field},\ }\href
  {https://doi.org/10.1088/0953-8984/3/1/005} {\bibfield  {journal} {\bibinfo
  {journal} {Journal of Physics: Condensed Matter}\ }\textbf {\bibinfo {volume}
  {3}},\ \bibinfo {pages} {69} (\bibinfo {year} {1991})}\BibitemShut {NoStop}%
\bibitem [{\citenamefont {Heinil{\"a}}\ and\ \citenamefont
  {Oja}(1993)}]{heinilaSelectionGroundState1993c}%
  \BibitemOpen
  \bibfield  {author} {\bibinfo {author} {\bibfnamefont {M.~T.}\ \bibnamefont
  {Heinil{\"a}}}\ and\ \bibinfo {author} {\bibfnamefont {A.~S.}\ \bibnamefont
  {Oja}},\ }\bibfield  {title} {\bibinfo {title} {Selection of the ground state
  in type-{{I}} fcc antiferromagnets in an external magnetic field},\ }\href
  {https://doi.org/10.1103/PhysRevB.48.7227} {\bibfield  {journal} {\bibinfo
  {journal} {Physical Review B}\ }\textbf {\bibinfo {volume} {48}},\ \bibinfo
  {pages} {7227} (\bibinfo {year} {1993})}\BibitemShut {NoStop}%
\bibitem [{\citenamefont {Griset}\ \emph {et~al.}(2011)\citenamefont {Griset},
  \citenamefont {Head}, \citenamefont {Alicea},\ and\ \citenamefont
  {Starykh}}]{grisetDeformedTriangularLattice2011}%
  \BibitemOpen
  \bibfield  {author} {\bibinfo {author} {\bibfnamefont {C.}~\bibnamefont
  {Griset}}, \bibinfo {author} {\bibfnamefont {S.}~\bibnamefont {Head}},
  \bibinfo {author} {\bibfnamefont {J.}~\bibnamefont {Alicea}},\ and\ \bibinfo
  {author} {\bibfnamefont {O.~A.}\ \bibnamefont {Starykh}},\ }\bibfield
  {title} {\bibinfo {title} {Deformed triangular lattice antiferromagnets in a
  magnetic field: {{Role}} of spatial anisotropy and {{Dzyaloshinskii-Moriya}}
  interactions},\ }\href {https://doi.org/10.1103/PhysRevB.84.245108}
  {\bibfield  {journal} {\bibinfo  {journal} {Physical Review B}\ }\textbf
  {\bibinfo {volume} {84}},\ \bibinfo {pages} {245108} (\bibinfo {year}
  {2011})}\BibitemShut {NoStop}%
\bibitem [{\citenamefont {Koutroulakis}\ \emph {et~al.}(2015)\citenamefont
  {Koutroulakis}, \citenamefont {Zhou}, \citenamefont {Kamiya}, \citenamefont
  {Thompson}, \citenamefont {Zhou}, \citenamefont {Batista},\ and\
  \citenamefont {Brown}}]{koutroulakisQuantumPhaseDiagram2015}%
  \BibitemOpen
  \bibfield  {author} {\bibinfo {author} {\bibfnamefont {G.}~\bibnamefont
  {Koutroulakis}}, \bibinfo {author} {\bibfnamefont {T.}~\bibnamefont {Zhou}},
  \bibinfo {author} {\bibfnamefont {Y.}~\bibnamefont {Kamiya}}, \bibinfo
  {author} {\bibfnamefont {J.~D.}\ \bibnamefont {Thompson}}, \bibinfo {author}
  {\bibfnamefont {H.~D.}\ \bibnamefont {Zhou}}, \bibinfo {author}
  {\bibfnamefont {C.~D.}\ \bibnamefont {Batista}},\ and\ \bibinfo {author}
  {\bibfnamefont {S.~E.}\ \bibnamefont {Brown}},\ }\bibfield  {title} {\bibinfo
  {title} {Quantum phase diagram of the {{$S = \frac{1}{2}$}}
  triangular-lattice antiferromagnet {{Ba$_3$CoSb$_2$O$_9$}}},\ }\href
  {https://doi.org/10.1103/PhysRevB.91.024410} {\bibfield  {journal} {\bibinfo
  {journal} {Physical Review B}\ }\textbf {\bibinfo {volume} {91}},\ \bibinfo
  {pages} {024410} (\bibinfo {year} {2015})}\BibitemShut {NoStop}%
\bibitem [{\citenamefont
  {Zhitomirsky}(2015)}]{zhitomirskyRealspacePerturbationTheory2015}%
  \BibitemOpen
  \bibfield  {author} {\bibinfo {author} {\bibfnamefont {M.~E.}\ \bibnamefont
  {Zhitomirsky}},\ }\bibfield  {title} {\bibinfo {title} {Real-space
  perturbation theory for frustrated magnets: Application to magnetization
  plateaus},\ }\href {https://doi.org/10.1088/1742-6596/592/1/012110}
  {\bibfield  {journal} {\bibinfo  {journal} {Journal of Physics: Conference
  Series}\ }\textbf {\bibinfo {volume} {592}},\ \bibinfo {pages} {012110}
  (\bibinfo {year} {2015})}\BibitemShut {NoStop}%
\bibitem [{\citenamefont {Chen}\ \emph {et~al.}(2013)\citenamefont {Chen},
  \citenamefont {Ju}, \citenamefont {Jiang}, \citenamefont {Starykh},\ and\
  \citenamefont {Balents}}]{chenGroundStatesSpin12013}%
  \BibitemOpen
  \bibfield  {author} {\bibinfo {author} {\bibfnamefont {R.}~\bibnamefont
  {Chen}}, \bibinfo {author} {\bibfnamefont {H.}~\bibnamefont {Ju}}, \bibinfo
  {author} {\bibfnamefont {H.-C.}\ \bibnamefont {Jiang}}, \bibinfo {author}
  {\bibfnamefont {O.~A.}\ \bibnamefont {Starykh}},\ and\ \bibinfo {author}
  {\bibfnamefont {L.}~\bibnamefont {Balents}},\ }\bibfield  {title} {\bibinfo
  {title} {Ground states of spin-1/2 triangular antiferromagnets in a magnetic
  field},\ }\href {https://doi.org/10.1103/PhysRevB.87.165123} {\bibfield
  {journal} {\bibinfo  {journal} {Physical Review B}\ }\textbf {\bibinfo
  {volume} {87}},\ \bibinfo {pages} {165123} (\bibinfo {year}
  {2013})}\BibitemShut {NoStop}%
\bibitem [{\citenamefont {Maksimov}\ \emph {et~al.}(2019)\citenamefont
  {Maksimov}, \citenamefont {Zhu}, \citenamefont {White},\ and\ \citenamefont
  {Chernyshev}}]{maksimovAnisotropicExchangeMagnetsTriangular2019}%
  \BibitemOpen
  \bibfield  {author} {\bibinfo {author} {\bibfnamefont {P.~A.}\ \bibnamefont
  {Maksimov}}, \bibinfo {author} {\bibfnamefont {Z.}~\bibnamefont {Zhu}},
  \bibinfo {author} {\bibfnamefont {S.~R.}\ \bibnamefont {White}},\ and\
  \bibinfo {author} {\bibfnamefont {A.~L.}\ \bibnamefont {Chernyshev}},\
  }\bibfield  {title} {\bibinfo {title} {Anisotropic-{{Exchange Magnets}} on a
  {{Triangular Lattice}}: {{Spin Waves}}, {{Accidental Degeneracies}}, and
  {{Dual Spin Liquids}}},\ }\href {https://doi.org/10.1103/PhysRevX.9.021017}
  {\bibfield  {journal} {\bibinfo  {journal} {Physical Review X}\ }\textbf
  {\bibinfo {volume} {9}},\ \bibinfo {pages} {021017} (\bibinfo {year}
  {2019})}\BibitemShut {NoStop}%
\bibitem [{\citenamefont {Rousochatzakis}\ \emph {et~al.}(2016)\citenamefont
  {Rousochatzakis}, \citenamefont {R{\"o}ssler}, \citenamefont {{van den
  Brink}},\ and\ \citenamefont
  {Daghofer}}]{rousochatzakisKitaevAnisotropyInduces2016}%
  \BibitemOpen
  \bibfield  {author} {\bibinfo {author} {\bibfnamefont {I.}~\bibnamefont
  {Rousochatzakis}}, \bibinfo {author} {\bibfnamefont {U.~K.}\ \bibnamefont
  {R{\"o}ssler}}, \bibinfo {author} {\bibfnamefont {J.}~\bibnamefont {{van den
  Brink}}},\ and\ \bibinfo {author} {\bibfnamefont {M.}~\bibnamefont
  {Daghofer}},\ }\bibfield  {title} {\bibinfo {title} {Kitaev anisotropy
  induces mesoscopic {{$\mathbb{Z}_2$}} vortex crystals in frustrated hexagonal
  antiferromagnets},\ }\href {https://doi.org/10.1103/PhysRevB.93.104417}
  {\bibfield  {journal} {\bibinfo  {journal} {Physical Review B}\ }\textbf
  {\bibinfo {volume} {93}},\ \bibinfo {pages} {104417} (\bibinfo {year}
  {2016})}\BibitemShut {NoStop}%
\bibitem [{\citenamefont {Becker}\ \emph {et~al.}(2015)\citenamefont {Becker},
  \citenamefont {Hermanns}, \citenamefont {Bauer}, \citenamefont {Garst},\ and\
  \citenamefont {Trebst}}]{beckerSpinorbitPhysicsJ12015}%
  \BibitemOpen
  \bibfield  {author} {\bibinfo {author} {\bibfnamefont {M.}~\bibnamefont
  {Becker}}, \bibinfo {author} {\bibfnamefont {M.}~\bibnamefont {Hermanns}},
  \bibinfo {author} {\bibfnamefont {B.}~\bibnamefont {Bauer}}, \bibinfo
  {author} {\bibfnamefont {M.}~\bibnamefont {Garst}},\ and\ \bibinfo {author}
  {\bibfnamefont {S.}~\bibnamefont {Trebst}},\ }\bibfield  {title} {\bibinfo
  {title} {Spin-orbit physics of {{$j=\frac{1}{2}$}} {{Mott}} insulators on the
  triangular lattice},\ }\href {https://doi.org/10.1103/PhysRevB.91.155135}
  {\bibfield  {journal} {\bibinfo  {journal} {Physical Review B}\ }\textbf
  {\bibinfo {volume} {91}},\ \bibinfo {pages} {155135} (\bibinfo {year}
  {2015})}\BibitemShut {NoStop}%
\bibitem [{\citenamefont {Li}\ \emph {et~al.}(2019)\citenamefont {Li},
  \citenamefont {Perkins},\ and\ \citenamefont
  {Rousochatzakis}}]{liCollectiveSpinDynamics2019}%
  \BibitemOpen
  \bibfield  {author} {\bibinfo {author} {\bibfnamefont {M.}~\bibnamefont
  {Li}}, \bibinfo {author} {\bibfnamefont {N.~B.}\ \bibnamefont {Perkins}},\
  and\ \bibinfo {author} {\bibfnamefont {I.}~\bibnamefont {Rousochatzakis}},\
  }\bibfield  {title} {\bibinfo {title} {Collective spin dynamics of
  {{$\mathbb{Z}_2$}} vortex crystals in triangular {{Kitaev-Heisenberg}}
  antiferromagnets},\ }\href {https://doi.org/10.1103/PhysRevResearch.1.013002}
  {\bibfield  {journal} {\bibinfo  {journal} {Physical Review Research}\
  }\textbf {\bibinfo {volume} {1}},\ \bibinfo {pages} {013002} (\bibinfo {year}
  {2019})}\BibitemShut {NoStop}%
\end{thebibliography}%

\end{document}


\title{Supplemental Material for "$\mathbb{Z}_2$ Vortex Crystal Candidate in the Triangular $S=1/2$ Quantum Antiferromagnet"}
	
	\author{Jakob~Nagl}
	\email{jnagl@ethz.ch}
	\affiliation{Laboratory for Solid State Physics, ETH Z{\"u}rich, 8093 Z{\"u}rich, Switzerland}
	
	\author{Kirill~Yu.~Povarov}
	\affiliation{Dresden High Magnetic Field Laboratory (HLD-EMFL) and W\"urzburg-Dresden Cluster of Excellence ct.qmat, Helmholtz-Zentrum Dresden-Rossendorf (HZDR), 01328 Dresden, Germany}
	
	\author{Benjamin~Duncan}
	\affiliation{Laboratory for Solid State Physics, ETH Z{\"u}rich, 8093 Z{\"u}rich, Switzerland}
	\author{Catharina~N{\"a}ppi}
	\affiliation{Laboratory for Solid State Physics, ETH Z{\"u}rich, 8093 Z{\"u}rich, Switzerland}
	
	\author{Dmitry~Khalyavin}
	\affiliation{ISIS Facility, Rutherford Appleton Laboratory, Chilton, Didcot, Oxon OX11 0QX, United Kingdom}
	\author{Pascal~Manuel}
	\affiliation{ISIS Facility, Rutherford Appleton Laboratory, Chilton, Didcot, Oxon OX11 0QX, United Kingdom}
	\author{Fabio~Orlandi}
	\affiliation{ISIS Facility, Rutherford Appleton Laboratory, Chilton, Didcot, Oxon OX11 0QX, United Kingdom}
	
	\author{Jeremy~Sourd}
	\affiliation{Dresden High Magnetic Field Laboratory (HLD-EMFL) and W\"urzburg-Dresden Cluster of Excellence ct.qmat, Helmholtz-Zentrum Dresden-Rossendorf (HZDR), 01328 Dresden, Germany}
	\author{Beat~Valentin~Schwarze}
	\affiliation{Dresden High Magnetic Field Laboratory (HLD-EMFL) and W\"urzburg-Dresden Cluster of Excellence ct.qmat, Helmholtz-Zentrum Dresden-Rossendorf (HZDR), 01328 Dresden, Germany}
	\author{Freya~Husstedt}
	\affiliation{Dresden High Magnetic Field Laboratory (HLD-EMFL) and W\"urzburg-Dresden Cluster of Excellence ct.qmat, Helmholtz-Zentrum Dresden-Rossendorf (HZDR), 01328 Dresden, Germany}
	\affiliation{Institut f\"ur Festk\"orper- und Materialphysik, Technische Universit\"at Dresden, 01062 Dresden, Germany}
	\author{Sergei~A.~Zvyagin}
	\affiliation{Dresden High Magnetic Field Laboratory (HLD-EMFL) and W\"urzburg-Dresden Cluster of Excellence ct.qmat, Helmholtz-Zentrum Dresden-Rossendorf (HZDR), 01328 Dresden, Germany}
	
	\author{Oksana~Zaharko}
	\affiliation{PSI Center for Neutron and Muon Sciences, Forschungsstrasse 111, 5232 Villigen, PSI, Switzerland}
	
	\author{Paul~Steffens}
	\affiliation{Institut Laue-Langevin, 71 avenue des Martyrs, CS 20156, 38042 Grenoble Cedex 9, France}
	\author{Arno~Hiess}
	\affiliation{Institut Laue-Langevin, 71 avenue des Martyrs, CS 20156, 38042 Grenoble Cedex 9, France}
	\affiliation{European Spallation Source ERIC, P.O. Box 176, 22100 Lund, Sweden}
	
	\author{David~R.~Allan}
	\affiliation{Diamond Light Source, Harwell Science and Innovation Campus, Didcot, Oxfordshire OX11 0DE, UK}
	\author{Sarah~A.~Barnett}
	\affiliation{Diamond Light Source, Harwell Science and Innovation Campus, Didcot, Oxfordshire OX11 0DE, UK}
	
	\author{Zewu~Yan}
	\affiliation{Laboratory for Solid State Physics, ETH Z{\"u}rich, 8093 Z{\"u}rich, Switzerland}
	
	\author{Severian~Gvasaliya}
	\affiliation{Laboratory for Solid State Physics, ETH Z{\"u}rich, 8093 Z{\"u}rich, Switzerland}
	
	\author{Andrey~Zheludev}
	\email{zhelud@ethz.ch}
	\homepage{http://www.neutron.ethz.ch/}
	\affiliation{Laboratory for Solid State Physics, ETH Z{\"u}rich, 8093 Z{\"u}rich, Switzerland}
	
	\date{\today}

	\begin{abstract}
		In this supplement, we provide supporting information about the low-temperature crystal structure, the single-ion fitting procedures \& ESR data, the magnetoelastic coupling, our inelastic neutron spectroscopy experiment and the proposed magnetic structures.
	\end{abstract}
	
	\maketitle
	
	\section{Crystal Structure \& Spin Hamilonian}
	
	\dmnrc adopts a trigonal $P\bar{3}m1$ structure at room temperature \cite{vishnoiStructuralDiversityMagnetic2020}, realizing an ideal triangular lattice of Ru$^{3+}$ ions in the spin-orbital entangled $j_{\rm eff} = 1/2$ state.
	The local point symmetry at the magnetic site amounts to $.\bar{3}m$.
	In this case, the most general 2D spin Hamiltonian for nearest neighbors, including all anisotropy terms allowed by symmetry, contains {\it four} exchange parameters.
	In the global Cartesian frame ($\hat{x} \parallel a$, $\hat{z} \parallel c$) it may be written as
	\begin{gather}
		\mathcal{H}_\parallel = \sum_{\langle i,j \rangle} J (S^x_i S^x_j + S^y_i S^y_j + \Delta S^z_i S^z_j) \label{eq:Hamilt}\\ \nonumber
		+ 2J_{\pm\pm} [(S^x_i S^x_j - S^y_i S^y_j) \cos(\varphi_\alpha) - (S^x_i S^y_j + S^y_i S^x_j) \sin(\varphi_\alpha)] \\ \nonumber
		+ J_{z\pm} [(S^y_i S^z_j + S^z_i S^y_j) \cos(\varphi_\alpha) - (S^x_i S^z_j + S^z_i S^x_j) \sin(\varphi_\alpha)].
	\end{gather}
	Here, the phases $\varphi_\alpha \in (0,2\pi/3,4\pi/3)$ associated with the primitive translation vectors $\mathbf{a, b, a+b}$ give rise to a bond dependence on the anisotropic $J_{\pm\pm}, J_{z\pm}$ couplings.
	These triangular layers exhibit a direct AA-stacking, resulting in a simple inter-plane coupling restricted to XXZ form
	\begin{gather}
		\mathcal{H}_\perp = \sum_{\langle i,j \rangle} J_c (S^x_i S^x_j + S^y_i S^y_j + \Delta_c S^z_i S^z_j).
	\end{gather}
	In both cases, a superexchange between next-nearest neighbors would involve significantly longer bond lengths (12.37~\AA~vs. 7.14~\AA~ in the triangular plane, 9.86~\AA~vs. 6.80~\AA~between layers) and is unlikely to play an important role.
	
\begin{figure}[tbp]
		\includegraphics[scale=1]{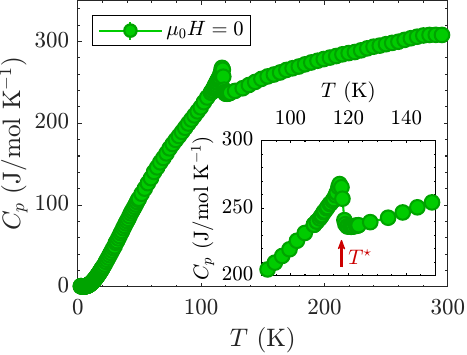}
		\caption{Heat capacity in \dmnrc against temperature, depicting a structural transition at $T^\star$. The inset shows a zoom-in close to the lambda anomaly.}
		\label{fig:HC}
	\end{figure}

	Towards lower temperatures, \dmnrc goes through a structural transition. 
	This is best seen in additional heat capacity measurements depicted in Fig.~\ref{fig:HC}, carried out on a 4.5~mg single crystal sample in a 9~T Physical Property Measurement System (PPMS).
	The data show a clear lambda anomaly at $T^\star \approx 118$~K, indicative of a continuous phase transition.
	An entropy release of order $\Delta S \sim R \log(2)$ would suggest a spontaneous choice between \textit{two} structural configurations.
	
	To characterize the structural changes near $T^\star$, we performed high-resolution single crystal x-ray experiments at the I19 synchrotron beamline (Diamond, UK).
	Shelx-based refinement results on a representative sample are summarized in \Cref{tab:Diamond,tab:StructureHT,tab:StructureLT}.
	A high-temperature dataset collected at 150~K ($T > T^\star$) yields excellent agreement to a model based on the nominal $P\bar{3}m1$ crystal structure, with $R_1 = 1.3$\% and less than 1~e/\AA$^3$ difference in electronic density.
	Due to their weak scattering, the light deuterium atoms are constrained with a riding model, assuming an ideal CH$_3$ group with variable distance and thermal parameters fixed by the non-riding atom.
	Repeating an analogous refinement on a 30~K ($T < T^\star$) dataset results in a significantly worse $R_1 = 9.68$\%, indicating some change in the chemical structure.
	None of the space groups consistent with the apparent Laue class $\bar{3}m1$ improve the fit.
	
	\begin{table*}[tbp]
		\caption{Summary of single crystal x-ray refinements of the chemical structure \dmnrc at 150~K ($T > T^\star$) and 30~K ($T < T_\star$).}
		\begin{tabularx}{0.8\textwidth}{YYYY}
			\midrule\midrule
			Characteristics & $T = 150$~K & $T = 30$~K, High Sym. & $T = 30$~K, Low Sym. \\
			\midrule
			Chemical formula & \dmnrc & \dmnrc & \dmnrc \\
			Wavelength (\AA) & 0.6889 & 0.6889 & 0.6889 \\
			Crystal System & Trigonal & Trigonal & Trigonal \\
			Space Group & $P\bar{3}m1$ & $P\bar{3}m1$ & $P\bar{3}$ \\
			$a, b$~(\AA) & 7.1964(1) & 7.1435(1) & 7.1435(1) \\
			$c$~(\AA) &  6.7939(1) & 6.7984(1) & 6.7984(1) \\
			$\alpha, \beta$ (deg) & 90 & 90 & 90 \\
			$\gamma$ (deg) & 120 & 120 & 120 \\
			$V$~(\AA$^3$) & 304.71(1) & 300.44(1) & 300.44(1) \\
			$Z$ & 1 & 1 & 1 \\
			$\rho_{calc}$ (g/cm$^3$) & 2.25 & 2.28 & 2.28 \\
			$2\theta_{\rm max}$ (deg) & 116.74 & 116.74 & 116.74 \\
			Measured Reflections & 37577 & 31016 & 31016 \\
			Independent Reflections & 1750 & 1796 & 1796 \\
			Refined Parameters & 18 & 18 & 23 \\
			$R_1$ (\%) & 1.33 & 9.68 & 2.45 \\
			$wR_2$ (\%) & 3.69 & 38.53 & 6.73 \\
			Goodness of Fit $S$ & 1.075 & 3.372 & 1.217 \\
			Twinning Ratio $f$ & - & - & 1.62 : 1 \\
			Diff. density (e/\AA$^3$) & 0.65 / -0.91 & 15.35 / -13.45  & 2.11 / -3.27 \\
			\midrule\midrule
		\end{tabularx}
		\label{tab:Diamond}
	\end{table*}
	
	\begin{figure}[tbp]
		\includegraphics[scale=1]{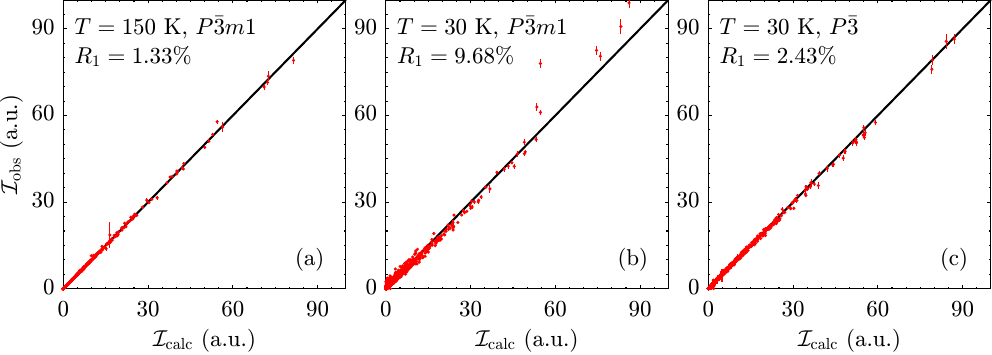}
		\caption{Observed versus calculated x-ray intensities at 150~K in $P\bar{3}m1$ space group (a), at 30~K with the nominal $P\bar{3}m1$ structure (b) and at 30~K for a lower symmetry $P\bar{3}$ model including 2 merohedral twins. The solid line shows the 1:1 agreement in all panels.}
		\label{fig:Refine}
	\end{figure}
	
	This tension may be resolved by considering a transition into a low-symmetry $P\bar{3}$ structure.
	Such a model allows for 2-fold merohedral twinning, which can disguise the "true" Laue class and is consistent with the observed release of entropy.
	Here the mirror plane is lost, allowing the RuCl$_6$ octahedra and CD$_3$ND$_3$ molecules to rotate by $\pm \psi$ around the $z$-axis passing through the high-symmetry site ($1b$ and $2d$ respectively).
	Equivalently, this corresponds to a loss of 2-fold rotational symmetry around $\mathbf{a}$, $\mathbf{b}$ or $\mathbf{a+b}$.
	A possible twinning matrix would be $[-1~0~0;~1~1~0;~0~0~1]$.
	The real (and reciprocal) space lattices of both twin domains overlap perfectly and the indexing of peaks does not change with respect to the parent structure.
	This model yields excellent agreement with the data, where $R_1 = 2.4$\%.
	Again, the deuterium positions are constrained by a riding model, though now the rotation angle of the CH$_3$ group also becomes a free parameter.
	
\begin{figure}[tbp]
		\includegraphics[scale=1]{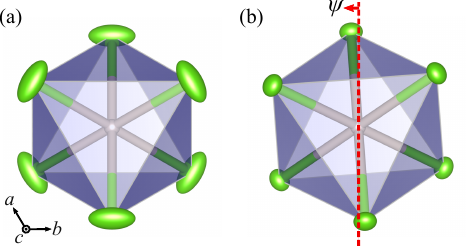}
		\caption{RuCl$_6$ octahedra for single-crystal x-ray refinements at 30~K against different structural models.
		(a) Nominal $P\bar{3}m1$ structure and (b) lower-symmetry $P\bar{3}$ structure with parameters in \cref{tab:Diamond,tab:StructureHT,tab:StructureLT}.
		The green ellipsoids illustrate the anisotropic thermal displacements, while the red line in (b) emphasizes the rotation of RuCl$_6$ octahedra by $\psi$.}
		\label{fig:Octahedra}
	\end{figure}

	Final refinement results and structural differences between both models are visualized in Fig.~\ref{fig:Refine} and Fig.~\ref{fig:Octahedra}.
	The rotation angle of the RuCl$_6$ octahedra at 30~K - controlling the degree of symmetry breaking - amounts to $\psi = \pm 2.95(1)^\circ$.
	Hints of the structural change are seen even in the high-symmetry $P\bar{3}m1$ model: The thermal parameters on chlorine are highly anisotropic, elongated in the triangular plane to cover both $\pm \psi$.
	
	We note that a non-centrosymmetric space group $P3$ would also be consistent with the merohedral twinning scenario discussed above.
	However, our refinement attempts yield no improvement in $R$-factors despite the extra parameters.
	Analogous refinements were carried out on several single crystal samples, always pointing to the $P\bar{3}m1 \rightarrow P\bar{3}$ transition with analogous $R$-factors ($8-10\%$ vs. $1.5-2.5\%$ for $P\bar{3}m1$ and $P\bar{3}$ models respectively).
	Similar structural changes have been reported in the triangular lattice systems RbFe(MoO$_4$)$_2$ \cite{waskowskaTemperaturePressuredependentLattice2010a} and Na$_2$BaCo(PO$_4$)$_2$ \cite{kajitaFerroaxialTransitionsGlaseriteType2024, woodlandContinuumExcitationsSharp2025}, making this a relatively common structural motif.
	
	\begin{table*}[tbp]
		\caption{Fractional atomic coordinates and thermal displacement parameters deduced from a refinement of single crystal x-ray diffraction data at $T = 150$~K against the nominal structural model with space group $P\bar{3}m1$ (No. 164). The light deuterium atoms are constrained with a riding model (CH$_3$ group with variable distance).}
		
		\begin{tabularx}{0.8\textwidth}{YYYYYY}
			\midrule\midrule
			Atom & Wyckoff & $x$ & $y$ & $z$ & Equiv. $B_\mathrm{iso}$ \\
			\midrule
			Na & 1a & 0 & 0 & 0 & 1.596(7) \\
			Ru & 1b & 0 & 0 & 0.5 & 0.814(2) \\
			Cl & 6i & 0.69054(2) & 0.84527(2) & 0.29605(2) & 2.002(2) \\
			N & 2d & 0.33333 & 0.66667 & 0.6592(1) & 2.09(1) \\
			C & 2d & 0.33333 & 0.66667 & 0.8741(2) & 2.32(2) \\
			D(1) & 6i & 0.26714 & 0.73811 & 0.61452 & 3.13 \\
			D(2) & 6i & 0.40828 & 0.59336 & 0.92220 & 3.48 \\
			\midrule
			$B_{11}$ & $B_{22}$ & $B_{33}$ & $B_{23}$ & $B_{31}$ & $B_{12}$ \\
			\midrule
			1.95(1) & 1.95(1) & 0.89(1) & 0 & 0 & 0.974(6) \\
			0.875(2) & 0.875(2) & 0.692(2) & 0 & 0 & 0.437(1) \\
			1.077(2) & 2.002(3) & 1.123(3) & -0.114(1) & -0.227(2) & 0.539(2) \\
			2.487(2) & 2.487(2) & 1.287(2) & 0 & 0 & 1.24(1) \\
			2.835(2) & 2.835(2) & 1.303(2) & 0 & 0 & 1.42(1) \\
			\midrule\midrule
		\end{tabularx}
		\label{tab:StructureHT}
	\end{table*}
	
	\begin{table*}[tbp]
		\caption{Same as Table~\ref{tab:StructureHT}, but for a refinement of the $T = 30$~K data against the reduced-symmetry space group $P\bar{3}$ (No. 147). Here the mirror plane is lost, allowing the Cl octahedra and CD$_3$ND$_3$ molecules to rotate. The light deuterium atoms are constrained with a riding model (CH$_3$ group with variable distance \& rotation angle).}
		
		\begin{tabularx}{0.8\textwidth}{YYYYYY}
			\midrule\midrule
			Atom & Wyckoff & $x$ & $y$ & $z$ & Equiv. $B_\mathrm{iso}$ \\
			\midrule
			Na & 1a & 0 & 0 & 0 & 0.70(1) \\
			Ru & 1b & 0 & 0 & 0.5 & 0.345(2) \\
			Cl & 6g & 0.68921(4) & 0.82456(4) & 0.29564(3) & 0.575(3) \\
			N & 2d & 0.33333 & 0.66667 & 0.3401(2) & 0.86(1) \\
			C & 2d & 0.33333 & 0.66667 & 0.1222(2) & 0.99(2)) \\
			D(1) & 6g & 0.22734 & 0.69112 & 0.61527 & 1.26 \\
			D(2) & 6g & 0.44432 & 0.63561 & 0.92585 & 1.50 \\
			\midrule
			$B_{11}$ & $B_{22}$ & $B_{33}$ & $B_{23}$ & $B_{31}$ & $B_{12}$ \\
			\midrule
			0.821(2) & 0.821(2) & 0.474(2) & 0 & 0 & 0.411(9) \\
			0.341(2) & 0.341(2) & 0.353(3) & 0 & 0 & 0.166(2) \\
			0.473(6) & 0.644(7) & 0.531(6) & -0.038(4) & -0.066(4) & 0.220(5) \\
			0.963(2) & 0.963(2) & 0.655(3) & 0 & 0 & 0.48(1) \\
			1.145(3) & 1.145(3) & 0.687(3) & 0 & 0 & 0.57(2) \\
			\midrule\midrule
		\end{tabularx}
		\label{tab:StructureLT}
	\end{table*}
	
	Let's discuss the potential impact of this structural phase transition on the magnetism.
	Anti-symmetric Dzyaloshinskii-Moriya terms remain forbidden by symmetry, at least down to $T_{\rm N}$ (the same goes for anything akin to a $J-J'$ isosceles distortion).
	However, the lower point symmetry $.\bar{3}$ at the Ru$^{3+}$ site allows for two new symmetric anisotropy terms in the triangular lattice Hamiltonian of Eq.~\ref{eq:Hamilt}.
	We can parameterize these by adding separate phase shifts to the bond angles $\varphi_\alpha \rightarrow \varphi_\alpha + \delta \phi$, one for each of the bond-dependent terms $J_{\pm\pm}$ and $J_{z\pm}$.
	
	\begin{figure}[tbp]
		\includegraphics[scale=1]{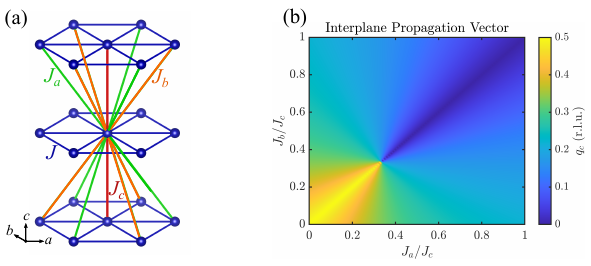}
		\caption{Effects of structural transition on the inter-plane coupling.
		(a) The diagonal next-nearest neighbor interactions become inequivalent by symmetry, i.e. $J_a \neq J_b$.
		(b) Out-of-plane propagation vector component $q_c$ obtained by minimizing the classical energy in Eq.~\ref{eq:E_OOP}. An $J_a \neq J_b$ should result in IC stacking along the $c$-axis.}
		\label{fig:OOP}
	\end{figure}

	As for the inter-plane coupling, the main effect of the transition would be to allow for two inequivalent diagonal terms $J_a \neq J_b$ [see Fig.~\ref{fig:OOP}(a)].
	This has been explored previously in both RbFe(MoO$_4$)$_2$ \cite{hearmonElectricFieldControl2012, zelenskiyMagneticPhasesS52021} and Na$_2$BaCo(PO$_4$)$_2$ \cite{woodlandContinuumExcitationsSharp2025}.
	The mean-field inter-plane exchange energy is given as
	\begin{equation}
		E(q_c) = -\frac{3}{2}JS^2 - \frac{1}{2}[3(J_a + J_b) - 2J_c]S^2 \cos(2\pi q_c) + \frac{3\sqrt{3}}{2}[J_b - J_b]S^2 \sin(2\pi q_c) \label{eq:E_OOP}
	\end{equation}
	per site.
	In practice, any $J_a \neq J_b$ will entail an incommensurate (IC) \textit{inter}-plane stacking $q_c \neq 1/2$, incompatible with experiment.
	We solve Eq.~\ref{eq:E_OOP} numerically and plot the resulting inter-plane propagation vector in Fig.~\ref{fig:OOP}(b).
	By setting $q_c \equiv 1/2$ we can restrict $J_a \approx J_b < J_c/3$, so the main physics is captured purely by $J_c$.
	
	Finally, we remark on our angular resolution in the XRD experiment.
	We have collected reflections at scattering angles up to $2\theta \approx 116^\circ$, corresponding to an experimental resolution of order $\Delta d/d \sim 0.05\%$.
	We can compare this value to the monoclinic lattice distortion required to explain the zero-field incommensurate reflections in terms of a "static" $J - J'$ isosceles model.
	Given the scale of magnetoelastic coupling determined in Sec.~\ref{sec:mag_el}, we estimate that a distortion at least of order $\gtrsim 0.1\%$ would be necessary, even after accounting for quantum renormalization of the classical ordering wavevector $q_{\rm IC} = \arccos(-J'/2J)/2\pi$.
	Therefore, such effects should be clearly resolved as peak splitting and can be ruled out by experiment.
	We note that structural changes even an order of magnitude smaller than the experimental resolution can typically be identified by a systematic increase in $R$-factors below $T^\star$ (not observed).

	\section{Single-Ion Fits \& Spin-Orbital Wavefunctions}
	
	\begin{figure}[tbp]
		\includegraphics[scale=1]{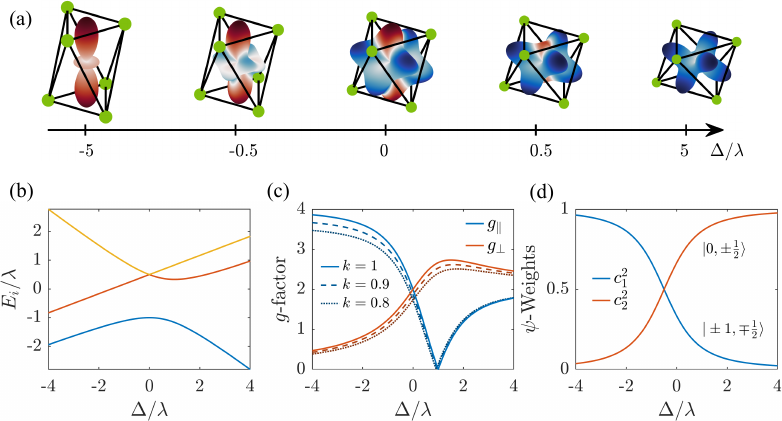}
		\caption{Effects of trigonal distortion on the spin-orbital wavefunctions.
			(a) Sketch of the spatial shape of the pseudospin wavefunctions for different ratios $\Delta / \lambda$.
			Red/blue color indicates spin up/down polarization of the hole.
			In (b,c,d) we show the dependence of the energy spectrum, $g$-factors and ground state weights on the trigonal distortion.
			Results of different $k$ due to covalency effects are also depicted.}
		\label{fig:Single_Ion}
	\end{figure}

	Here we summarize our fitting procedures of the single-ion + mean-field model discussed in the main text.
	We keep only the low-energy $\hat{\ell}_{\rm eff} = 1$ (orbital) and $\hat{S} = 1/2$ (spin) degrees of freedom, corresponding to a single hole in the $t_{2g}$ manifold.
	This results in the on-site Hamiltonian
	\begin{equation}
		\mathcal{H} = \lambda \hat{\boldsymbol{\ell}} \cdot \mathbf{\hat{S}} + \Delta \hat{\ell}^2_z - \mu_{\rm B} \mathbf{H} \cdot \hat{\mathbf{M}}
	\end{equation}
	where $\mathbf{\hat{M}} = 2\mathbf{\hat{S}} - k \boldsymbol{\hat{\ell}}$ is the magnetization and $k \lesssim 1$ accounts for a small reduction in orbital moment due to hybridization with Cl$^-$ anions (i.e. covalency effects \cite{abragamElectronParamagneticResonance, kubotaSuccessiveMagneticPhase2015}).
	By diagonalizing $\mathcal{H}$, we obtain the energy spectrum $E_i$ and eigenstates $\ket{\psi_i}$ (see Fig.~\ref{fig:Single_Ion}).
	For dominant spin-orbit coupling $|\Delta/\lambda| \lesssim 1$, the ground state is a spin-orbital entangled $j_{\rm eff} = 1/2$ Kramers doublet, written as
	\begin{gather}
		\ket{\pm} = c_1 \ket{\pm 1, \mp 1/2} + c_2 \ket{0, \pm 1/2} \\
		{\rm where} \qquad c_{1,2} = \pm \sqrt{\frac{1}{2} \mp \frac{A}{2\sqrt{A^2 + 1}}} \qquad {\rm and} \qquad A = \frac{2 \Delta/\lambda - 1}{2 \sqrt{2}}. \nonumber
	\end{gather}	
	Larger trigonal distortions $|\Delta| \gg \lambda$ result in a more trivial state with no interesting orbital physics (the spin polarization of the hole is synonymous with that of the pseudospins, cf. red/blue colors in Fig.~\ref{fig:Single_Ion}(a)), purely composed of $\ket{\pm 1, \mp \frac{1}{2}}$ or $\ket{0, \pm \frac{1}{2}}$ states (depending on the sign of $\Delta$).
	In order to determine the Hamiltonian parameters, we use wavefunctions to calculate several experimentally accessible properties.
	The longitudinal/transverse $g$-factors are given as
	\begin{equation}
		g_{\parallel} = 2|(1+k)c_1^2 - c_2^2| \qquad {\rm and} \qquad g_{\perp} = 2c_2 (c_2 - \sqrt{2}k c_1).
	\end{equation}
	The single-ion magnetization can be calculated as
	\begin{gather}
		\label{eq:Magnetiz}
		M^{\rm SI}_\alpha (\mathbf{H},T) = \mu_{\rm B} \sum_i p_i \bra{\psi_i} 2\hat{S}_\alpha - k \hat{\ell}_\alpha \ket{\psi_i},
	\end{gather}
	where $p_i$ are Boltzmann weights and the uniform susceptibility
	\begin{gather}
		\chi_{\alpha \beta} = \frac{\partial M_\alpha}{\partial H_\beta}
	\end{gather}
	is simply evaluated numerically.	
	For a direct comparison to experiment, we include also a temperature independent susceptibility contribution $\chi^0_{\alpha\beta}$ as well as a mean-field Heisenberg exchange parameter $J_{\rm MF}$.	
	Each ion experiences a local field $\mathbf{H}_{\rm MF}$ proportional to the magnetization of its neighbors, determined self-consistently using the relation
	\begin{align}
	\qquad\qquad \mathbf{M}^{\rm MF} &= \mathbf{M}^{\rm SI} (\mathbf{H} + \mathbf{H}_{\rm MF},T) \qquad {\rm where} \qquad \mathbf{H}_{\rm MF} = -\frac{z J_{\rm MF}}{({\bf g} \mu_{\rm B})^2} \mathbf{M}^{\rm MF}.
	\end{align}
	Here $z = 6$ is the number of nearest neighbors, while ${\bf g} = {\rm diag}(g_\perp, g_\perp, g_\parallel)$ is the $g$-tensor.
	By simultaneously fitting the magnetization, susceptibility and ESR data for all field directions we obtain the parameter estimates summarized in \Cref{tab:SI_fit}.
	We note that in experiment we are highly sensitive to the ratio $\Delta / \lambda$, while the overall energy scale only appears in the high-$T$ curvature of inverse susceptibilities and becomes muddled together with $\chi^0_\alpha$.
	Therefore, one can still obtain reasonable agreement by changing both spin-orbit coupling and trigonal crystal field simultaneously within $\sim 20\%$, but $\Delta / \lambda \simeq -0.51$ remains tightly constrained.
	
	\begin{table*}[tbp]
		\caption{Fit results of the single-ion + mean-field model against susceptibility, magnetization and ESR data.}
		\begin{tabularx}{0.6\textwidth}{YY}
			\midrule\midrule
			Parameter & Value \\
			\midrule
			$\lambda$ & $153.26 \pm 0.03$~meV \\
			$\Delta$ & $-77.94 \pm 0.01$~meV \\
			$k$ & $0.9661 \pm 0.0001$ \\
			$J_{\rm MF}$ & $1.71 \pm 0.01$~K \\
			$\chi^0_a$ & $-9.2 \pm 4.8 \times 10^{-5}$~emu/mol \\
			$\chi^0_c$ & $-6.5 \pm 1.2 \times 10^{-4}$~emu/mol \\
			\midrule\midrule
		\end{tabularx}
		\label{tab:SI_fit}
	\end{table*}

	\section{Electron Spin Resonance}
	
\begin{figure}[tbp]
		\includegraphics[scale=1]{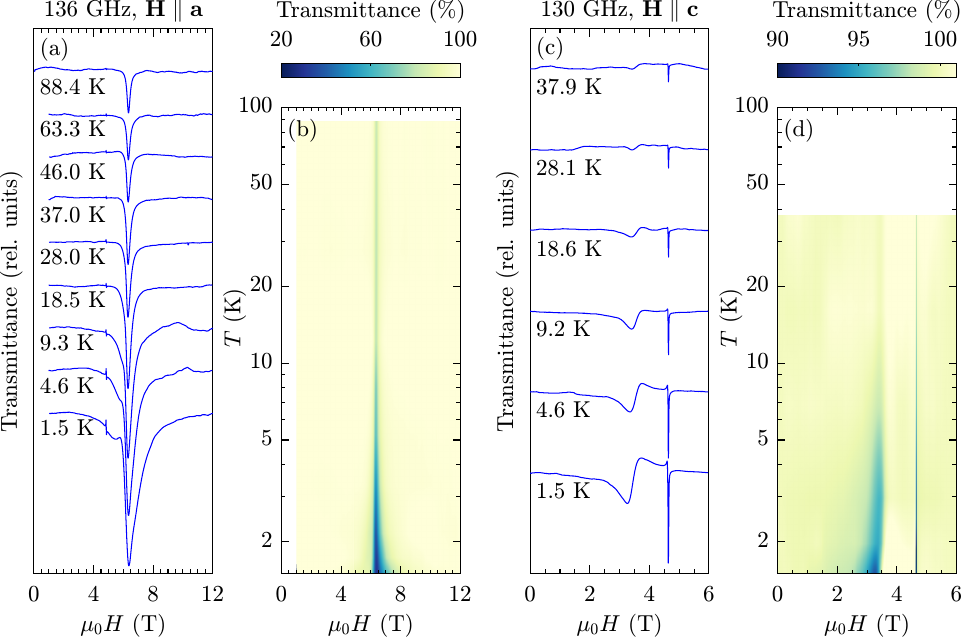}
		\caption{Temperature dependence of electron spin resonance excitations.
		 (a,c) Fixed frequency ESR spectra at various temperatures in $\mathbf{H \parallel a}$ and $\mathbf{H \parallel c}$ experimental configurations.
		 (b,d) False colorplots of the optical transmittance against temperature and magnetic field.}
		\label{fig:ESR_temp}
	\end{figure}

	In Fig.~\ref{fig:ESR_temp} we present additional ESR data, characterizing the temperature dependence of the spectra discussed in the main text.
	For both $\mathbf{H \parallel a}$ and $\mathbf{H \parallel c}$ configurations, the resonance line is visible as a dip in optical transmission, which becomes broader and more intense upon cooling.
	In both cases the Larmor frequency, i.e. the $g$-factor, does not seem to change with temperature.

	\section{Magnetoelastics \& Exchange Striction}
	\label{sec:mag_el}
	
	Let us recall the exchange striction mechanism commonly responsible for magnetoelastic effects near phase transitions in magnetic insulators \cite{linesElasticPropertiesMagnetic1979}.
	Exchange integrals critically depend on the distances between magnetic ions, as well as their angles with the ligands mediating the bonding.
	For small changes, this can be linearized as
	\begin{equation}
		J(|\mathbf{r}_i - \mathbf{r}_j|) = J(|\mathbf{r}^0_{ij}|) + \rho_{ij} \frac{\partial J}{\partial \rho_{ij}} \bigg|_{\rho_{ij}=0} + \mathcal{O}(\rho^2_{ij})
	\end{equation}
	where $\rho_{ij} = |\delta \mathbf{r}_{ij}|/|\mathbf{r}^0_{ij}|$ is a normalized bond displacement induced by some lattice strain $\varepsilon_\Gamma$.
	This allows us to write down a simple "bond-phonon" \cite{pencHalfMagnetizationPlateauStabilized2004, bergmanModelsDegeneracyBreaking2006} magnetoelastic Hamiltonian
	\begin{gather}
		\mathcal{H}_{\rm m-e} = \sum_{\braket{i,j}} \left[ (J_0 - g \rho_{ij}) \mathbf{S}_i \cdot \mathbf{S}_j + \frac{k}{2} \rho_{ij}^2 \right]
	\end{gather}	
	where $k$ is an elastic modulus and the coupling constant $g \equiv - \frac{\partial J}{\partial \rho_{ij}}$ sets the strength of spin-lattice effects.
	In equilibrium we have $\partial \mathcal{H} / \partial \rho_{ij} = 0$, which yields $ \dbraket{\rho_{ij}} = \frac{g}{k} \braket{\mathbf{S}_i \cdot \mathbf{S}_j}$ for the normalized changes in bond length.
	As spin correlations develop below $T_{\rm N}$, each bond acquires an additional stiffness contribution $\propto \braket{\mathbf{S}_i \cdot \mathbf{S}_j}$ and the lattice contracts to accommodate this.
	In principle one can "integrate out" the phonon terms in the Hamiltonian, which produces an effective bi-quadratic term $ \sim b (\mathbf{S}_i \cdot \mathbf{S}_j)^2$ in the spin model where $b_{\rm eff} = -g^2/k$.
	
	\begin{table}[h!]
		\centering
		\begin{tabular}{cccc}
			\hline
			\textbf{Mode} & $\delta r_{12}$ & $\delta r_{13}$ & $\delta r_{23}$ \\
			\hline
			A$_1$ & $-1$ & $-1$ & $-1$ \\
			E(1) & $1$ & $-\frac{1}{2}$ & $-\frac{1}{2}$ \\
			E(2) & $0$ & $\frac{\sqrt{3}}{2}$ & $-\frac{\sqrt{3}}{2}$ \\
			\hline
		\end{tabular}
		\caption{Bond elongation $\delta r_{ij}$ for $D_3$ symmetry strains (i.e. vibrational modes) A$_1$ and E of an equilateral triangle.}
	\end{table}
	
	\subsection{Estimating the Coupling Constants}

	Based on the lattice strains $\Delta l/l$ determined in zero field, we attempt to estimate the magnetoelastic coupling constants.
	To check which type of distortions may be realized, we look at the symmetry strains, irreps of the fundamental vibrational modes that may deform the crystal \cite{tchernyshyovOrderDistortionString2002a}.
	Restricting ourselves to the case of $\mathbf{k} = 0$ distortions, the problem simplifies to that of the $D_3$ point group for an equilateral triangle.
	There are two relevant irreps: The trivial A$_1$ "breathing" mode, expanding/contracting the entire triangle, and a two-fold degenerate E mode allowing e.g. for a $J-J'$ isosceles distortion.
	
	The simple A$_1$ mode should be always active, since the magnetic energy depends linearly on the strain $\varepsilon_{\rm A_1}$, e.g.
	\begin{equation}
		E_{\rm mag}(\varepsilon) \approx -\frac{3}{2} (J - g_{\rm A_1} \dbraket{\varepsilon_{\rm A_1}}) S^2 + \mathcal{O}(\varepsilon^2)
	\end{equation}
	for a classical Heisenberg model on the triangular lattice.
	As for the "real" low-temperature structure, we take $\braket{\mathbf{S}_i \cdot \mathbf{S}_j} \approx -\frac{1}{2} \dbraket{S}^2$ with a reduced ordered moment $\dbraket{S} \approx 0.21$ \cite{capriottiLongRangeNeelOrder1999, zhengExcitationSpectraSpin$frac12$2006, whiteNeelOrderSquare2007a} and ignore the small incommensurability (the angles between adjacent spins remain close to $120^\circ$).
	The relevant stiffness constant is $k = \frac{V_0(c_{11} + c_{12})}{2}$, where $V_0$ is the unit cell volume, the longitudinal sound velocity $v_{11} = \sqrt{c_{11}/\rho} \simeq 2.1$~km/s is determined from ultrasound and $\rho$ represents the mass density.
	Here we make the rough assumption $k \sim \frac{3V_0 c_{11}}{4}$ based on the elastic stability conditions $0 < c_{12} < c_{11}$ for hexagonal crystals \cite{mouhatNecessarySufficientElastic2014}.
	Given the strain $\dbraket{\varepsilon_{\rm A_1}} \equiv \Delta l/l \sim -5 \times 10^{-6}$ observed in our zero field thermal expansion data, we estimate the spin-lattice coupling constant
	\begin{equation}
		g_{\rm A_1} = \frac{k \Delta l/l}{\dbraket{\mathbf{S}_i \cdot \mathbf{S}_j}} \sim 40~{\rm K}.
	\end{equation}

	\subsection{$J - J'$ Magnetoelastic Distortion}
	
	While our x-ray data confirm the trigonal crystal structure down to $T_{\rm N}$, they cannot rule out the possibility of a "dynamic" $J-J'$ isosceles distortion coupled to the magnetic order parameter.
	Such a monoclinic lattice distortion would correspond to the E(1) vibrational mode and may coexist with A$_1$ below $T_{\rm N2}$ on a similar scale $\lesssim 10^{-6}$.
	Starting from a simple Heisenberg picture, the triangular lattice would be deformed into inequivalent "chain" and "zig-zag" couplings with renormalized exchanges $J_0 \rightarrow J = J_0(1 - \epsilon)$ and $J_0 \rightarrow J' = J_0(1 + \frac{\epsilon}{2})$ respectively (here we define $\epsilon \equiv g_{\rm E_1} \dbraket{\varepsilon_{\rm E_1}} / J_0$ in the above exchange-striction framework).
	This affects both the classical propagation vector and the ground state energy, resulting in
	\begin{gather}
		|\mathbf{q}(\epsilon)| = 2\cos^{-1} \left( -\frac{J'}{2J} \right) = 
		2\cos^{-1} \left( -\frac{1 + \epsilon/2}{2(1 - \epsilon)} \right) = \frac{4\pi}{3} + \sqrt{3}\epsilon + \mathcal{O}(\epsilon^2) \\
		E_{\rm mag}(\epsilon) = J\cos(q) + 2J' \cos \left( \frac{q}{2} \right) = -\frac{3}{2} J_0 S^2 - \frac{9}{8} J_0 S^2 \epsilon^2 + \mathcal{O}(\epsilon^3).
	\end{gather}
	Qualitatively, this approach can explain the IC peak positions observed in zero field as due to three structural domains with an underlying spiral ground state.
	However, there are several problems with such a theory.
	First off, a lattice distortion $\dbraket{\varepsilon_{\rm E(1)}} \sim 10^{-6}$ consistent with experiment would require a \textit{huge} coupling constant $g_{\rm E(1)} \sim 10^3 g_{\rm A_1}$ to explain the IC shift $\delta q/q = 3 \sqrt{3} g_{\rm E(1)} \dbraket{\varepsilon_{\rm E(1)}} / 4\pi J_0 \sim 2\%$ of magnetic Bragg peaks observed experimentally.
	This in turn would necessitate a stiffness constant $k_{\rm E(1)} = \frac{v(c_{11} - c_{12})}{2}$ three orders of magnitude smaller, i.e. $c_{11} - c_{12} \sim 10^{-3} c_{11}$.
	What's even more problematic is that the classical magnetic energy gain from the distortion is \textit{quadratic} in the strain at lowest order, not linear.
	The latter is directly competing against the cost in elastic energy (also quadratic), which means an E-type distortion is generally not expected to appear.
	
	\begin{figure}[tbp]
		\includegraphics[scale=1]{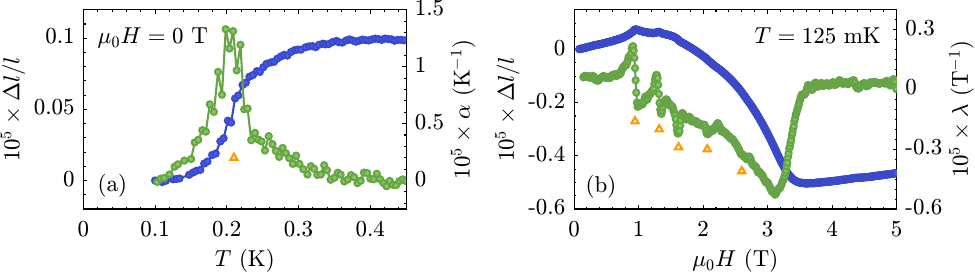}
		\caption{Magnetoelastic effects in \dmnrc perpendicular to the triangular planes.
		(a) Temperature dependent sample dilation $\Delta l/l$ along the $c$-axis, together with its numerical derivative, the thermal expansion coefficient $\alpha = \frac{1}{l} \frac{\partial \Delta l}{\partial T}$.
		(b) Changes in sample length in a magnetic field $\mathbf{H \parallel c}$, along with the magnetostriction coefficient $\lambda = \frac{1}{\mu_0 l} \frac{\partial \Delta l}{\partial H}$.
		Orange triangles in both panels denote the anomalies in $\alpha(T)$ and $\lambda(H)$, in agreement with the phase transitions discussed in the main text.}
		\label{fig:Striction}
	\end{figure}
	
	\subsection{Dilatometry in $\mathbf{E \parallel c}$ Geometry}

	In Fig.~\ref{fig:Striction} we present additional dilatometry data in the $\mathbf{H \parallel E \parallel c}$ configuration, showing the length changes \textit{perpendicular} to the triangular planes.
	We observe the same phase transitions discussed in the main text, both against temoerature (thermal expansion) and magnetic field (magnetostriction).
	Note that the overall scale of these magnetoelastic effects is roughly an order of magnitude weaker than within the triangular plane, as expected for a quasi-2D system.

	\section{Inelastic Neutron Scattering}
	
	\begin{figure}[tbp]
		\includegraphics[scale=1]{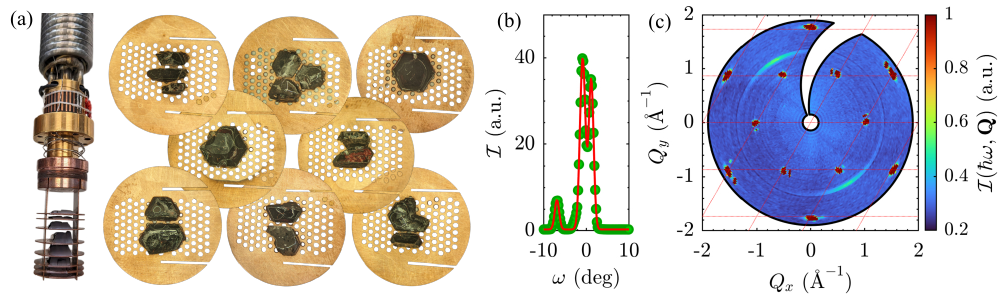}
		\caption{(a) The probe used for neutron spectroscopy, containing 16 single crystals coaligned in the $ab$ scattering plane.
		(b) A cut through the $\mathbf{Q} = (1,0,0)$ nuclear bragg peak.
		(c) Elastic neutron scattering intensity in the $(hk0)$ plane taken at $T \simeq 1$~K. The powder line visible at $|\mathbf{Q}| \sim 1.3$~\AA$^{-1}$ originates from teflon.}
		\label{fig:INS}
	\end{figure}
	
	\begin{figure}[tbp]
		\includegraphics[scale=1]{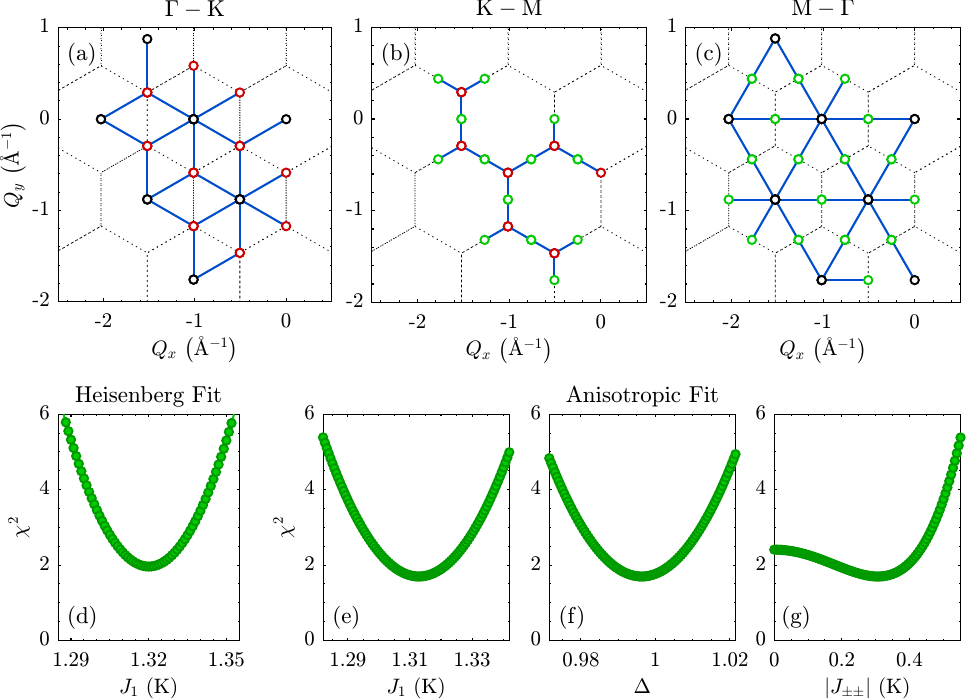}
		\caption{(a-c) Schematic showing the equivalent reciprocal space paths along $\Gamma$ - K, K - M and M - $\Gamma$, used to plot the averaged excitation spectrum in the main text. Brillouin zone boundaries are indicated as black dotted lines.
		(d) $\chi^2$ goodness-of-fit metric of the extracted dispersion relation to a nearest-neighbor Heisenberg model on the 2D triangular lattice.
		(e-g) Analogous $\chi^2$ plots for the general anisotropic spin model of Eq.~\ref{eq:Hamilt}.}
		\label{fig:INS_fits}
	\end{figure}

	In the following, we provide further details about the sample and fitting procedure employed for the neutron spectroscopy experiment carried out at THALES (ILL).
	Fig.~\ref{fig:INS}(a) shows a picture of the coaligned probe, composed of 16 crystals fixed to a copper holder with cytop glue and teflon.
	The total sample mass amounts to $m = 1.9$~g.
	The quality of the alignment can be seen in Fig.~\ref{fig:INS}(b,c), with a sample mosaic of roughly $3^\circ$ FWHM.
	One can also spot a small crystallite, slightly misaligned by $6.9^\circ$ in the $(hk0)$ plane and contributing about $7(2)\%$ to the total scattering intensity.
	
	To reconstruct the spin-wave dispersion relation presented in the main text, we masked out spurious intensity from inelastically re-scattered Bragg peaks ("Currat-Axe spurions"), subtracted a constant background and normalized the data by the magnetic form factor.
	We then averaged over the various equivalent reciprocal space paths presented in Fig.~\ref{fig:INS_fits}(a-c) and performed Gaussian fits to constant-$\mathbf{Q}$ cuts in order to extract the dispersion points $\hbar\omega_\mathbf{Q}$.
	
	These observations are compared to the magnon dispersion of a fully polarized ferromagnetic state in a least squared fitting procedure.
	There is only one branch (propagation vector $\mathbf{q} = \Gamma$) and spin-wave theory is exact \cite{coldeaDirectMeasurementSpin2002}, making this the perfect regime to extract exchange parameters.
	For longitudinal fields $\mathbf{H \parallel c}$, the magnon dispersion for the general anisotropic spin Hamiltonian of Eq.~\ref{eq:Hamilt} is given as
	\begin{gather}
		\hbar\omega(\mathbf{Q}) = 2S\sqrt{A_\mathbf{Q}^2 - |B_\mathbf{Q}|^2} \label{eq:SW_disp} \\ \nonumber
		A_\mathbf{Q} = g_c \mu_B B_z - 3J\Delta + J\gamma_1(\mathbf{Q}) \\ \nonumber
		B_\mathbf{Q} = J_{\pm\pm} \left( \gamma_2(\mathbf{Q}) + \sqrt{3}i [\cos(2\pi k) - \cos(2\pi (h+k))] \right),
	\end{gather}
	where
	\begin{gather}
		\gamma_1(\mathbf{Q}) = \cos(2\pi h) + \cos(2\pi k) + \cos(2\pi(h+k)) \\ \nonumber
		\gamma_2(\mathbf{Q}) = -2\cos(2\pi h) + \cos(2\pi k) + \cos(2\pi(h+k)).
	\end{gather}
	In this geometry, $J_{z\pm}$ is absent entirely from the spin-wave dispersion. Furthermore, at high fields we have $|A_\mathbf{Q}| \gg |B_\mathbf{Q}|$, i.e. $\hbar\omega(\mathbf{Q}) \approx 2S|A_\mathbf{Q}|$ and we are mainly sensitive to the XXZ-type couplings with only a weak dependence on $|J_{\pm\pm}|$.
	In the Heisenberg limit ($\Delta = 1, J_{\pm\pm} = J_{z\pm} = 0$), Eq.~\ref{eq:SW_disp} simplifies to
	\begin{equation}
		\hbar\omega(\mathbf{Q}) = 2S \left( g_c \mu_B B_z - 3J + J\gamma_1(\mathbf{Q}) \right).
	\end{equation}
	The $\chi^2$ goodness-of-fit metrics, obtained by fitting the experimental dispersion relation to the Heisenberg and the anisotropic model, are depicted in Fig.~\ref{fig:INS_fits}(d-g).

	\section{Commensurate Magnetic Structures}

	Here we discuss the nature of the commensurate magnetic structures in \dmnrc.
	Their essential physics can be understood through a Heisenberg model on the triangular lattice, with the addition of a small AFM inter-plane coupling $J_c > 0$ \cite{yamamotoMicroscopicModelCalculations2015}.
	Given the observed propagation vector $\mathbf{q} = (1/3,1/3,1/2)$, we find the ground state by considering all possible six-sublattice spin configurations and resorting to semi-classical energy minimization.
	At $T = 0$, the classical degeneracy is lifted only by quantum fluctuations (i.e. order-by-disorder selection), favoring the "most collinear" states with the softest excitation spectrum \cite{villainOrderEffectDisorder1980, chubukovQuantumTheoryAntiferromagnet1991a}.
	On a phenomenological level, one can emulate this by adding a small bi-quadratic coupling $\sim b(\mathbf{S}_i \cdot \mathbf{S}_j)^2$ to the Hamiltonian \cite{heinilaSelectionGroundState1993c, grisetDeformedTriangularLattice2011, koutroulakisQuantumPhaseDiagram2015}. 
	In fact, such a term can be derived rigorously through real space perturbation theory (RSPT), by calculating the leading order corrections to the "mean field" (i.e. classical) ground state energy. For the triangular lattice Heisenberg antiferromagnet, the second order RSPT correction is given as \cite{zhitomirskyRealspacePerturbationTheory2015}
	\begin{equation}
		\delta E_{2} = \sum_n \frac{\bra{0} \hat{V} \ket{n} \bra{n} \hat{V} \ket{0}}{E_0 - E_n} = \\
		-\frac{J^2 S^2}{8 H_{\rm loc}} \sum_{\braket{i,j}} (1 - \cos \theta_{ij})^2
	\end{equation}
	where $H_{\rm loc} = 3JS$ is the local field acting on each spin, resulting in $b_{\rm eff} \equiv -J/(24S^3)$. Since this energy correction amounts to an effective Hamiltonian operating on the ground state manifold of classical vector spins $\mathbf{n}_i = \mathbf{S}_i/S$, the bi-quadratic term $\sim \cos^2\theta_{ij} = (\mathbf{n}_i \cdot \mathbf{n}_j)^2$ can show up even for $S=1/2$ quantum systems.
	
\begin{figure}[tbp]
		\includegraphics[scale=1]{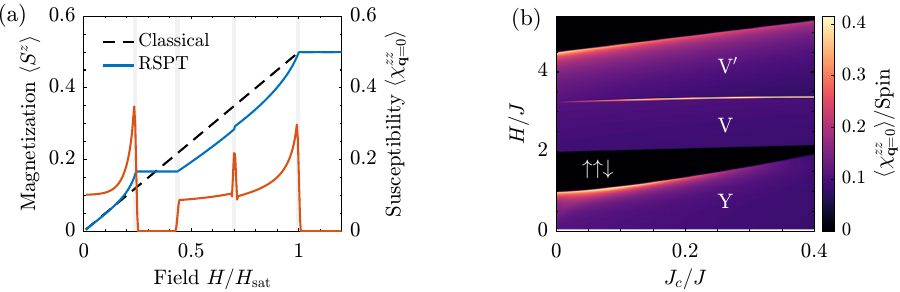}
		\caption{Effects of inter-plane coupling on the magnetic structures for the Heisenberg triangular lattice antiferromagnet.
		(a) Magnetization curve and its numerical derivative for $J_c = 0.1J$, obtained from semi-classical model discussed in the text.
		(b) $T = 0$ ground state phase diagram in the same approach, visualized as false color plot of the uniform magnetic susceptibility $\chi(J_c,H)$.
		Both panels clearly show the presence of an additional phase transition between the V and V$^\prime$ states.}
		\label{fig:Interplane}
	\end{figure}

	We find the optimal six-sublattice ground state by minimizing the following semi-classical Hamiltonian
	\begin{gather}
		\mathcal{H} = 
		\sum_{\langle i,j \rangle \in \Delta} \left[ J \mathbf{S}_i \cdot \mathbf{S}_j + b_{\rm eff} (S^2 - \mathbf{S}_i \cdot \mathbf{S}_j)^2 \right] +
		J_c \sum_{\langle i,j \rangle \parallel c} \mathbf{S}_i \cdot \mathbf{S}_j - 
		\sum_i \mathbf{S}_i \cdot \mathbf{H}
	\end{gather}
	as a function of the inter-plane coupling $J_c$ and magnetic field $H$, resulting in the spin configurations discussed in the main text.
	The phase diagram, as well as magnetization curves for a reasonable parameter choice $J_c = 0.1J$, are shown in Fig.~\ref{fig:Interplane}.
	There are two differences compared to the pure 2D model:
	Firstly, the inter-plane term becomes frustrated by the field, resulting in a mixed FM-AFM stacking with both $\mathbf{q} = (1/3,1/3,1/2)$ and $2\mathbf{q} = (1/3,1/3,0)$ Fourier components.
	Furthermore, the stacking preferred just above the plateau differs from the one close to saturation, which gives rise to a first-order transition around $H_c \approx 0.7 H_{\rm sat}$, above/below which the choice of minority spin is uniform/staggered between layers.
	The evolution between phases Y - $\uparrow\uparrow\downarrow$ - V - V$^\prime$ appears to be highly robust with respect to $J_c$, until the collinear magnetization plateau eventually becomes suppressed for large $J_c \gtrsim 0.5 J$.
	
	Finally, we remark on the effects of other perturbing terms on these magnetic structures.
	The presence of several IC phases in \dmnrc clearly points to significant bond-dependent interactions or magnetoelastic effects.
	But towards the center of the phase diagram, the energy gain through order-by-disorder is maximized, surpassing those other contributions and leading to a lock-in to the semi-classical commensurate states discussed above \cite{chenGroundStatesSpin12013, grisetDeformedTriangularLattice2011}.
	In absence of a lattice distortion, magnetoelastic effects can only contribute a tiny bi-quadratic term ($\sim 20$ times smaller than the one obtained from RSPT, see Sec.~\ref{sec:mag_el}) and should be irrelevant.
	The bond anisotropies would favor collinear four-sublattice stripe order \cite{maksimovAnisotropicExchangeMagnetsTriangular2019}.
	Below some critical exchange ratio, their main effect would be to renormalize the critical fields $H_c$ and break the $U(1)$ rotational symmetry, pinning the staggered in-plane ordered moment to some discrete orientation.

	\section{$\mathbb{Z}_2$ Vortex Crystal Phase}
	
\begin{figure}[tbp]
		\includegraphics[scale=1]{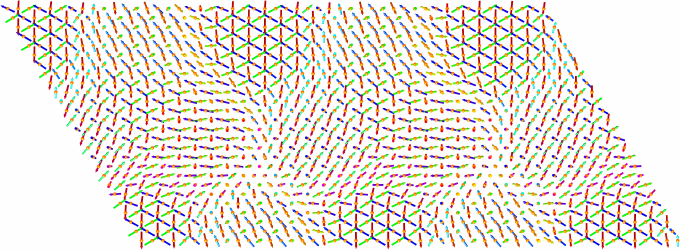}
		\caption{Sketch of the 2D spin configuration for a $\mathbb{Z}_2$ vortex crystal in the Heisenberg-Kitaev model, with parameters $K/J_0 = -0.3$.
		The $hsv$-color of each spin indicates its rotation angle in the triangular plane.
		}
		\label{fig:Vortex}
	\end{figure}

	For sake of completeness, we reproduce here the expected 2D spin configuration for the $\mathbb{Z}_2$ vortex crystal phase \cite{rousochatzakisKitaevAnisotropyInduces2016, beckerSpinorbitPhysicsJ12015, liCollectiveSpinDynamics2019}.
	Taking the cubic axes, one can equivalently re-write the spin Hamiltonian of Eq.~\ref{eq:Hamilt} as a $J_0 - K - \Gamma - \Gamma'$ model \cite{maksimovAnisotropicExchangeMagnetsTriangular2019}.
	The vortex crystal has been predicted to exist in the Heisenberg-Kitaev limit (i.e. $\Gamma = \Gamma' = 0$), where 
	\begin{gather}
		\mathcal{H} = J_0 \sum_{\langle i,j \rangle} \mathbf{S}_i \cdot \mathbf{S}_j + 
					  K \sum_{\gamma \parallel \langle i,j \rangle} S^\gamma_i S^\gamma_j
	\end{gather}
	and the coordinate transformation amounts to $J_0 = J + 6J_{\pm\pm}, K = -6J_{\pm\pm}$ with the constraints $\Delta = 1$ (confirmed experimentally in \dmnrc) and $J_{z\pm} = 2\sqrt{2}J_{\pm\pm}$ (yet to be determined).
	In momentum space, the classical HK-Hamiltonian has three eigenvalues at
	\begin{gather}
		\mathcal{J}^{\gamma\gamma} (\mathbf{q}) = K\cos(\mathbf{q \cdot \hat{e}_\gamma}) + 
		J \sum_{\alpha = x,y,z} \cos(\mathbf{q \cdot \hat{e}_\alpha}),
	\end{gather}
	i.e. each spin component $\gamma \in (x,y,z)$ prefers a different ordering vector.
	We obtain three sets of incommensurate positions $\pm\mathbf{q}^\gamma$, along the $\Gamma$-K direction in momentum space. These minima move towards the $\Gamma$ (M)-point for positive (negative) $K$, where
	\begin{equation}
		|\mathbf{q}| = 1 - \frac{1}{\pi} \cos^{-1} \left( \frac{J}{2(J + K)} \right)
		\quad \implies \quad |\delta q| \approx \frac{1}{\sqrt{3}\pi} \frac{|K|}{|J + \frac{7}{6} K|}.
	\end{equation}	
	A simple superposition of three such incommensurate spirals will not be a good ground state - the moment size would be strongly modulated.
	Instead the system would turn anharmonic, introducing additional Fourier components to retain a semi-classical equal moment structure.
	The resulting ground state preserves threefold symmetry and locally mimics the "parent" $120^\circ$ structure, but exhibits a superlattice of $\mathbb{Z}_2$ vortices on larger length scales.
	Up to second harmonics, the magnetic structure may be approximated as
	\begin{equation}
		S^\gamma(\mathbf{r}) = \frac{4S}{3\sqrt{3}} {\rm Re} \left[ e^{i [\mathbf{q_{\rm \gamma} \cdot r} + \phi]}  + 
		\frac{1}{4} \sum_{\alpha \neq \gamma} e^{i [\mathbf{(q_{\rm \gamma} + q_{\rm \alpha}) \cdot r} + \phi]} \right] \label{eq:VC_config}
	\end{equation}
	with a phase shift $\phi = 2\pi n/3$ between sublattices, visualized in Fig.~\ref{fig:Vortex}.
	
	\bibliography{Bibliography_npj25}